%% file: pkpm-theory.tex
\documentclass{jpp}

\usepackage{graphicx}
\usepackage[makeroom]{cancel}
\usepackage{comment}
\usepackage{natbib}
\usepackage{subcaption}
\usepackage{amsfonts}
\usepackage{amsmath}
\usepackage{amssymb}
\usepackage{color}
\usepackage{graphicx}
\usepackage{hyperref}
\usepackage{physics}
\usepackage{sidecap}
\usepackage{tabularx}
\usepackage{tensor}
\usepackage{xcolor}
\usepackage{tikz}
\usepackage{color}
\usepackage{bm}
\usepackage{authblk}

\ifCUPmtlplainloaded \else
  \checkfont{eurm10}
  \iffontfound
    \IfFileExists{upmath.sty}
      {\typeout{^^JFound AMS Euler Roman fonts on the system,
                   using the 'upmath' package.^^J}%
       \usepackage{upmath}}
      {\typeout{^^JFound AMS Euler Roman fonts on the system, but you
                   dont seem to have the}%
       \typeout{'upmath' package installed. JPP.cls can take advantage
                 of these fonts, if you use 'upmath' package.^^J}%
      }
  \else
  \fi
\fi

\ifCUPmtlplainloaded \else
  \checkfont{msam10}
  \iffontfound
    \IfFileExists{amssymb.sty}
      {\typeout{^^JFound AMS Symbol fonts on the system, using the
                'amssymb' package.^^J}%
       \usepackage{amssymb}%

      }{}
  \fi
\fi

\ifCUPmtlplainloaded \else
  \checkfont{msam10}
  \iffontfound
    \IfFileExists{amssymb.sty}
      {\typeout{^^JFound AMS Symbol fonts on the system, using the
                'amssymb' package.^^J}%
       \usepackage{amssymb}%

      }{}
  \fi
\fi

\ifCUPmtlplainloaded \else
\IfFileExists{amsbsy.sty}
             {\typeout{^^JFound the 'amsbsy' package on the system, using it.^^J}%
     \usepackage{amsbsy}}
    {\providecommand\boldsymbol[1]{\mbox{\boldmath $##1$}}}
\fi

\newsavebox{\astrutbox}
\sbox{\astrutbox}{\rule[-5pt]{0pt}{20pt}}

\newcommand{\eqr}[1]{Eq.\thinspace(#1)}
\newcommand{\pfrac}[2]{\frac{\partial #1}{\partial #2}}

\newcommand{\mvec}[1]{\mathbf{#1}}
\newcommand{\bmvec}[1]{\breve{\mathbf{#1}}}
\newcommand{\gvec}[1]{\boldsymbol{#1}}

\newcommand{\cbas}[1]{\gvec{\sigma}_{#1}}

\newcommand{\basis}[1]{\mvec{e}_{#1}}
\newcommand{\dbasis}[1]{\mvec{e}^{#1}}

\newcommand{\bbasis}[1]{\breve{\mvec{e}}_{#1}}

\newcommand{\tbasis}[1]{\tilde{\mvec{e}}_{#1}}
\newcommand{\tdbasis}[1]{\tilde{\mvec{e}}^{#1}}

\newcommand{\nbasis}[1]{\hat{\mvec{e}}_{#1}}
\newcommand{\veps}{\gvec{\varepsilon}}
\newcommand{\bdum}{\breve{\_}}
\newcommand{\delx}{\nabla_\mvec{x}}

\newcommand{\delxp}{\nabla_{\mvec{x}'}}
\newcommand{\delvp}{\nabla_{\mvec{v}'}}
\newcommand{\delxpp}{\nabla_{\mvec{x}''}}
\newcommand{\delvpp}{\nabla_{\mvec{v}''}}
\newcommand{\bdf}{\bar{f}}

\newcommand{\buni}{\mvec{b}}

\newcommand{\dvol}[1]{\thinspace d{#1}}

\newcommand{\gke}{\texttt{Gkeyll}}

\newcommand{\vecspace}{\mathcal{V}}

\newcommand{\AH}[1]{{\color{red} #1}}

\title[PKPM Theory]{A Parallel-Kinetic-Perpendicular-Moment Model for Magnetized Plasmas}

\author{James Juno\thanks{Email address
for correspondence: jjuno@pppl.gov}$^{1}$, Ammar Hakim$^{1}$, Jason~M.~TenBarge$^{2}$}
  
\affiliation{$^1$Princeton Plasma Physics Laboratory, Princeton, NJ 08543, USA\\
[\affilskip]
$^2$Department of Astrophysical Sciences, Princeton University, Princeton, NJ 08544, USA
}

\pubyear{2023}
\volume{?}
\pagerange{?}
\date{?; revised ?; accepted ?. - To be entered by editorial office}

\begin{document}
\maketitle

\begin{abstract}
We describe a new model for the study of weakly-collisional, magnetized plasmas derived from exploiting the separation of the dynamics parallel and perpendicular to the magnetic field.
This unique system of equations retains the particle dynamics parallel to the magnetic field while approximating the perpendicular dynamics through a spectral expansion in the perpendicular degrees of freedom, analogous to moment-based fluid approaches.
In so doing, a hybrid approach is obtained which is computationally efficient enough to allow for larger-scale modeling of plasma systems while eliminating a source of difficulty in deriving fluid equations applicable to magnetized plasmas.
We connect this system of equations to historical asymptotic models and discuss advantages and disadvantages of this approach, including the extension of this parallel-kinetic-perpendicular-moment beyond the typical region of validity of these more traditional asymptotic models. 
This paper forms the first of a multi-part series on this new model, covering the theory and derivation, alongside demonstration benchmarks of this approach including shocks and magnetic reconnection. 
\end{abstract}

\begin{PACS}
\end{PACS}

\input{intro}
\input{prelim-spec}
\input{pkpm-model-deriv}

\input{pkpm-history-ctx.tex}
\input{pkpm-demo.tex}
\input{summary.tex}

\section*{Data availability}
The input files for the \gke~ the simulations presented here are available in the following GitHub repository, \url{https://github.com/ammarhakim/gkyl-paper-inp}.
All plots in this paper utilize the \texttt{postgkyl} package \url{https://github.com/ammarhakim/postgkyl}.

\section*{Acknowledgements}
The authors thank the \gke~team for useful discussions and wish to especially thank P. Cagas for assistance in plotting distribution functions from PKPM data. 
J.~Juno acknowledges A. Philippov for insights from relativistic plasmas that motivated the PKPM approach, J.~Chan for helpful conversations on discretizing fluid equations with discontinuous Galerkin methods, and A.~Hoffmann for useful discussions on spectral gyrokinetic solvers and Laguerre resolution. 
This research is part of the Frontera computing project at the Texas Advanced Computing Center. Frontera is made possible by National Science Foundation (NSF) Award OAC-1818253.
\section*{Funding}
J.~Juno, A.~Hakim, J.~M.~TenBarge, and the development of \texttt{Gkeyll} were partly funded by the NSF-CSSI program, Award No. 2209471.
J.~Juno and A.~Hakim were also supported by the U.S. Department of Energy under Contract No. DE-AC02-09CH1146 via LDRD grants.
J.~M.~TenBarge was also supported by NASA grant 80NSSC23K0099.
\section*{Declaration of interests}
The authors report no conflict of interests.

\bibliography{abbrev.bib,pkpm.bib}

\appendix
\input{ap-notation}
\input{ap-lag-poly}
\input{ap-agyro}
\input{ap-lbo}

\end{document}

%% file: intro.tex
\section{Introduction}
Many plasma systems are observed to be in two distinct parameter regimes which can have profound impacts on the dynamics of these plasmas: weakly collisional and magnetized.
In astrophysical systems, plasmas are routinely observed to have plasma densities and temperatures which imply the collisional mean-free-path is large, while the measured or inferred magnetic field strengths imply the particle gyroradius is small. 
Across a diverse array of astrophysical environments: accretion disks around black holes \citep{Quataert:2001,EHT:2019a,EHT:2019b,EHT:2021}, plasma-filled pulsar magnetospheres \citep{Philippov:2022}, the hotter phases of the interstellar \citep{Cox:2005,Humphrey:2011}, circum-galactic \citep*{Tumlinson:2017}, intergalactic \citep{Nicastro:2005}, and intracluster media \citep*{Fabian:1994,Peterson:2006,Kunz:2022}, or the inner \citep{Marsch:2006,Borovsky:2018} and outer heliosphere \citep{Richardson:2022}, low collisionality, highly magnetized plasmas can be found.
Closer to home, a variety of terrestrial plasmas, such as the plasmas confined by diverse fusion reactor concepts \citep*{Spitzer:1958, Furth:1975, Wesson:2011, imbertgerard:2020}, are constructed to be in this parameter regime in which the effects of particle collisions happen on longer time scales and larger length scales than the dynamics of the particles gyrating around the magnetic field.

When the collisional mean-free-path is large and the particle gyroradius is small, the plasma will adjust its dynamics such that motion parallel to the magnetic field is uninhibited compared to motion perpendicular to the magnetic field.
In effect, the plasma has an easier time transporting momentum and heat parallel to the local magnetic field while the transport perpendicular to the field is more tightly constrained.
The particles can traverse large distances before collisions deflect their trajectories off field-lines, while the particle motion perpendicular to the field is mostly limited to the combination of cyclotron motion as the particle gyrates about the local magnetic field and a set of bulk, fluid, drifts which arise due to the local geometry of the magnetic field, such as the magnetic field curvature \citep{Northrop:1961, Northrop:1963}. 

This dichotomy between parallel and perpendicular motion has further dramatic impacts on the plasma's dynamics.
In general, the plasma can develop different temperatures parallel and perpendicular to the magnetic field as a result of these distinct motions parallel and perpendicular to the magnetic field \citep{Gary:2001, Kasper:2002, Hellinger:2006, Schekochihin:2006, Bale:2009, Chen:2016}, and a zoo of micro-scale instabilities can be launched which further modify the plasma's dynamics \citep{Schekochihin:2008a, Rosin:2011, Kunz:2014a}, causing viscous stresses \citep{Kunz:2016, Melville:2016, Squire:2019, Kempski:2019, StOnge:2020, Squire:2023} and limiting thermal conduction \citep{Riquelme:2016, Komarov:2016, RobergClark:2016, RobergClark:2018, Komarov:2018}.
Thus, a proper accounting of this separation between parallel and perpendicular dynamics is required for an accurate treatment of the transport of momentum and heat in many plasma systems.

On one hand, modeling these weakly collisional, magnetized plasmas can be achieved in a straightforward manner with kinetic models that solve the Boltzmann, or Vlasov, equation for the particle dynamics in a six-dimensional phase space.
However, phase-space dynamics are computationally demanding due to both the high-dimensionality of the equation system and the multi-scale nature of kinetic plasma dynamics, which often require a computational model to resolve very small length scales and very fast time scales for stability of the scheme. 
On the other hand, one can attempt to incorporate the modified parallel versus perpendicular dynamics in hydrodynamic models.
While extensions of magnetohydrodynamics to include pressure anisotropy and parallel heat conduction are possible \citep*{CGL:1956, Braginskii:1965}, these approaches suffer from the difficulty all fluid modeling must address: the development of a proper closure which truncates the number of equations being solved and provides an accurate approximation of the dynamics in the higher moments \citep{Grad:1949}.
An accurate treatment of the feedback of micro-scale instabilities on the pressure anisotropy and parallel heat conduction is particularly challenging, and more rigorous fluid models may be computationally prohibitive in their own right due to approximations of these dynamics leading to ``non-local'' models of the fast parallel transport along the magnetic field \citep{Hammett:1990, Hammett:1992, Snyder:1997, Ramos:2003, Ramos:2005} or require some sort of artificial limiting of the instabilities in their feedback on the plasma \citep{Sharma:2006}. 

It is the purpose of this study to demonstrate an alternative approach to previous modeling efforts which exploits the dichotomy between parallel and perpendicular dynamics to construct a new hybrid system of equations, one which remains kinetic parallel to the magnetic field, while reducing the dynamics perpendicular to the field in a fluid-like hierarchy.
This new parallel-kinetic-perpendicular-moment, or PKPM, model has natural connections to a variety of historical asymptotic approaches, including Kulsrud's kinetic magnetohydrodynamics (kinetic MHD or KMHD) \citep{Kulsrud:1964, Kulsrud:1983} and the work of Ramos on finite Larmor radius (FLR) corrections to drift-kinetics \citep{Ramos:2008, Ramos:2016}. 
But, while the connection to these approaches which leverage the magnetized nature of the plasma will be made apparent, we emphasize now that our model is not an asymptotic one involving an expansion in a small parameter or parameters, and instead shares a similar construction to recent work in spectral decompositions of the kinetic equation \citep{Delzanno:2015, Parker:2015, Vencels:2016, Roytershteyn:2018, Koshkarov:2021, Pagliantini:2023, Issan:2024}, especially approaches which combine spectral expansions in a subset of the velocity degrees of freedom \citep{Schween:2024, Schween:2025}. 
In particular, the PKPM model performs a spectral expansion of the distribution function in the perpendicular degrees of freedom, reducing the six-dimensional Vlasov equation to a set of four-dimensional equations. 
The exact number of four-dimensional equations is set by the desired physics fidelity perpendicular to the magnetic field. 

This reduction of the dynamics perpendicular to the field from the spectral expansion gives two key benefits.
Firstly, in many magnetized plasma environments, the plasma remains mostly gyrotropic as the plasma evolves \citep[see, e.g.,][]{Marsch:1982, Marsch:2006}.
If the plasma remains gyrotropic, the distribution function may be well approximated by a small number of spectral coefficients in the perpendicular degrees of freedom, so that significant computational gains can be realized by reducing the six-dimensional Vlasov equation to just a few four-dimensional kinetic equations.
Of course, a magnetized plasma can develop a certain amount of agyrotropy due to cyclotron resonances and other finite Larmor radius effects, but the framework in which a plasma is gyrotropic with agyrotropic corrections is the exact framework in which spectral expansions excel.
There have been a number of demonstrations of spectral decompositions achieving high accuracy with coarse resolution in cases when the deviations from, e.g., a Maxwellian using a Hermite expansion, are small \citep{Roytershteyn:2019, Vega:2023}, and spectral decompositions are quickly becoming one of the most efficient ways to solve classic asymptotic models of magnetized plasma dynamics such as gyrokinetics as applied to the tokamak core \citep{Mandell:2018, Frei:2023a, Frei:2023b, Hoffmann:2023a, Hoffmann:2023b, Mandell:2024} and edge \citep{Frei:2020, Frei:2024}.

Secondly, in cases where we expect such spectral expansions to fail, such as the presence of sharp gradients in phase space, we can instead choose a different numerical method designed to handle complex phase-space structures because we have specifically derived the PKPM model to only expand the distribution function in the perpendicular degrees of freedom.
In particular, parallel to the magnetic field we can expect the plasma to develop a variety of non-trivial phase-space structures which are typically difficult to represent numerically with a spectral basis such as trapped-passing boundaries, beams, and power-law tails, but which can be easily handled with previous developments in discontinuous Galerkin methods for kinetic equations \citep{Juno:2018, HakimJuno:2020, Mandell:2020}. 
Of course, these sorts of phase-space structures can arise perpendicular to the magnetic field, such as the loss-cone in the perpendicular degrees of freedom of a magnetic mirror, but these kinetic structures are so ubiquitous parallel to the local magnetic field, from field-aligned beams in collisionless shocks to parallel energization mechanisms in collisionless magnetic reconnection, we are well motivated to take a hybrid approach which can resolve the plasma's evolution in phase space with as few degrees of freedom as possible. 
Thus, the reduction in resolution requirements perpendicular to the field, combined with a numerical method optimized to handle the plasma's phase-space dynamics parallel to the field, generates a unique, computationally efficient model for describing a diverse array of plasmas in weakly collisional, magnetized regimes.

The rest of the paper is organized as follows. In Section~\ref{sec:prelim}, we detail the necessary prerequisites for understanding the PKPM derivation, including a discussion of spectral expansions and coordinate transformation, and difficulties typically encountered when constructing spectral expansions that motivate the particular form of the PKPM model. 
In Section~\ref{sec:pkpm}, we derive the PKPM model through a series of coordinate transformations, first to the local fluid flow frame and then to a coordinate system aligned with the local magnetic field, followed by a spectral expansion in the perpendicular degrees of freedom. 
We discuss a number of historical connections in Section~\ref{sec:history} and attempt to contextualize the physics content of the PKPM model in comparison to other models applied to weakly collisional, magnetized plasmas while also drawing attention to unique features of the model compared to asymptotic models and other approaches which leverage spectral expansions of the kinetic equation.
Most importantly, Section~\ref{sec:history}'s principal goal is to give context to the reader for why the PKPM model has been formulated the way it \AH{has} been: a model which contains the same physics as many of the historical magnetized, collisionless plasma models, but which ameliorates many numerical difficulties which prevented easily solving some of these models with a computer, in addition to other breakthroughs in the spectral expansion formulation, such as the inclusion of the normalization of the perpendicular spectral coefficients without the need for auxiliary equations.
We then give a brief demonstration of the model in Section~\ref{sec:demo} with test cases beyond the regime of applicability for traditional asymptotic, magnetized plasma models, including a parallel, electrostatic collisionless shock and moderate-guide field magnetic reconnection.
Finally, we conclude the discussion of the theory of the PKPM model in Section~\ref{sec:conclusions}.

This paper forms the first of a multi-part series, with the principal focus of this paper being the motivation, theory, properties, and historical context of this new system of equations.
We defer an in-depth description of the numerical discretization of the PKPM model to a future paper, both to keep focus on the details of the continuous system of equations and its properties, and because of the complexities which arise in the discrete system to maintain the properties of the continuous system of equations.

%% file: prelim-spec.tex
\section{Preliminaries on Spectral Expansions and Coordinate Transformations}\label{sec:prelim}

Consider the non-relativistic Vlasov equation, 
\begin{align}
    \pfrac{f}{t} + \mvec{v} \cdot \nabla_{\mvec{x}} f + \frac{q}{m} (\mvec{E} + \mvec{v} \times \mvec{B}) \cdot \nabla_{\mvec{v}} f = C[f], \label{eq:nrvlasov}
\end{align}
where $f = f(\mvec{x}, \mvec{v}, t)$ is the particle distribution function for the plasma species, $q$ and $m$ are the charge and mass of the plasma species respectively, $\mvec{E} = \mvec{E}(\mvec{x}, t)$ and $\mvec{B} = \mvec{B}(\mvec{x}, t)$ are the electric and magnetic fields respectively, and $C[f]$ is a general operator for modeling discrete particle effects, such as inter-particle collisions \footnote{We note that elsewhere in the literature, the inclusion of the collision operator for modeling discrete particle effects changes the name of this equation to the Boltzmann equation or the Vlasov-Fokker-Planck equation depending on the form of the collision operator---see \citet{Henon:1982} for a discussion of this linguistic history.}.
Combined with Maxwell's equations, the Vlasov-Maxwell system of equations provides a near first-principles description of the vast majority of plasma systems in the universe.
However, this equation system is a formidable nonlinear system of equations in a six-dimensional, position-velocity, phase space, and direct modeling of this equation system not only requires handling the high-dimensional nature of the Vlasov equation, but also the vast array of spatial and temporal scales contained in said equation.
Thus, there is a long history of exploring reductions and simplification to the Vlasov equation, both analytically and numerically, to make the study of plasmas more tractable.

One approach to constructing such approximations to the Vlasov equation is to assume the solutions are a small deviation from some \emph{symmetry} or near-symmetry. 
For example, in a highly collisional plasma where the effects of $C[f]$ on the right-hand-side of the Vlasov equation are dominant, the plasma will be close to thermodynamics equilibrium, and the distribution function will deviate only slightly from a Maxwellian. 
In this limit, one can use fluid equations to evolve the lower, thermodynamic moments, i.e., number density, momentum, and pressure, constructing closure relations to determine the higher moments by asymptotic analysis. 
This approach leads to the well-known Braginskii equations \citep{Braginskii:1965}, in which non-thermodynamic moments are expressed in terms of thermodynamic moments and their gradients, and in the magnetized limit, the direction of the local magnetic field. 
The Braginskii equations, or forms of them, are implemented in major codes both for astrophysical \citep{Hopkins:2016, Berlok:2019, Stone:2020} and fusion applications \citep{Sovinec:2004, Gunter:2004, Ferraro:2006, Gunter:2007, Breslau:2009, Meier:2010}.
The generalization of this approach beyond the near-equilibrium approximation, in which a larger set of fluid moments is evolved and then some truncation of the fluid approximation is employed, is well-trod ground \citep{Hakim:2006, Hakim:2008, Wang:2015, Miller:2016, AllmannRahn:2018, Ng:2018, Wang:2020, Ng:2020, AllmannRahn:2024, Kuldinow:2024a} but runs into the classical closure problem of expressing the highest moment in the fluid truncation in terms of the lower fluid moments \citep{Grad:1949}.
Along these lines of inquiry, there is a rich and growing history of leveraging the connection between spectral representation of the distribution function in Hermite moments and the fluid hierarchy \citep{Holloway:1996, Delzanno:2015, Parker:2015, Vencels:2016}.
This approach discretizes the Vlasov equation directly with Hermite moments of some order, in effect creating a more general fluid hierarchy which converges to the kinetic description and achieving a computationally efficient approach for problems where only a few Hermite moments are needed to represent the solution. 

Alternative asymptotic approaches which seek to directly reduce the kinetic equation instead of expressing the kinetic equation as a hierarchy of fluid moments are also a mature approach to approximating the kinetic dynamics of a plasma.
In the limit the plasma is magnetized, one can obtain a variety of kinetic models which reduce the full six-dimensional phase-space dynamics, such as kinetic MHD \citep{Kulsrud:1964, Kulsrud:1983} and other flavors of drift-kinetics \citep{Frieman:1966, HintonWong:1985, Ramos:2008, Ramos:2016}, and gyrokinetics \citep{Antonsen:1980, Catto:1981, FriemanChen:1982, BrizardHahm:2007, Cary:2009}. 
All of these approaches average over the fast cyclotron motion to produce a five-dimensional kinetic equation, with gyrokinetics specifically considering an ordering in which FLR effects on the fluctuating quantities are retained and thus one can think of reducing the evolution of the particular particle motion to the evolution of ``rings'' of charge.

In particular, gyrokinetic theory and the corresponding numerical discretizations of the gyrokinetic equation are one of the major breakthroughs in plasma physics in the last few decades \citep{Krommes:2012}, and this breakthrough forms the backbone of modern research in a variety of contexts, including astrophysical plasmas \citep{Howes:2006, Schekochihin:2009} and magnetized fusion and turbulence theory \citep{Abel:2013}.
We draw particular attention to the numerical discretization of gyrokinetics in recent works, which has likewise leveraged the connection between spectral representations of the kinetic equation and fluid hierarchies.
For the case of gyrokinetics, a Hermite-Laguerre representation of the kinetic equation is directly connected to the gyrofluid moments of the gyrokinetic equation \citep{Dorland:1993}. 
Similar to the works discretizing the Vlasov equation directly with Hermite moments, these spectral expansions have created very computationally efficient methods for solving the gyrokinetic equation \citep{Mandell:2018, Frei:2020, Hoffmann:2023a, Frei:2024}, as well as other magnetized models of plasma dynamics, such as the Kinetic Reduced Electron Heating Model (KREHM) \citep{Zocco:2011, Loureiro:2013, Zocco:2015, Loureiro:2016} and drift-kinetics \citep{Parker:2016}. 

The success of spectral methods along with the perspective afforded by the theoretical foundations of magnetized asymptotic models such as kinetic MHD, drift-kinetics, and gyrokinetics provides a convenient lens for establishing the basis for the parallel-kinetic-perpendicular-moment model.
Consider the velocity space coordinate system employed in these magnetized plasma models:
\begin{align}
    \mvec{v} = v_\parallel \mvec{b}(\mvec{x}, t) + v_\perp \cos\theta {\gvec{\tau}}_1 (\mvec{x}, t) + v_\perp \sin\theta {\gvec{\tau}}_2 (\mvec{x}, t), \label{eq:vparvperp}
\end{align}
a cylindrical coordinate system, where the unit-vector $\mvec{b}$ is the axial direction aligned with the local magnetic field, the unit-vectors ${\gvec{\tau}}_1, {\gvec{\tau}}_2$ define the plane perpendicular to the local magnetic field, and $\theta$ is the velocity gyroangle.
Before proceeding further, we make two important, related, points on nomenclature. 
The first is that this velocity coordinate system is also employed by magnetized models such as gyrokinetics, but in magnetized models such as gyrokinetics, there is a further configuration space transformation to gyrocenter coordinates. 
We emphasize that we are not transforming configuration space here and are maintaining the configuration space coordinates to be the particle position\footnote{Note that by particle position, we mean position in an Eulerian sense; we will focus exclusively on distribution function dynamics so the configuration space coordinates correspond to the probability of finding particles at those specific positions, unlike Eulerian approaches to models such as gyrokinetics, which construct a distribution function of gyrocenters and thus determine the probability of a particles' gyrocenter being at a particular position.}.
And, because we are not transforming configuration space to gyrocenter coordinates, we choose to call the angular coordinate in our cylindrical velocity space coordinates the velocity gyroangle to distinguish from derivations of gyrokinetics in which this angle is the gyrophase of the particle along its cyclotron orbit \emph{because} the configuration space coordinates are at gyrocenters. 

In this coordinate system, an exact spectral representation of the distribution function could take the form, 
\begin{align}
    & f(\mvec{x}, \mvec{v}, t) =  \sum_{j=0}^{\infty} \sum_{k=0}^{\infty}  \sum_{\ell=-\infty}^{\infty} \mathcal{F}_{jk\ell}(\mvec{x}, t)\exp \left ( -\frac{ m\left [v_\parallel - u_\parallel (\mvec{x},t)\right ]^2}{2 T_\parallel (\mvec{x}, t)} - \frac{m \left [v_\perp - u_\perp (\mvec{x},t)\right ]^2}{2 T_\perp (\mvec{x}, t)}  \right ) \notag \\
    & \times \sqrt{\frac{m}{2 \pi T_\parallel(\mvec{x},t)}} H_j \left ( \sqrt{\frac{m}{2 T_\parallel (\mvec{x}, t)}} \left [v_\parallel - u_\parallel(\mvec{x}, t) \right ] \right ) \frac{m}{2 \pi T_\perp(\mvec{x}, t)} L_k \left (\frac{ m \left [v_\perp - u_\perp(\mvec{x}, t) \right ]^2}{2 T_\perp(\mvec{x}, t)} \right ) e^{i\ell\theta}, \label{eq:HLF}
\end{align}
where $H_{j}$ are the physicists' Hermite polynomials,
\begin{align}
    H_m(x) = (-1)^m e^{x^2} \frac{d^m}{dx^m} e^{-x^2}, 
\end{align}
$L_k$ are the Laguerre polynomials,
\begin{align}
    L_m(x) = \frac{e^x}{m!} \frac{d^m}{dx^m} \left (e^{-x} x^m \right ),
\end{align}
$e^{i\ell\theta}$ are the Fourier basis, and $\mathcal{F}_{jk\ell}(\mvec{x}, t)$ are the spectral coefficients encoding the representation of the distribution function in this spectral expansion.
We have utilized $u_\parallel, u_\perp, T_\parallel$ and $T_\perp$ as the notation for a general set of shift and normalization factors. 
We emphasize the choice of notation that $u_\parallel, u_\perp, T_\parallel$ and $T_\perp$ share symbols with the flow velocities and temperatures parallel and perpendicular to the local magnetic field is deliberate; while these are general shift and normalization factors in the Hermite-Laguerre expansion, the optimal Hermite-Laguerre expansion---the spectral expansion whose zeroth, first, and second velocity moments are the lowest order fluid equations---is the one in which $u_\parallel, u_\perp, T_\parallel$ and $T_\perp$ are the local parallel and perpendicular flow velocities and temperatures.

This basis representation of Hermite in $v_\parallel$, Laguerre in $v_\perp$, and Fourier in $\theta$ for the distribution function is exact, provided of course that one retains an infinite number of these Hermite, Laguerre, and Fourier basis functions. 
In the discrete limit, a truncation of this series expansion is performed to Hermite polynomials, Laguerre polynomials, and Fourier harmonics of some order. 
As mentioned previously, a particular choice of shift and normalization factors such that the shifts, $u_\parallel$ and $u_\perp$, are the parallel and perpendicular flow velocities and normalizations, $T_\parallel$ and $T_\perp$, are the parallel and perpendicular temperatures, gives spectral expansion which can be directly related to the fluid hierarchy. 
Thus, the spectral expansion is optimized by this choice of shift and normalization factors in terms of flow velocities and temperatures because we can now guarantee an exact representation of our lowest order fluid equations by our truncated spectral expansion. 
We note that this truncation to some order corresponding to a fluid hierarchy of some order is the same irrespective of the details of the spectral expansion so long as the spectral expansion has these optimized shift and normalization factors. 
In other words, a complete Hermite decomposition of the Vlasov equation or a Hermite-Laguerre decomposition of the gyrokinetic equation can just as easily be connected to fluid or gyrofluid equations provided one makes the right choice of shift and normalization factors.
We also note that in general, a spectral representation need only be applied to a subset of the degrees of freedom, e.g.,
\begin{align}
    f(\mvec{x}, \mvec{v}, t) =  \sum_{k=0}^{\infty}  \sum_{\ell=-\infty}^{\infty} & \mathcal{F}_{k\ell}(\mvec{x}, v_\parallel, t)\exp \left (-\frac{m \left [v_\perp - u_\perp (\mvec{x},t)\right ]^2}{2 T_\perp (\mvec{x}, t)}  \right ) \notag \\
    & \times \frac{m}{2 \pi T_\perp(\mvec{x}, t)} L_k \left (\frac{ m \left [v_\perp - u_\perp(\mvec{x}, t) \right ]^2}{2 T_\perp(\mvec{x}, t)} \right ) e^{i\ell\theta},
\end{align}
for an expansion in only the perpendicular degrees of freedom, $v_\perp, \theta$ or
\begin{align}
    f(\mvec{x}, \mvec{v}, t) =\sum_{\ell=-\infty}^{\infty} \frac{1}{2 \pi } \mathcal{F}_{\ell}(\mvec{x}, v_\parallel, v_\perp, t) e^{i\ell\theta},
\end{align}
for an expansion only in the velocity gyroangle, $\theta$.

Two important points are worth emphasizing immediately.
Firstly, the success of magnetized models such as drift-kinetics and gyrokinetics reveals that with the right coordinate system about the local magnetic field, certain truncations of these spectral expansions already contain a significant amount of the physics of magnetized plasmas. 
These models, in effect, retain only the $\ell = 0$ Fourier harmonic due to the gyroaveraging and thus evolve only the gyrotropic part of the distribution function, or in the case of gyrokinetics the gyrocenter distribution function.
In particular, spectral gyrokinetic codes evolve the gyrotropic component of the gyrocenter distribution function, the $\ell = 0$ Fourier harmonic, and a few Laguerre polynomials in $v_\perp$, have been utilized to study a diverse array of laboratory, space, and astrophysical plasmas \citep{Frei:2023b, Hoffmann:2023b, Mandell:2024, Hoffmann:2025, Frei:2024}.

Secondly however, the connection between this Hermite-Laguerre-Fourier spectral expansion and the classical fluid hierarchy of Braginskii and Grad is valid only for specific choices of shift and normalization factors. 
To properly connect our truncated spectral expansion to the fluid hierarchy, the shift factors must be the bulk flow velocities of the plasma and the normalization factors must be the temperatures of the plasma. 
Thus, the optimized shift and normalization factors must be temporally and spatially dependent.
Unsurprisingly, the use of shift and normalization factors which themselves depend on space and time introduces significant complexity in constructing the discrete spectral representation, and the deployment of these methods up to this point has principally focused on constant shift and normalization factors.
Without a proper accounting for these factors, the spectral representation will degrade in quality, and more spectral coefficients may be required to accurately represent the dynamics of the plasma.

The principal challenge in extending spectral methods to these more general use cases with time-dependent shift and normalization factors leads us to a necessary prerequisite discussion on coordinate transformations. 
Discrete spectral representations are constructed from orthogonality conditions, i.e., in one dimension, 
\begin{align}
    \int_{-\infty}^\infty e^{-x^2} H_m(x) H_n(x) \thinspace dx = \delta_{mn} 
\end{align}
for Hermite polynomials, or 
\begin{align}
    \int_{0}^\infty e^{-x} L_m(x) L_n(x) \thinspace dx = \delta_{mn} 
\end{align}
for Laguerre polynomials.
Utilizing these orthogonality conditions, one can derive evolution equations for each coefficient of the spectral expansion up to some desired order.

However, the introduction of time and spatially dependent shift and normalization factors complicates our ability to utilize these orthogonality relations to construct the discrete spectral expansion.
These orthogonality relations only hold for fixed input variables, i.e., two Laguerre polynomials with different shift and normalization factors are not orthogonal to each other, so one must be careful when the basis expansion itself varies in space and time. 
But, if we instead consider a transformation to a new variable,
\begin{align}
    \zeta(\mvec{x}, \mvec{v}, t) = \mvec{v} - \mvec{u}(\mvec{x}, t)
\end{align}
along with a careful absorption of the normalization into the variable of integration, we can rearrange the spectral expansion's orthogonality condition.
For example, the Laguerre orthonormality condition for the perpendicular velocity in \eqr{\ref{eq:HLF}} can be rearranged as
\begin{align}
    \int_0^\infty \frac{m}{T_\perp} e^{-m\zeta_\perp^2/2 T_\perp} L_m \left (\frac{m\zeta_\perp^2}{2 T_\perp} \right )  L_n \left (\frac{m\zeta_\perp^2}{2 T_\perp} \right ) \zeta_\perp d\zeta_\perp & = \notag \\ 
    \int_0^\infty e^{-m\zeta_\perp^2/2 T_\perp} L_m \left (\frac{m\zeta_\perp^2}{2 T_\perp} \right )  L_n \left (\frac{m\zeta_\perp^2}{2 T_\perp} \right ) d \left ( \frac{m\zeta_\perp^2}{2 T_\perp} \right ) & = \delta_{mn}. 
\end{align}
Note that in this rearrangement, we have utilized the fact that the velocity coordinate transformation in \eqr{\ref{eq:vparvperp}} modifies the volume element
\begin{align}
    d^3\mvec{v} = v_\perp dv_\parallel dv_\perp d\theta,
\end{align}
but this modification allows us to succinctly construct an orthogonality condition for our new spatially and temporally dependent coordinate.

Thus, we are well motivated to seek coordinate transformations which themselves contain the necessary modifications to the spectral basis to optimize the spectral expansion. 
While the use of spatially and temporally dependent coordinates is itself a non-trivial endeavor, through the careful construction of a particular coordinate system, we can shift the challenge in model discretization back to the continuum limit and derive the necessary modifications to the Vlasov equation that can be optimally discretized by a spectral expansion. 
Importantly, we may transform the velocity-space coordinates and configuration-space coordinates independently of each other in the Vlasov equation, and in fact we need only transform the velocity-space coordinates to optimize our spectral expansions. 
In general, such a transformation of the velocity coordinate takes the form,
\begin{align}
  \mvec{v} = \mvec{v}(\mvec{v}',\mvec{x}',t'),
\end{align}
where $\mvec{x} = \mvec{x}(\mvec{x}') = \mvec{x}$ and $t = t(t') = t$ remain unchanged by the new coordinate transformation.
The time and spatial derivatives then transform as
\begin{align}
  \frac{\partial}{\partial t} &=
  \frac{\partial}{\partial t'} \label{eq:t-trans}
  +
  \pfrac{\mvec{v}'}{t}\cdot\delvp \\
  \delx &= \delxp + \left(\delx\otimes\bmvec{v}' \right)\cdot\delvp
  \label{eq:x-trans}
\end{align}
Note that the breve notation indicates with which index the dot product is performed. 
In Einstein summation notation, for Cartesian tensors, the spatial derivative transformation can be written as
\begin{align}
    \partial_{x_i} = \partial_{x'_i} + \partial_{x_i} v'_j \partial_{v_j'}. \label{eq:cartesian-transform}
\end{align}
We emphasize that the expression in \eqr{\ref{eq:cartesian-transform}} is only valid for Cartesian coordinates, but the general expressions \thinspace\ref{eq:t-trans} and \ref{eq:x-trans} are coordinate independent. 

The transformations presented below require careful use of coordinate independent notation to ensure that the derivations and final expressions are applicable in arbitrary coordinate systems. 
This generality is particularly important, for example, when using field-aligned or other non-orthogonal coordinates, as commonly employed in simulations of fusion reactors, or when applying the expressions to plasma problems in non-Cartesian coordinates. 
Hence, to ensure our expressions remain coordinate independent, we will use an extended form of vector and tensor notation throughout the remainder of the manuscript. 
For a brief description of our notation, see Appendix \thinspace\ref{app:notation}.

%% file: pkpm-model-deriv.tex
\section{Coordinate Transformations}\label{sec:coordinate-transform}

The following two subsections detail the necessary algebraic steps to manipulate the Vlasov equation into a form with equivalent physical content, but which is amenable to our stated goal of an optimized spectral basis in a subset of the velocity degrees of freedom. 
These manipulations of our velocity-space coordinate system to depend upon space and time are well-known and utilized extensively in the derivation of e.g., Kulsrud's KMHD \citep{Kulsrud:1964, Kulsrud:1983} or Ramos' FLR Kinetic Theory \citep{Ramos:2008, Ramos:2016}.
Nevertheless, we repeat these derivations within the text in our preferred notation so that the subsequent section on the optimized spectral expansion need not repeat any definitions to clarify our notation. 
Readers familiar with these kinds of coordinate transformations may comfortably skip to the next section, Section~\ref{sec:pkpm}, on the spectral expansions which produce the final model we discretize.

\subsection{Transformation to a moving frame}

Consider a velocity-space coordinate transformation to a frame moving with velocity $\mvec{u}$:
\begin{align}
  \mvec{v} = \mvec{v}' + \mvec{u}(\mvec{x},t).
\end{align}
For this transformation, we have
\begin{align}
  \pfrac{\mvec{v}'}{t} &= -\pfrac{\mvec{u}}{t} \\
  \left(\delx\otimes\bmvec{v}' \right)\cdot\delvp
  &= -\left(\delx\otimes\bmvec{u} \right)\cdot\delvp.
\end{align}
Using these in the Vlasov equation, we obtain
\begin{align}
  \pfrac{\bdf}{t} -\pfrac{\mvec{u}}{t}\cdot\delvp\bdf
  +
  \delxp\cdot\left[ (\mvec{v}'+\mvec{u})\bdf \right]
  -
  \left[\left(\delx\otimes\bmvec{u} \right)\cdot\delvp\right]
  \cdot
  \left[ (\mvec{v}'+\mvec{u})\bdf \right]
  +
  \delvp\cdot(\mvec{a}'\bdf) = 0,
\end{align}
where $\bdf = \bdf(\mvec{x}',\mvec{v}',t')$, and
\begin{align}
  \mvec{a}' = \frac{q}{m}(\mvec{E} + \mvec{u}\times\mvec{B})
  + \frac{q}{m} \mvec{v}'\times\mvec{B}.
\end{align}
Since $\mvec{u}(\mvec{x},t)$ does not depend on $\mvec{v}'$, we can rearrange various terms above to obtain
\begin{align}
  \pfrac{\bdf}{t}
  +
  \delxp\cdot\left[ (\mvec{v}'+\mvec{u})\bdf \right]
  +
  \delvp\cdot
  \left[
  \left(
  -\pfrac{\mvec{u}}{t}
  -
  \mvec{u}\cdot\delx\mvec{u}
  -\mvec{v}'\cdot\delx\mvec{u}
  + \frac{q}{m}\left(\mvec{E}+\mvec{u}\times\mvec{B}\right)
  +
  \frac{q}{m} \mvec{v}'\times\mvec{B}
  \right) \bdf
  \right]
  =
  0. \label{eq:flow-frame-step1}
\end{align}
We can simplify this system by choosing $\mvec{u}$ to be the fluid velocity
\begin{align}
  \mvec{u} = \frac{1}{n}\int \mvec{v} f \thinspace d^3\mvec{v},
\end{align}
where $n$ is the number density. 
With this choice, we can use the equations for mass and momentum conservation,
\begin{align}
    & \pfrac{\rho \mvec{u}}{t} + \nabla_{\mvec{x}} \cdot \left (\rho \mvec{u}\mvec{u} + \mvec{P} \right ) = \frac{q}{m} \rho \left ( \mvec{E} + \mvec{u} \times \mvec{B} \right ) \quad \rightarrow \label{eq:momentum} \\
    & \rho \left ( \pfrac{\mvec{u}}{t} + \mvec{u} \cdot \nabla_{\mvec{x}} \mvec{u} \right ) + \nabla_{\mvec{x}} \cdot \mvec{P}  + \mvec{u} \underbrace{\left( \pfrac{\rho}{t} + \nabla_{\mvec{x}} \cdot \left ( \rho \mvec{u} \right ) \right) }_{= 0} = \frac{q}{m} \rho \left ( \mvec{E} + \mvec{u} \times \mvec{B} \right ),
\end{align}
to obtain an equation for the velocity evolution
\begin{align}
  \pfrac{\mvec{u}}{t} + \mvec{u}\cdot\delx\mvec{u}
  + \frac{1}{\rho}\delx\cdot\mvec{P}
  =
  \frac{q}{m}\left(\mvec{E}+\mvec{u}\times\mvec{B}\right),
  \label{eq:vel}
\end{align}
where $\rho = m n$ is the mass density, and $\mvec{P}$ is the pressure tensor defined as
\begin{align}
  \mvec{P} =
  m \int (\mvec{v}-\mvec{u})\otimes (\mvec{v}-\mvec{u}) f \thinspace d^3\mvec{v}
  =
  m \int \mvec{v}'\otimes \mvec{v}' \bdf \thinspace d^3\mvec{v}'.
    \label{eq:pten}
\end{align}

We can then substitute \eqr{\ref{eq:vel}} into \eqr{\ref{eq:flow-frame-step1}} to simplify our transformed Vlasov equation to
\begin{align}
  \pfrac{\bdf}{t}
  +
  \delxp\cdot\left[ (\mvec{v}'+\mvec{u})\bdf \right]
  +
  \delvp\cdot
  \left[
  \left(
  \frac{1}{\rho} \delx\cdot\mvec{P}
  -\mvec{v}'\cdot\delx\mvec{u}
  +
  \frac{q}{m} \mvec{v}'\times\mvec{B}
  \right) \bdf
  \right]
  =
  0.
  \label{eq:vlasov-mov}
\end{align}
We note that because the velocity coordinates of \eqr{\ref{eq:vlasov-mov}} have been transformed to move with the local fluid velocity, we must have the condition that
\begin{align}
  \int \mvec{v}' \bdf \thinspace d^3\mvec{v}' = 0. \label{eq:fp-vel-cond}
\end{align}
Multiplying \eqr{\ref{eq:vlasov-mov}} by $\mvec{v}'$ and integrating over all velocity space shows that if \eqr{\ref{eq:fp-vel-cond}} is satisfied at $t=0$, then it will be satisfied for all $t>0$, showing that this is an initial condition constraint on the distribution function.

In addition to solving the transformed Vlasov equation, to close the system of equations we also need to evolve the equation for the velocity, \eqr{\ref{eq:vel}}, or momentum \eqr{\ref{eq:momentum}}, with the pressure tensor determined by the moments of the distribution function, \eqr{\ref{eq:pten}}. 
The transformed Vlasov equation coupled to the momentum/velocity equation has the same physical content as the Vlasov equation in the laboratory frame.

For further reference, we derive the equation for the evolution of the
pressure tensor by multiplying \eqr{\ref{eq:vlasov-mov}} with
$m \mvec{v}'\otimes \mvec{v}'$, integrate the velocity term by parts,
and then use the fact that $\delvp\otimes\mvec{v}'$ is the metric
tensor in velocity space, to find
\begin{align}
  \pfrac{\mvec{P}}{t}
  +
  \delx\cdot
  [
  \mvec{Q}
  +
  \mvec{P}\otimes\bmvec{u}
  ]
  +
  \delx\cdot
  [
  \mvec{u}\otimes\mvec{P}(\breve{\_},\_) +
  \mvec{P}(\breve{\_},\_)\otimes\mvec{u}
  ]
  -
  [
  \mvec{u}\otimes(\delx\cdot\mvec{P})
  +
  (\delx\cdot\mvec{P})\otimes\mvec{u}
  )]  = \notag \\
  \frac{q}{m}
  [
  \mvec{B}\times \mvec{P}(\breve{\_},\_)
  +
  \mvec{B}\times \mvec{P}(\_,\breve{\_})
  ]
  .
\end{align}
We can take the trace of this equation and note that
$\Tr(\mvec{P}) = 3 p$ to derive an equation for the total scalar
internal energy, $3p/2$, as
\begin{align}
  \frac{\partial }{\partial t}
  \frac{3}{2} p
  +
  \delxp\cdot
  \left[
  \mvec{q}
  +
  \frac{3}{2} p \mvec{u}
  +
  \mvec{u}\cdot\mvec{P}
  \right]
  =
  \mvec{u}\cdot[\delx\cdot\mvec{P}].
  \label{eq:int-er}
\end{align}
where the heat-flux vector is defined as
\begin{align}
  \mvec{q}
  =
  \int
  \frac{1}{2} m v'^2 \mvec{v}' \bdf
  \thinspace d^3\mvec{v}'
  =
  \frac{1}{2}
  \Tr
  (
  \mvec{Q}
  ),
  \label{eq:heat-q-vec}
\end{align}
and the trace is over any two of the three slots of the fully symmetric
$\mvec{Q}$. Here, $\mvec{Q}$ is the third-order heat-flux tensor
defined as
\begin{align}
  \mvec{Q} = m
  \int
  \mvec{v}'\otimes\mvec{v}'\otimes\mvec{v}' \bdf
  \thinspace d^3\mvec{v}'.
\end{align}
An alternate form of the pressure evolution equation can be derived by
moving the $\delxp\cdot[\mvec{u}\cdot\mvec{P}]$ term to the right-hand
to write instead
\begin{align}
  \frac{\partial }{\partial t}
  \frac{3}{2} p
  +
  \delxp\cdot
  \left[
  \mvec{q}
  +
  \frac{3}{2} p \mvec{u}
  \right]
  =
  -\mvec{P} : \delxp\otimes\mvec{u}.
  \label{eq:int-er-2}  
\end{align}
In this form, we now recognize that the second moment of the Vlasov equation in the moving frame, \eqr{\ref{eq:vlasov-mov}}, gives the evolution of the internal energy of the plasma, and the total energy of the plasma can be reconstructed from the appropriate manipulation of the momentum equation \eqr{\ref{eq:momentum}} for the evolution of the total kinetic energy,
\begin{align}
    & \mvec{u} \cdot \left [ \pfrac{\rho \mvec{u}}{t} + \nabla_{\mvec{x}} \cdot \left (\rho \mvec{u}\mvec{u} + \mvec{P} \right ) \right] = \mvec{u} \cdot \left[ \frac{q}{m} \rho \left ( \mvec{E} + \mvec{u} \times \mvec{B} \right ) \right] \notag \\
    & \frac{1}{2} \pfrac{\rho |\mvec{u}|^2}{t} + \nabla_{\mvec{x}} \cdot \left (\frac{1}{2} \rho |\mvec{u}|^2 \mvec{u} \right ) = -\mvec{u} \cdot [\nabla_{\mvec{x}} \cdot \mvec{P}] + \frac{q}{m} \rho \mvec{u} \cdot \mvec{E}, 
\end{align}
which when summed with \eqr{\ref{eq:int-er}} gives the identical total energy equation obtained from the untransformed Vlasov equation. 
\subsection{Transformation to CGL frame}

We now consider a magnetized plasma. The pressure tensor $\mvec{P}$ can be split into two parts
\begin{align}
  \mvec{P} = \mvec{P}^C + \gvec{\Pi}^a,
\end{align}
where $\mvec{P}^C$ is the CGL pressure part of the tensor given by
\begin{align}
  \mvec{P}^C = p_\parallel \buni\otimes\buni
  + p_\perp (\mvec{g}-\buni\otimes\buni)
\end{align}
and $\gvec{\Pi}^a$ is the trace-free, agyrotropic part and $\mvec{g}$ is the metric tensor in configuration space. 
We have defined $p_\parallel \equiv \mvec{P} : \buni\otimes\buni = \mvec{P}(\mvec{b},\mvec{b})$, and because $\gvec{\Pi}^a$ is trace-free, we have $\Tr(\mvec{P}) = \Tr(\mvec{P}^C) = p_\parallel + 2 p_\perp \equiv 3p$, where $p$ is the scalar pressure. Now, for any second-order tensor $\mvec{A}$, $\lambda$ and $\mvec{r}$ are an eigenvalue/eigenvector pair if $\mvec{A}(\_,\mvec{r}) = \lambda\mvec{r}$. 
For the CGL part of the pressure tensor, it is straightforward to see that $\mvec{P}^C(\_,\buni) = p_\parallel\buni$ and $\mvec{P}^C(\_,\gvec{\tau}) = p_\perp\gvec{\tau}$ for all $\buni\cdot\gvec{\tau} = 0$. 
Hence, in a frame with orthonormal unit vectors $(\buni, \gvec{\tau}_1, \gvec{\tau}_2)$, where $\gvec{\tau}_{1,2}$ are orthogonal to $\buni$ and $\gvec{\tau}_{1}\times \gvec{\tau}_{2} = \buni$, the CGL part of the pressure-tensor will be diagonal\footnote{An example coordinate systems in the case that the field is not uniformly pointing in one direction is given by considering $\gvec{\tau}_1$ aligned with the magnetic curvature, $\buni\cdot\delx\buni$. Another example: in the case when $\buni$ is a constant vector, then one can locally align the $z$-axis with the total magnetic field and take the Cartesian basis vectors as eigenvectors.}. 
However, in general these vectors are \emph{not} the eigenvectors of the full pressure tensor $\mvec{P}$. 
Nevertheless, we can always utilize this separation into gyrotropic and agyrotropic pressure tensor, and we shall call the frame in which the CGL part of the pressure tensor is diagonal the \emph{CGL frame}.

We will now transform the velocity coordinates to a local frame aligned with the $(\buni, \gvec{\tau}_1, \gvec{\tau}_2)$ basis vectors. 
In this frame, we can write the transform
\begin{align}
  \mvec{v}'
  =
  \mvec{v}'(\mvec{v}'',\mvec{x}'',t'')
  =
  v_\parallel \buni(\mvec{x}',t) + \mvec{v}_\perp
  =
  v_\parallel \buni(\mvec{x}',t) +
  v_\perp\cos\theta \gvec{\tau}_1(\mvec{x}',t) + v_\perp\sin\theta \gvec{\tau}_2(\mvec{x}',t),
\end{align}
where $(w^1, w^2, w^3) \equiv (v_\parallel,v_\perp,\theta)$ form cylindrical velocity coordinates in the CGL frame. 
In general, the basis vectors depend both on position and time. 
For this transform, we can compute the tangent vectors in velocity space, $\tbasis{i}$, as
\begin{align}
  \tbasis{i} = \pfrac{\mvec{v}'}{w^i}.
\end{align}
These are $\tbasis{\parallel} = \buni$,
\begin{align}
  \tbasis{\perp} &= \cos\theta \gvec{\tau}_1 + \sin\theta \gvec{\tau}_2 \\
  \tbasis{\theta} &= -v_\perp \sin\theta \gvec{\tau}_1 + v_\perp \cos\theta \gvec{\tau}_2.
\end{align}
The dual basis $\tdbasis{i}$ are $\tdbasis{\parallel} = \tbasis{\parallel}$, $\tdbasis{\perp} = \tbasis{\perp}$, and
\begin{align}
  \tdbasis{\theta} &= -\frac{1}{v_\perp} \sin\theta \gvec{\tau}_1
                    + \frac{1}{v_\perp} \cos\theta \gvec{\tau}_2.
\end{align}
The Jacobian of the transform is simply $\mathcal{J}_v = v_\perp$. 
In terms of $\tbasis{\perp}$, we can write $\mvec{v}_\perp = v_\perp \tbasis{\perp}$.

We can again use Eqns.\thinspace\ref{eq:t-trans} and \ref{eq:x-trans} to derive the Vlasov equation in the CGL frame. 
First notice that
\begin{align}
  \delvpp\cdot\left( \pfrac{\mvec{v}''}{t} \right)
  =
  \frac{\partial}{\partial t}
  (\delvpp\cdot\mvec{v}'') = 0,
\end{align}
and
\begin{align}
  \delvpp\cdot\left[ \delxp\otimes\bmvec{v}'' \right]
  =
  \delxp (\delvpp\cdot \mvec{v}'')
  = 0,
\end{align}
because $\delvpp\cdot\mvec{v}'' = 3$. 
With these identities, we can write the Vlasov equation in the CGL frame as
\begin{align}
  \pfrac{\bdf}{t}
  +
  \delxpp\cdot\left[ (v_\parallel \buni + \mvec{v}_\perp+\mvec{u})\bdf \right]
  +
  \frac{1}{v_\perp}\frac{\partial}{\partial w^i}
  \left( v_\perp a^i \bdf \right)
  = 0,
  \label{eq:vlasov-cgl}  
\end{align}
where now $\bdf = \bdf(\mvec{x}'',v_\parallel,v_\perp,\theta,t)$, and
$a^i = \tdbasis{i}\cdot\mvec{a}''$ are the components of the
acceleration in the CGL frame
\begin{align}
  \mvec{a}'' =
  \pfrac{\mvec{v}''}{t}
  +
  (\mvec{v}''+\mvec{u})\cdot\delx\mvec{v}''
  +
  \frac{1}{\rho}\delx\cdot\mvec{P}
  -\mvec{v}''\cdot\delx\mvec{u}
  +
  \frac{q}{m} \mvec{v}''\times\mvec{B},
\end{align}
and $\mvec{v}'' = v_\parallel \buni + \mvec{v}_\perp$.

In the CGL frame, the phase-space incompressibility can be read off
from \eqr{\ref{eq:vlasov-cgl}} as
\begin{align}
  \delxpp\cdot\left[ v_\parallel \buni + \mvec{v}_\perp+\mvec{u}\right]
  +
  \frac{1}{v_\perp}\frac{\partial}{\partial w^i}
  \left( v_\perp a^i\right)
  =
  0.
  \label{eq:phase-incomp-cgl}
\end{align}
At this point, the physics content of \eqr{\ref{eq:vlasov-cgl}} combined with the momentum equation (and Maxwell equations) is the same as the Vlasov-Maxwell system.
No information has been lost in the process of the transformations to the local flow and CGL frames. 

\section{Spectral Expansions in Gyroangle and Perpendicular Velocity} \label{sec:pkpm}

With the Vlasov equation now obtained in the velocity coordinate system moving with the flow velocity and aligned with the local magnetic field, we are ready to deploy spectral expansions in velocity gyroangle and perpendicular velocity. 
As described above, these velocity coordinates are ideal coordinates to perform the expansions: the flow velocity is separated out of the coordinates, and the coordinate system is locally aligned with the magnetic field. 
The combination of these transformations allows arbitrary flows, without any need to split the flow velocity into individual components parallel and perpendicular to the local magnetic field. 
Further, we can accurately capture streaming along field-lines while also allowing for a rapidly converging spectral expansion in the perpendicular degrees of freedom without any need for a time or spatially dependent shift vector.

\subsection{Fourier Expansion in the CGL Frame}

We first expand the distribution function in velocity gyroangle using the
Fourier series (dropping primes on $\mvec{x}$)
\begin{align}
  \bdf(\mvec{x},v_\parallel,v_\perp,\theta,t)
  =
  \bdf_0(\mvec{x},v_\parallel,v_\perp,t)
  +
  \sum_{n=1}^\infty
  \left[
  \bdf_n^c(\mvec{x},v_\parallel,v_\perp,t)\cos n\theta
  +
  \bdf_n^s(\mvec{x},v_\parallel,v_\perp,t)\sin n\theta
  \right].
\end{align}
Here $\bdf_0$ is the \emph{gyrotropic} 
part, and $\bdf_n^{c,s}$ are the \emph{agyrotropic} parts of the
distribution function. These are given by
\begin{align}
  \bdf_0 = \frac{1}{2\pi}
  \int_0^{2\pi} \bdf(\mvec{x},v_\parallel,v_\perp,\theta,t) \thinspace d\theta
\end{align}
and
\begin{align}
  \bdf_n^c(\mvec{x},v_\parallel,v_\perp,t) &= \frac{1}{\pi}
  \int_0^{2\pi} \bdf(\mvec{x},v_\parallel,v_\perp,\theta,t) 
  \cos n\theta \thinspace d\theta \\
  \bdf_n^s(\mvec{x},v_\parallel,v_\perp,t) &= \frac{1}{\pi}
  \int_0^{2\pi} \bdf(\mvec{x},v_\parallel,v_\perp,\theta,t) 
  \sin n\theta \thinspace d\theta
\end{align}
for $n>0$.

We can derive equations for each of the Fourier harmonics. 
However, to illustrate the approach, we will focus on the gyrotropic distribution function $\bdf_0$. 
We can derive the equation for the zeroth harmonic by integrating \eqr{\ref{eq:vlasov-cgl}} over gyroangles to get
\begin{align}
  \pfrac{\bdf_0}{t}
  +
  \delx\cdot\left[ (v_\parallel \buni + \mvec{u})\bdf_0 \right]
  +
  &\delx\cdot\mvec{M}_\perp + \notag \\
  &\frac{1}{v_\perp}
  \frac{\partial}{\partial v_\parallel}
  \left( v_\perp
    \frac{1}{2\pi}
    \int_0^{2\pi} 
    a^\parallel \bdf d{\theta}
    \right)
  +
  \frac{1}{v_\perp}
  \frac{\partial}{\partial v_\perp}
  \left( v_\perp
    \frac{1}{2\pi}
  \int_0^{2\pi} 
  a^\perp \bdf d{\theta}
  \right)
  = 0.
  \label{eq:gyro-f}
\end{align}
Here,
\begin{align}
  \mvec{M}_\perp(\mvec{x},v_\parallel,v_\perp,t)
  =
  \frac{1}{2\pi}
  \int_0^{2\pi}
  \mvec{v}_\perp
  \bdf (\mvec{x},v_\parallel,v_\perp,\theta,t)
  \thinspace d{\theta}
  =
  \frac{v_\perp}{2}
  \left[
  \bdf_1^c(\mvec{x},v_\parallel,v_\perp,t) \gvec{\tau}_1 + \bdf_1^s(\mvec{x},v_\parallel,v_\perp,t) \gvec{\tau}_2
  \right], \label{eq:Mperp}
\end{align}
is the contribution to the streaming term from agyrotropy of the
distribution function.

We can compute
$a^\parallel = \tdbasis{\parallel}\cdot\mvec{a}'' =
\buni\cdot\mvec{a}''$ as
\begin{align}
  a^\parallel
  =
  \mvec{v}_\perp\cdot
  \left[
  \pfrac{\buni}{t}
  +
  (v_\parallel\buni + \mvec{v}_\perp + \mvec{u})\cdot\delx\buni
  \right]
  +
  \frac{1}{\rho} \buni\cdot\left[\delx\cdot\mvec{P}\right]
  -
  \buni\cdot
  \left[
  (v_\parallel\buni + \mvec{v}_\perp)\cdot\delx\mvec{u}
  \right],
\end{align}
from which we find
\begin{align}
  \frac{1}{2\pi}
  \int_0^{2\pi} 
  a^\parallel \bdf d{\theta}
  =
  \buni\cdot
  \left[
  \frac{1}{\rho} \delx\cdot\mvec{P}
  -
  v_\parallel\buni\cdot\delx\mvec{u}
  \right] \bdf_0
  + 
  \mvec{M}_\perp\cdot
  \left[
  \pfrac{\buni}{t}
  +
  (v_\parallel\buni + \mvec{u})\cdot\delx\buni
  \right] \notag \\
  - \buni\cdot\left[\mvec{M}_\perp\cdot\delx\mvec{u}\right]
  + \mvec{F}_{\perp\perp} : \delx\otimes\mvec{b},
\end{align}
where
\begin{align}
  \mvec{F}_{\perp\perp}(\mvec{x},v_\parallel,v_\perp,t)
  \equiv
  \frac{1}{2\pi}
  \int_0^{2\pi} 
  (\mvec{v}_\perp\otimes\mvec{v}_\perp)
  \bdf (\mvec{x},v_\parallel,v_\perp,\theta,t)
  \thinspace d{\theta},
\end{align}
is a second-order symmetric tensor. 
Using the Fourier expansion of the distribution function, we get
\begin{align}
  \mvec{F}_{\perp\perp}(\mvec{x},v_\parallel,v_\perp,t)
  =
  \frac{v_\perp^2\bdf_0}{2}
  \left(
  \mvec{g} - \buni\otimes\buni
  \right)
  +
  \frac{v_\perp^2\bdf_2^c}{4}
  \left( 
  \gvec{\tau}_1\otimes\gvec{\tau}_1 - \gvec{\tau}_2\otimes\gvec{\tau}_2
  \right)
  +
  \frac{v_\perp^2\bdf_2^s}{4}
  \left( 
  \gvec{\tau}_1\otimes\gvec{\tau}_2 + \gvec{\tau}_2\otimes\gvec{\tau}_1
  \right),
  \label{eq:F-perp-perp}
\end{align}
where we used $\mvec{g} = \gvec{\tau}_1\otimes\gvec{\tau}_1 + \gvec{\tau}_2\otimes\gvec{\tau}_2 + \buni\otimes\buni$ to rewrite the first term. 
Using this result, we can compute
\begin{align}
  \mvec{F}_{\perp\perp} : \delx\otimes\mvec{b}
  = 
  \frac{v_\perp^2\bdf_0}{2} \delx\cdot\buni
  -
  \frac{v_\perp^2\bdf_2^c}{4}
  \buni\cdot\left( 
  \gvec{\tau}_1\cdot\nabla\gvec{\tau}_1 - \gvec{\tau}_2\cdot\nabla\gvec{\tau}_2
  \right)
  -
  \frac{v_\perp^2\bdf_2^s}{4}
  \buni\cdot\left( 
  \gvec{\tau}_1\cdot\nabla\gvec{\tau}_2 + \gvec{\tau}_2\cdot\nabla\gvec{\tau}_1
  \right).
\end{align}
The first term in this expression is the contribution to the parallel acceleration from the gyrotropic part of the distribution function, and the remaining two terms are the contributions from the agyrotropy of the distribution function. 
An alternate form of the first term can be obtained, since we can replace
\begin{align}
  \delx\cdot\buni = \delx\cdot\frac{\mvec{B}}{B} =
  \frac{1}{B}\delx\cdot\mvec{B} + \mvec{B}\cdot\delx{\frac{1}{B}}
  =
  -\frac{1}{B} \nabla_\parallel B,
\end{align}
where $\nabla_\parallel \equiv \buni\cdot\delx$.

Similarly, we can compute $a^\perp = \tdbasis{\perp}\cdot\mvec{a}''$ as
\begin{align}
  a^\perp
  &=
  \tdbasis{\perp}
  \cdot
  \left[
  -v_\parallel \pfrac{\buni}{t}
  -v_\parallel(v_\parallel\buni + \mvec{v}_\perp + 
  \mvec{u})\cdot\delx\buni 
  + \frac{1}{\rho} \delx\cdot\mvec{P}
  - (v_\parallel\buni + \mvec{v}_\perp)\cdot\delx\mvec{u}
  \right].
\end{align}
Since $v_\perp\tdbasis{\perp} = \mvec{v}_\perp$, we find
\begin{align}
  v_\perp
  \frac{1}{2\pi}
  \int_0^{2\pi} 
  a^\perp \bdf d{\theta}
  =
  \mvec{M}_\perp
  \cdot
  \left[
  -v_\parallel \pfrac{\buni}{t}
  -v_\parallel(v_\parallel\buni + \mvec{u})\cdot\delx\buni
  +
  \frac{1}{\rho} \delx\cdot\mvec{P}
  - v_\parallel\buni\cdot\delx\mvec{u}
  \right] \notag \\
  - \mvec{F}_{\perp\perp} : (v_\parallel\delx\otimes\buni + \delx\otimes\mvec{u}).
\end{align}
Though the first term of the velocity gyroangle integrated perpendicular acceleration has only agyrotropic contributions arising from the coupling to the first Fourier harmonic, $\mvec{M}_\perp$, the $\mvec{F}_{\perp\perp}$ term in $a^\perp$ does have a gyrotropic contribution as well---see \eqr{\ref{eq:F-perp-perp}} and the term proportional to $\bdf_0$. 
Putting all these expressions together, we finally obtain the equation for the gyrotropic distribution function
\begin{align}
  \pfrac{\bdf_0}{t}
  +
  \delx\cdot\left[ (v_\parallel \buni + \mvec{u})\bdf_0 \right]
  +
  \frac{\partial}{\partial v_\parallel}
  \left[
  \buni\cdot
  \left(
  \frac{1}{\rho} \delx\cdot\mvec{P}
  -
  v_\parallel\buni\cdot\delx\mvec{u}
  \right)\bdf_0
  +
  \frac{v_\perp^2}{2}\delx\cdot\buni
  \bdf_0
  \right] \notag \\
  + \frac{1}{v_\perp}
  \frac{\partial}{\partial v_\perp}
  \left[
  \frac{v_\perp^2}{2}
  \left(
  \buni\cdot[\buni\cdot \delx\mvec{u}]
  -
  \delx\cdot(v_\parallel\buni+ \mvec{u})
  \right)\bdf_0
  \right]
  + \mathrm{AG}
  =
  0,
  \label{eq:gyro-f-f0}
\end{align}
where $\mathrm{AG}$ are the agyrotropic terms, which involve the first and second Fourier harmonics of the distribution function. 
The expressions for these terms are given in Appendix~\ref{app:f0-agyro}.
We note that to determine the coupling of $\bdf_0$ to the agyrotropic terms, further equations would be needed; for example, evolution equations for $\mvec{M}_\perp$ and $\mvec{F}_{\perp\perp}$ can be derived by multiplying the Vlasov equation in our transformed coordinates, \eqr{\ref{eq:vlasov-cgl}}, by $\mvec{v}_\perp$ and $\mvec{v}_\perp \otimes \mvec{v}_\perp$ and integrating over velocity gyroangle.
In fact, this procedure of determining equations for e.g., $\mvec{M}_\perp$ and $\mvec{F}_{\perp\perp}$, provides a natural procedure for the determination of higher Fourier harmonics, at least through the second Fourier harmonic, because the coupling to the $f^{c,s}_{1,2}$ coefficients of the Fourier expansion appears only through these vector and tensor combinations of Fourier coefficients and plane perpendicular to the magnetic field, $\gvec{\tau}_{1,2}$.

In terms of the Fourier expansion, various velocity moments can be determined through integration over $v_\parallel$ and $v_\perp$.
For example, the number density is
\begin{align}
  n
  &=
  2\pi 
  \int_0^\infty v_\perp \dvol{v_\perp} \int_{-\infty}^\infty \dvol{v_\parallel}
  \bdf_0, \label{eq:num-f0}
\end{align}
and the CGL pressure tensor components can be computed as
\begin{align}
  p_\parallel
  &=
  2\pi 
  \int_0^\infty v_\perp \dvol{v_\perp} \int_{-\infty}^\infty \dvol{v_\parallel}
  m v_\parallel^2 \bdf_0, \label{eq:p-par-f0} \\
  p_\perp
  &=
  2\pi 
  \int_0^\infty v_\perp \dvol{v_\perp} \int_{-\infty}^\infty \dvol{v_\parallel}
    \frac{1}{2} m v_\perp^2 \bdf_0. \label{eq:p-perp-f0}
\end{align}
Thus, the gyrotropic part of the distribution function contains the total density and total internal energy. 
The agyrotropic part of the pressure tensor is
\begin{align}
  \mvec{\Pi}^a
  =
  2&\pi m
  \int_0^\infty v_\perp \dvol{v_\perp} \int_{-\infty}^\infty \dvol{v_\parallel}
  \times \notag \\
  &\left[
  v_\parallel (\buni\otimes\mvec{M}_\perp + \mvec{M}_\perp\otimes\buni)
  +
  \frac{v_\perp^2\bdf_2^c}{4}
  \left( 
  \gvec{\tau}_1\otimes\gvec{\tau}_1 - \gvec{\tau}_2\otimes\gvec{\tau}_2
  \right)
  +
  \frac{v_\perp^2\bdf_2^s}{4}
  \left( 
  \gvec{\tau}_1\otimes\gvec{\tau}_2 + \gvec{\tau}_2\otimes\gvec{\tau}_1
  \right)
  \right]. \label{eq:Pi-agyro}
\end{align}
Clearly, both the first and the second Fourier harmonics contribute to $\mvec{\Pi}^a$. 
In the gyrotropic limit, it follows that we must have $\mvec{\Pi}^a = 0$.

The heat-flux vector defined in \eqr{\ref{eq:heat-q-vec}} is
\begin{align}
  \mvec{q}
  =
  (q_\parallel + q_\perp) \buni 
  +
  2\pi \int_0^\infty v_\perp \dvol{v_\perp} \int_{-\infty}^\infty \dvol{v_\parallel}
  \thinspace
  \frac{1}{2} m (v_\parallel^2 + v_\perp^2) \mvec{M}_\perp, \label{eq:vec_q}
\end{align}
where
\begin{align}
  q_\parallel &= 2\pi \int_0^\infty v_\perp \dvol{v_\perp} \int_{-\infty}^\infty \dvol{v_\parallel}
  \thinspace
  \frac{1}{2} m v_\parallel^3 \bdf_0, \\
  q_\perp &= 2\pi \int_0^\infty v_\perp \dvol{v_\perp} \int_{-\infty}^\infty \dvol{v_\parallel}
  \thinspace
  \frac{1}{2} m v_\parallel v_\perp^2 \bdf_0,
\end{align}
are the parallel components of the parallel and perpendicular heat
fluxes respectively. 
These components are determined fully from the gyrotropic distribution function. 
On the other hand, the components of the heat-flux vector perpendicular to the magnetic field are determined from the first Fourier harmonic.

The characteristic velocities for the evolution of the gyrotropic
distribution function can be read off from \eqr{\ref{eq:gyro-f-f0}} with
$\mathrm{AG}=0$ as
\begin{align}
  \dot{\mvec{x}} &= v_\parallel \buni + \mvec{u}, \label{eq:xdot-g} \\
  \dot{v}_\parallel
  &=
  \buni\cdot
  \left(
  \frac{1}{\rho} \delx\cdot\mvec{P}
  -
  v_\parallel\buni\cdot\delx\mvec{u}
  \right)
  +
  \frac{v_\perp^2}{2}\delx\cdot\buni, \label{eq:vpar-dot-g} \\
  \dot{v}_\perp
  &=
  \frac{v_\perp}{2}
  \left[
  \buni\cdot(\buni\cdot \delx\mvec{u})
  -
  \delx\cdot(v_\parallel\buni+ \mvec{u})
  \right], \label{eq:vperp-dot-g}
\end{align}
Further, integrating the phase-space incompressibility condition
\eqr{\ref{eq:phase-incomp-cgl}} over all angles gives
\begin{align}
  \delx\cdot \dot{\mvec{x}}
  +
  \pfrac{\dot{v}_\parallel}{v_\parallel}
  +
  \frac{1}{v_\perp}
  \frac{\partial}{\partial v_\perp}\big(v_\perp \dot{v}_\perp \big) = 0.
\end{align}
Hence, in the gyrotropic limit, the evolution of the gyrotropic distribution function does not violate phase-space incompressibility.
For reference, the portion of the parallel acceleration coming from the pressure tensor can be rewritten in a more familiar form as
\begin{align}
  \frac{1}{\rho} \buni\cdot(\delx\cdot \mvec{P})
  & =
  \frac{1}{\rho}\buni\cdot\delx p_{\parallel}
  +
  \frac{(p_{\parallel}-p_{\perp})}{\rho}\delx\cdot\buni
  +
  \frac{1}{\rho} \buni\cdot\left( \delx\cdot \mvec{\Pi}^a \right), \notag \\
  & = \frac{1}{\rho} \left [ \delx \cdot \left (  p_{\parallel} \buni \right ) - p_\perp \delx\cdot\buni \right ]   
  +
  \frac{1}{\rho} \buni\cdot\left( \delx\cdot \mvec{\Pi}^a \right). 
\end{align}

\subsection{Reduction to the Parallel-Kinetic Perpendicular-Moment
  System}

We now expand the gyrotropic distribution function in a Laguerre
series in $v_\perp$ as
\begin{align}
  \bdf_0(\mvec{x},v_\parallel,v_\perp,t)
  =
  \sum_{n=0}^\infty
  F_{n}(\mvec{x},v_\parallel,t) G_M(v_\perp,T_\perp)
  L_n\left(\frac{m v_\perp^2}{2 T_\perp}\right),
  \label{eq:f0-expand}
\end{align}
where $L_n(x)$ are Laguerre polynomials of order $n$, and
\begin{align}
  G_M(v_\perp,T_\perp) = \frac{m}{2\pi T_\perp}e^{-m v_\perp^2/2 T_\perp}.
\end{align}
In these expressions, $T_\perp = T_\perp(\mvec{x},t)$ is the
perpendicular temperature, in terms of which the perpendicular
pressure is $p_\perp = n T_\perp$. The derivation of the equations for
the Laguerre coefficients $F_k$ is complicated by the fact that
$T_\perp$ depends on position as well as time. Typically in this
expansion, $T_\perp$ would need to be determined from another equation, such as an equation for the perpendicular pressure.

The $F_k(\mvec{x},v_\parallel,t)$
coefficient of this series is
\begin{align}
  F_{k}(\mvec{x},v_\parallel,t) =
  2\pi \int_0^\infty v_\perp \dvol{v_\perp}
  L_k\left(\frac{m v_\perp^2}{2 T_\perp}\right)
  \bdf_0(\mvec{x},v_\parallel,v_\perp,t).
  \label{eq:fk0-gen}
\end{align}
The number density and the parallel pressure (see
\eqr{\ref{eq:num-f0}} and \eqr{\ref{eq:p-par-f0}}) are completely
determined from the zeroth Laguerre coefficient of the gyrotropic
distribution function:
\begin{align}
  n &= \int_{-\infty}^\infty F_0 \dvol{v_\parallel}, \\
  p_\parallel
  &=
  \int_{-\infty}^\infty m v_\parallel^2 F_0 \dvol{v_\parallel}. \label{eq:p-par-F0}
\end{align}
The parallel components of the heat-flux are
\begin{align}
  q_\parallel &=
  \int_{-\infty}^\infty \frac{1}{2} m v_\parallel^3 F_0 \dvol{v_\parallel}, \\
  q_\perp &=
  T_\perp \int_{-\infty}^\infty v_\parallel (F_0-F_1) \dvol{v_\parallel},
\end{align}
which are determined from just the first two Laguerre coefficients of the gyrotropic distribution function. 
Finally, the constraint \eqr{\ref{eq:fp-vel-cond}} on the $\mvec{v}'$ moment of the distribution function shows that we must have
\begin{align}
  \int_{-\infty}^\infty v_\parallel F_0 \dvol{v_\parallel} &= 0, \\
  \int_0^\infty v_\perp \dvol{v_\perp} \int_{-\infty}^\infty \dvol{v_\parallel} 
  \thinspace
  \mvec{M}_\perp
  &=
  0.
\end{align}

\subsubsection{Some Preliminaries}

To derive the equations for $F_{k}$ for $k>0$ (recall $L_0 = 1$), we need to account for the fact that $T_\perp$ depends both on space and time. 
To do this derivation then, we write
\begin{align}
  L_k\left(\frac{m v_\perp^2}{2T_\perp}\right)
  \pfrac{\bdf_0}{t}
  =
  \frac{\partial}{\partial t}
  \left[
  L_k\left(\frac{m v_\perp^2}{2T_\perp}\right) \bdf_0
  \right]
  -
  \bdf_0 \frac{\partial}{\partial t}
  \left[
  L_k\left(\frac{m v_\perp^2}{2T_\perp}\right)
  \right].
\end{align}
Now, using \eqr{\ref{eq:d-lag}} we get
\begin{align}
  \frac{\partial }{\partial t} L_k\left(\frac{m v_\perp^2}{2T_\perp}\right)
  =
  -\frac{1}{T_\perp}\pfrac{T_\perp}{t}
  \left[
  k L_k\left(\frac{m v_\perp^2}{2T_\perp}\right)
  -
  k L_{k-1}\left(\frac{m v_\perp^2}{2T_\perp}\right)
  \right].
\end{align}
Also,
\begin{align}
  L_k\left(\frac{m v_\perp^2}{2T_\perp}\right)
  \delx\cdot\left[ (v_\parallel\buni+\mvec{u})\bdf_0 \right]
  =
  \delx\cdot
  \left[
  (v_\parallel\buni+\mvec{u}) L_k\left(\frac{m v_\perp^2}{2T_\perp}\right)  \bdf_0
  \right]
  - \bdf_0 \left \{
  (v_\parallel\buni+\mvec{u})
  \cdot \delx
  \left[
  L_k\left(\frac{m v_\perp^2}{2T_\perp}\right)
  \right] \right \},
\end{align}
where, as above, we can write
\begin{align}
  \delx
  \left[
  L_k\left(\frac{m v_\perp^2}{2T_\perp}\right)
  \right]
  =
  -\frac{1}{T_\perp}\delx T_\perp
  \left[
  k L_k\left(\frac{m v_\perp^2}{2T_\perp}\right)
  -
  k L_{k-1}\left(\frac{m v_\perp^2}{2T_\perp}\right)
  \right].
\end{align}

\subsubsection{The Equation for $F_k$}

We can multiply \eqr{\ref{eq:gyro-f-f0}} by $L_k(m v_\perp^2/2 T_\perp)$ and
integrate over all $v_\perp$ to obtain
\begin{align}
  \pfrac{F_{k}}{t}
  &+
  \delx\cdot\left[ (v_\parallel\buni+\mvec{u})F_{k} \right] \notag \\
  &+
  \frac{\partial}{\partial v_\parallel}
  \left[
  \buni\cdot
  \left(
  \frac{1}{\rho}\delx\cdot\mvec{P} - v_\parallel\buni\cdot\delx\mvec{u}
  \right)  F_{k}
  +
  \left[
  (2k+1) F_k
  -
  k F_{k-1} - (k+1) F_{k+1}
 \right]\frac{T_\perp}{m}\delx\cdot\buni
  \right] \notag \\
  &=
  S_k^T(\mvec{x},v_\parallel) +  S_k(\mvec{x},v_\parallel).
  \label{eq:fk-coeff}
\end{align}
The source $S^T_k$ appears because $T_\perp$ depends both on space and time. 
Using the expressions derived in the previous section, we can write this source term as
\begin{align}
  S^T_k(\mvec{x},v_\parallel)
  =
  -\frac{1}{T_\perp}
  \left[
  \pfrac{T_\perp}{t}
  +
  (v_\parallel\buni + \mvec{u})\cdot\delx T_\perp
  \right]
  (k F_{k} - k F_{k-1}).
\end{align}
The source $S_k$ results from the integration-by-parts of the perpendicular velocity derivative. 
Calculating this term, we get
\begin{align}
  S_k(\mvec{x},v_\parallel)
  =
  \left[
  \buni\cdot(\buni\cdot\delx\mvec{u})
  - \delx\cdot(v_\parallel\buni + \mvec{u})
  \right]
  (k F_{k} - k F_{k-1}).
\end{align}

At this point we have all the equations we need to evolve the Laguerre coefficients of the gyrotropic distribution function. 
The number of Laguerre coefficients one will need will depend on the structure of the distribution function in the perpendicular coordinate. 
One yet undetermined quantity in these equations is $T_\perp$. 
We can derive an explicit equation for this quantity; however, we can also pursue an alternate approach which will completely eliminate the need for an auxiliary equation for $T_\perp$.

\subsection{The First Two Laguerre Coefficients and Perpendicular Temperature}

The evolution of the first two Laguerre coefficients of the gyrotropic distribution can be determined by first setting $k=0$ in \eqr{\ref{eq:fk-coeff}}. 
This substitution gives
\begin{align}
  \pfrac{F_{0}}{t}
  &+
  \delx\cdot\left[ (v_\parallel\buni+\mvec{u})F_{0} \right] \notag \\
  &+
  \frac{\partial}{\partial v_\parallel}
  \left[
  \buni\cdot
  \left(
  \frac{1}{\rho}\delx\cdot\mvec{P} - v_\parallel\buni\cdot\delx\mvec{u}
  \right)  F_{0}
  + \mathcal{G}\delx\cdot\buni
  \right] = 0,
  \label{eq:F0-coeff}
\end{align}
where we have defined
\begin{align}
\mathcal{G} \equiv \frac{T_\perp}{m}(F_0 - F_1).   
\end{align}
Further, setting $k=1$ in \eqr{\ref{eq:fk-coeff}}
\begin{align}
  \pfrac{F_{1}}{t}
  &+
  \delx\cdot\left[ (v_\parallel\buni+\mvec{u})F_{1} \right] \notag \\
  &+
  \frac{\partial}{\partial v_\parallel}
  \left[
  \buni\cdot
  \left(
  \frac{1}{\rho}\delx\cdot\mvec{P} - v_\parallel\buni\cdot\delx\mvec{u}
  \right)  F_{1}
  +
  \left[ 3 F_1 - F_{0} - 2 F_{2} \right]\frac{T_\perp}{m}\delx\cdot\buni
  \right] \notag \\
  &=
  -\frac{1}{T_\perp}
  \left[
  \pfrac{T_\perp}{t}
  +
  (v_\parallel\buni + \mvec{u})\cdot\delx T_\perp
  \right]
  (F_{1} - F_{0})
  +  S_1(\mvec{x},v_\parallel).
  \label{eq:F1-coeff}
\end{align}
Subtracting the second equation from the first, and multiplying throughout by $T_\perp/m$, we get
\begin{align}
  \pfrac{\mathcal{G}}{t}
  &+
  \delx\cdot\left[ (v_\parallel\buni+\mvec{u})\mathcal{G} \right] \notag \\
  &+
  \frac{\partial}{\partial v_\parallel}
  \left[
  \buni\cdot
  \left(
  \frac{1}{\rho}\delx\cdot\mvec{P} - v_\parallel\buni\cdot\delx\mvec{u}
  \right)  \mathcal{G}
  +
  \left[
  4\mathcal{G} + \frac{2 T_\perp}{m}(F_2-F_0)
 \right]\frac{T_\perp}{m}\delx\cdot\buni
  \right] \notag \\
  &=
  \left[
  \buni\cdot(\buni\cdot\delx\mvec{u})
  - \delx\cdot(v_\parallel\buni + \mvec{u})
  \right] \mathcal{G}.
  \label{eq:G-coeff}
\end{align}
Instead of solving the equation for $F_1$, we can solve this equation for $\mathcal{G}$. 
The advantage of this approach is that the source $S^T_1(\mvec{x},v_\parallel)$ due to the temporal and spatial variation of $T_\perp$ does not appear, avoiding the need to
compute the time and spatial derivatives of $T_\perp$. 
Further, we claim that
\begin{align}
  p_\perp = n T_\perp  
  = T_\perp  \int_{-\infty}^\infty F_0 \dvol{v_\parallel}
  = \int_{-\infty}^\infty T_\perp(F_0-F_1) \dvol{v_\parallel}
  = m \int_{-\infty}^\infty \mathcal{G} \dvol{v_\parallel},
\end{align}
thus eliminating the need to evolve an explicit equation for the
perpendicular temperature. 

As a consistency check, we integrate \eqr{\ref{eq:G-coeff}} over all velocities to obtain
\begin{align}
  \pfrac{p_\perp}{t}
  +
  \delx\cdot
  \left[
  q_\perp \buni + p_\perp \mvec{u} 
  \right]
  = 
  p_\perp  \buni\cdot[\buni\cdot\delx\mvec{u}] 
  - q_\perp \delx\cdot\buni - p_\perp \delx\cdot\mvec{u}.
  \label{eq:p-perp-g}
\end{align}
Likewise, multiplying \eqr{\ref{eq:F0-coeff}} by $1/2 \thinspace m v_\parallel^2$ and integrating over all velocities we get an evolution equation for the parallel pressure:
\begin{align}
  \frac{1}{2} \pfrac{p_\parallel}{t}
  +
  \delx\cdot
  \left[
  q_\parallel \buni + \frac{1}{2} p_\parallel \mvec{u} 
  \right]
  =
  -p_\parallel \buni\cdot[\buni\cdot\delx\mvec{u}] + q_\perp \delx\cdot\buni.
  \label{eq:p-par-g}
\end{align}
If we add \eqr{\ref{eq:p-perp-g}} and \eqr{\ref{eq:p-par-g}}, we obtain
\begin{align}
    \pfrac{}{t} \left (\frac{1}{2} p_\parallel + p_\perp \right ) + \delx\cdot \left[ \left (q_\parallel + q_\perp \right ) \buni + \left (\frac{1}{2} p_\parallel + p_\perp \right ) \mvec{u} \right ] = (p_\perp - p_\parallel)  \buni\cdot[\buni\cdot\delx\mvec{u}] - p_\perp \delx\cdot\mvec{u}.
\end{align}
Recall that in the gyrotropic limit,
\begin{align}
  \mvec{P} = \mvec{P}^C = p_\parallel \buni\otimes\buni
  + p_\perp (\mvec{g}-\buni\otimes\buni) \notag,
\end{align}
which implies
\begin{align}
    (p_\perp - p_\parallel)  \buni\cdot[\buni\cdot\delx\mvec{u}] - p_\perp \delx\cdot\mvec{u} = -\mvec{P} : \delx\otimes\mvec{u}.
\end{align}
Thus, utilizing the fact that $3 p/2 = p_\parallel/2 + p_\perp$, we obtain the gyrotropic limit of \eqr{\ref{eq:int-er-2}} from the second moment of the $F_0$ equation and the zeroth moment of the $\mathcal{G}$ equation.

The statement that we evolve the perpendicular pressure $p_\perp$ by evolving $\mathcal{G}$ is non-trivial.
In effect, we have absorbed the $T_\perp$ normalization of the Laguerre expansion that appears in our fundamental ansatz \eqr{\ref{eq:f0-expand}} into the evolution of $\mathcal{G}$, even though this normalization must typically be determined independently and not depend on Laguerre coefficients.
Thus, our spectral expansion in Laguerres can fully leverage the optimized spatially and temporarily dependent $T_\perp$ normalization without the need for auxiliary equations.
Note this formulation of the equation for $\mathcal{G}$ indicates that we must have the constraint
\begin{align}
  \int_{-\infty}^\infty F_1 \dvol{v_\parallel} = 0.
\end{align}
From \eqr{\ref{eq:F1-coeff}}, can show that this is an initial-value constraint: that we must ensure  \eqr{\ref{eq:F1-coeff}} is true at $t=0$ for it to be satisfied for all $t>0$.

\subsection{The Lowest Order PKPM System of Equations}
We define the lowest order PKPM system as the zeroth Fourier harmonic and the first two Laguerre coefficients, with the truncation that all higher Fourier harmonics and all higher Laguerre coefficients are zero.
With these approximations, we are left with the following two coupled four dimensional kinetic equations
\begin{align}
  \pfrac{F_{0}}{t}
  &+
  \delx\cdot\left[ (v_\parallel\buni+\mvec{u})F_{0} \right] \notag \\
  &+
  \frac{\partial}{\partial v_\parallel}
  \left[
  \buni\cdot
  \left(
  \frac{1}{\rho}\delx\cdot\mvec{P} - v_\parallel\buni\cdot\delx\mvec{u}
  \right)  F_{0}
  + \mathcal{G}\delx\cdot\buni
  \right] = 0, \label{eq:reducedF0} \\
  \pfrac{\mathcal{G}}{t}
  &+
  \delx\cdot\left[ (v_\parallel\buni+\mvec{u})\mathcal{G} \right] \notag \\
  &+
  \frac{\partial}{\partial v_\parallel}
  \left[
  \buni\cdot
  \left(
  \frac{1}{\rho}\delx\cdot\mvec{P} - v_\parallel\buni\cdot\delx\mvec{u}
  \right)  \mathcal{G}
  +
  \left[
  4\mathcal{G} - 2 \frac{T_\perp}{m} F_0
 \right]\frac{T_\perp}{m}\delx\cdot\buni
  \right] \notag \\
  &=
  \left[
  \buni\cdot(\buni\cdot\delx\mvec{u})
  - \delx\cdot(v_\parallel\buni + \mvec{u})
  \right] \mathcal{G}.  \label{eq:reducedG}
\end{align}
The system is closed via the solution of an equation for the flow velocity, $\mvec{u}$, which comes from conservation of momentum,
\begin{align}
    & \pfrac{\rho \mvec{u}}{t} + \delx\cdot\left[\rho \mvec{u} \otimes \mvec{u} + \mvec{P} \right ] = \rho \frac{q}{m} \left [ \mvec{E} + \mvec{u} \times \mvec{B} \right ], \label{eq:finalMomentum} \\
    & \mvec{P} = p_\parallel \buni\otimes\buni
  + p_\perp (\mvec{g}-\buni\otimes\buni),  \label{eq:finalPressure} \\
    & p_\parallel = m \int v_\parallel^2 F_0 \thinspace dv_\parallel, \label{eq:finalppar} \\
    & p_\perp = m \int \mathcal{G} \thinspace dv_\parallel, \label{eq:finalpperp}
\end{align}
plus Maxwell's equations for the evolution of the electromagnetic fields,
\begin{align}
  \frac{\partial \mvec{B}}{\partial t} + \delx\times\mvec{E} &= 0, \label{eq:dbdt} \\
  \epsilon_0\mu_0\frac{\partial \mvec{E}}{\partial t} - \delx\times\mvec{B} &= -\mu_0\mvec{J}, \label{eq:dedt} \\
  \delx\cdot\mvec{E}&= \frac{\varrho_c}{\epsilon_0}, \label{eq:divE} \\
  \delx\cdot\mvec{B}&= 0, \label{eq:divB}
\end{align}
with the coupling to the electromagnetic fields arising directly from the momentum equation through the plasma currents,
\begin{align}
    \mvec{J} = \sum_s \frac{q_s}{m_s} \rho_s \mvec{u}_s.
\end{align}

%% file: pkpm-history-ctx.tex
\section{An Interlude on Historical Context}\label{sec:history}

Having derived the PKPM model utilizing velocity coordinate transformations to the local flow frame and frame aligned with the local magnetic field, followed by the spectral expansions in both velocity gyroangle and perpendicular velocity, it is worthwhile to place this system of equations in the larger historical context, with a focus on the lowest order PKPM system of equations consistenting of the zeroth Fourier Harmonic and the first two Laguerre coefficients.
Our principal goal throughout this discussion is two-fold: connect the physical content of the PKPM model to equation systems a reader may be more familiar with and simultaneously describe numerical difficulties which are ameliorated with the approach in the PKPM model. 
The discussion of numerical difficulties in certain asymptotic models is the principal motivation for this section, as our central ambition of the PKPM model is to leverage the same intuition that informed this long history of developing models for magnetized, collisionless plasmas, but with an eye towards avoiding the difficulties that have prevented some of these analytical breakthroughs from being easily numerically integrable and thus leveraged for efficient modeling of these systems. 

\subsection{Connection to Kulsrud's Kinetic MHD and Ramos' Finite Larmor Radius Kinetic Theory}

One of the most foundational models in the theory of magnetized plasmas is Kinetic Magnetohydrodynamics, often referred to as Kulsrud's Kinetic MHD or KMHD \citep{Kulsrud:1964, Kulsrud:1983}. 
KMHD is a hybrid fluid-kinetic model which transforms the kinetic equation in an identical fashion as the PKPM derivation outlined here, first transforming velocity coordinates to a bulk fluid frame, specifically the $\mvec{E} \times \mvec{B}$ frame, and then transforming the velocity coordinates to a field-aligned coordinate system $(v_\parallel, v_\perp, \theta)$. 
However, following these transformations, an asymptotic expansion is performed with small parameter $\epsilon = m/e$ or simply $\epsilon = 1/e$, where $m$ is the species mass and $e$ is the elementary charge. 
Because the bulk fluid flow being utilized is only the $\mvec{E} \times \mvec{B}$ velocity, this asymptotic expansion yields an immediate dynamical consequence: the parallel electric field is also $\mathcal{O}(\epsilon)$ smaller than the perpendicular electric field, and to lowest order the distribution function is gyrotropic. 

In this limit, the gyrotropic distribution function evolves as
\begin{align}
    \pfrac{\bdf_0}{t} 
    & + \delx \cdot \left [ \left (v_\parallel \buni + \mvec{U}_E \right ) \bdf_0 \right ] \notag \\ 
    & + \pfrac{}{v_\parallel} \left [ \left (-\buni \cdot \pfrac{\mvec{U}_e}{t} - \buni \cdot \left \{\mvec{U}_e \cdot \delx \mvec{U}_e \right \} + \frac{q_s}{m_s} E_\parallel - \buni \cdot \left \{v_\parallel \buni \cdot \delx \mvec{U}_e \right \} + \frac{v_\perp^2}{2} \delx \cdot \buni \right ) \bdf_0 \right ] \notag \\ 
    & + \pfrac{}{v_\perp^2/2} \left [ \frac{v_\perp^2}{2} \left ( \buni \cdot \left \{ \buni \cdot \delx\mvec{U}_e \right \} - \delx \cdot \left \{ v_\parallel\buni + \mvec{U}_e \right \} \right ) \bdf_0 \right ] = 0, \label{eq:KMHD}
\end{align}
where 
\begin{align}
    \mvec{U}_E = \frac{\mvec{E} \times \mvec{B}}{|\mvec{B}|^2},
\end{align}
is the $\mvec{E} \times \mvec{B}$ velocity, and we have transformed the perpendicular velocity derivative from $1/v_\perp \partial_{v_\perp}$ to a derivative on the variable $v_\perp^2/2$ for convenience since everywhere the perpendicular velocity appears in this equation, it appears as $v_\perp^2/2$.
Note that typically, a further coordinate transformation is performed from $v_\perp^2/2$ to the magnetic moment $\mu = v_\perp^2/(2B)$, where $B = |\mvec{B}|$ is the magnitude of the magnetic field, to further simplify the kinetic equation by eliminating the derivative in $v_\perp^2/2$ since the magnetic moment is conserved by the dynamics of this equation. 
Irrespective of whether one utilizes $v_\perp$, $v_\perp^2/2$, or $\mu$ as a velocity coordinate, \eqr{\ref{eq:KMHD}} is often referred to as the drift-kinetic equation \citep{Frieman:1966, HintonWong:1985}.

At first glance, it would seem that the only difference between this approach and the PKPM approach, at least when keeping only the zeroth Fourier harmonic, i.e., the gyrotropic component of the distribution function, is which particular hydrodynamic velocity is utilized for the initial velocity coordinate transformation. 
The characteristics of the particles in all the dimensions are identical save for a variable substitution of $\mvec{U}_E \rightarrow \mvec{u}$ and a subsequent manipulation of the parallel forces to utilize conservation of momentum, 
\begin{align}
    -\buni \cdot \pfrac{\mvec{u}}{t} - \buni \cdot \left \{\mvec{u} \cdot \delx \mvec{u} \right \} + \frac{q}{m} E_\parallel = \buni \cdot \left ( \frac{1}{\rho} \delx \cdot \mvec{P} \right ).
\end{align}
Indeed, we expect on scales larger than the gyroradius and in plasmas where the bulk velocity is dominated by the $\mvec{E} \times \mvec{B}$ velocity that the PKPM model contains all the physics of kinetic MHD.

The differences between the PKPM approach and the classical approach of Kulsrud are made more salient when considering how the system is closed. 
To close the KMHD system, the kinetic equation in \eqr{\ref{eq:KMHD}}, for each species, is coupled to a set of bulk fluid equations for the evolution of the total mass density and total momentum,
\begin{align}
    \pfrac{\rho}{t} + \delx \cdot \left ( \rho \mvec{U} \right ) = 0, \\
    \pfrac{\rho \mvec{U}}{t} +  \delx \cdot \left ( \rho \mvec{U}\otimes \mvec{U} + \sum_s \gvec{\mathcal{P}}_s \right ) = \mvec{J} \times \mvec{B},
\end{align}
where $\rho$ is the total mass density, often approximated as simply the ion mass density, and $\mvec{U}$ is the center-of-mass velocity, often approximated as simply the ion bulk velocity. 
Here, $\gvec{\mathcal{P}}_s$ is the pressure tensor of species $s$, 
\begin{align}
  \gvec{\mathcal{P}}_s = p_{\parallel_s} \buni\otimes\buni
  + p_{\perp_s} (\mvec{g}-\buni\otimes\buni),  \\
  p_{\parallel_s} = \int m_s \left (v_\parallel - u_{\parallel_s} \right )^2 f_s dv_\parallel d \left (\frac{v_\perp^2}{2} \right ),  \\
  p_{\perp_s} = \int m_s \frac{v_\perp^2}{2} f_s dv_\parallel d \left (\frac{v_\perp^2}{2} \right ), 
\end{align}
where we have reintroduced the species index to the mass, parallel velocity, and distribution function for clarity. 
We can see that $\gvec{\mathcal{P}}_s$ is the same gyrotropic pressure tensor we defined earlier for the lowest order PKPM system, but with a different means of computing the parallel pressure and assuming the only bulk perpendicular velocity is $\mvec{E} \times \mvec{B}$; in other words, there is an implicit ordering that the remaining guiding center drifts, such as $\delx B$ and curvature drifts, are all smaller than $\mvec{E} \times \mvec{B}$.
The forces on the bulk motion after summing over all species only require $\mvec{J}$, the current density of the plasma, usually given simply by
\begin{align}
    \mvec{J} = \frac{1}{\mu_0} \delx \times \mvec{B},
\end{align}
if one also assumes non-relativistic flows, though generalizations of the KMHD approach to relativity are possible via the relativistic generalizations of guiding center theory \citep{Vandervoort:1960, Ripperda:2018, Bacchini:2020, Trent:2023, Trent:2024}. 
Finally, the time evolution of the magnetic field is given by Faraday's equation, 
\begin{align}
    \pfrac{\mvec{B}}{t} = - \delx \times \mvec{E}
\end{align}
and in the commonly employed non-relativistic limit an Ohm's Law for the electric field gives the electric field
\begin{align}
    \mvec{E} = \mvec{U} \times \mvec{B}.
\end{align}
So, up to the coupling to the pressure tensor of each evolved species, the fluid equations, Faraday's Law, and the Ohm's Law for the electric field are exactly the same equations as those of ideal MHD. 
It is only the closure---how the pressure is determined for the evolution of the momentum---that is affected by the kinetic response of the plasma. 

Immediately, we note two subtleties to the evolution of the KMHD system: the total momentum equation permits the development of perpendicular motions which are not $\mvec{E} \times \mvec{B}$, but these bulk motions do not explicitly feedback on the kinetic response of the plasma, and the parallel electric field is as-yet-unspecified by the Ohm's Law. 
Firstly, the bulk motions drive perpendicular currents such as
\begin{align}
    \mvec{J}_{\perp_{\gvec{\mathcal{P}}}} & = -\frac{1}{|\mvec{B}|^2} \left [ \delx \cdot \left ( \sum_s \gvec{\mathcal{P}}_s \right ) \times \mvec{B} \right ] \notag \\
    & = -\underbrace{\frac{\delx \left (\sum_s p_{\perp_s} \right ) \times \mvec{B}}{|\mvec{B}|^2}}_{\textrm{diamagnetic}} - \underbrace{\sum_s \left (p_{\perp_s} - p_{\parallel_s} \right ) \frac{\left (\buni \cdot \delx \buni \right ) \times \mvec{B}}{|\mvec{B}|^2}}_{\textrm{curvature}}
\end{align}
which we have labeled as the diamagnetic and curvature-driven perpendicular currents respectively. 
However, as we noted earlier, the form of the kinetic equation in KMHD assumes there are no further contributions of perpendicular bulk flows to the computation of the perpendicular pressure. 
Thus for example, any diamagnetic currents which should modify the equilibrium pressure profile are not correctly captured by KMHD in its current form, at least dynamically, because the response of the kinetic equation does not include how the particles react to the presence of other perpendicular flows. 
These perpendicular currents do implicitly modify the local magnetic field, which itself may modify the plasma pressure through, e.g., adiabatic heating and cooling, but the interplay between the development of perpendicular flows and heating of the plasma is not contained in the KMHD system as written.
Importantly, these sorts of time-dependent interplays between different components of the system, while sub-dominant in the asymptotics, may be a critical component of well-behaved numerical discretizations of time-dependent partial differential equations, as including the interplay allows the system to relax to the desired equilibrium instead of trying to satisfy potentially stiff constraint equations. 

Further, to obtain an equation for the parallel electric field, we must go to higher order in our expansion to find the parallel force balance equation from the evolution of the electron momentum
\begin{align}
    E_\parallel = -\frac{m_e}{e n_e} \buni \cdot \left [ \pfrac{n_e \mvec{u}_e}{t} + \delx \cdot \left (n_e \mvec{u}_e\otimes \mvec{u}_e \right ) + \frac{1}{m_e} \delx \cdot \gvec{\mathcal{P}}_e \right ] \approx -\frac{1}{e n_e} \buni \cdot \delx \cdot \gvec{\mathcal{P}}_e, \label{eq:KMHD-Epar}
\end{align}
assuming the electron mass is small and thus the electron inertia is ignorable. 
Unfortunately, \eqr{\ref{eq:KMHD-Epar}} is only a steady-state equation, and thus its application to solving a time-dependent partial differential equation poses a nontrivial numerical difficulty. 
It need not be the case that \eqr{\ref{eq:KMHD-Epar}} holds instantaneously, and enforcing that it does may be impossible without some sort of numerical operator that diffuses spurious fluctuations in the parallel electric field. 

Similar to the feedback of perpendicular currents on the particle dynamics, while the equilibrium is contained at the desired order in the derivation of the equation system, the relaxation to this equilibrium is not. 
In the case of solving this constraint equation for the parallel electric field, in the physical system there will be the development of parallel currents due to a parallel electric field, which then must relax to a state of $J_\parallel \approx 0$ and a parallel electric field supported by the parallel pressure force.
But then in a time-dependent partial differential equation, we may develop spurious fluctuations in the parallel electric field caused by non-zero parallel currents since this equation system contains a kinetic equation for every evolved species.
We are thus not guaranteed to satisfy the identity $J_\parallel = 0$ and \eqr{\ref{eq:KMHD-Epar}} instantaneously in time as each species independently develops parallel flows. 

These two subtleties, the lack of feedback of the other guiding center drifts on the kinetic response of the plasma, and the parallel electric field equation only arising due to a steady-state parallel force balance equation, were principal motivations for the work of \citet{Ramos:2008} and \citet{Ramos:2016} and the derivation of what \citet{Ramos:2008} referred to as Finite Larmor Radius (FLR) Kinetic Theory. 
By transforming to the total bulk flow frame, irrespective of the make up of that bulk flow, whether large parallel flows exist or all the guiding center drifts are comparable in magnitude, all of the benefits of the KMHD formalism can be realized while eliminating the ambiguities of how to couple the fluid-kinetic system of equations. 
Not only the perpendicular electric field, but the parallel electric field too, are both eliminated from the kinetic equation via conservation of momentum. 
And with a Fourier expansion in velocity gyroangle, we can clearly identify both the evolution of the gyrotropic component of the distribution function and how the gyrotropic component couples to each subsequent Fourier harmonic as well as the impact of FLR effects on the plasma's evolution. 

In this regard, the PKPM formalism is the realization of the formalism outlined in \citet{Ramos:2008} to transform the velocity coordinates to the total flow frame but, to our knowledge, never numerically implemented anywhere. 
By transforming the velocity coordinates of each distribution function to their particular bulk flow frame and determining that bulk flow from conservation of momentum of that particular plasma species, the difficulties in handling the macroscopic parallel electric fields, parallel currents, and the evolution of the equilibrium magnetic field due to, e.g., diamagnetic effects, are all self-consistently captured by the interplay and coupling of Maxwell's equations with each species' momentum equations.
Where the PKPM approach diverges from the approach outlined in \citet{Ramos:2008} is the avoidance of any asymptotic ordering of our transformed Vlasov equation, and we instead reduce the number of velocity degrees of freedom compared to the full Vlasov-Maxwell system through the spectral expansions in velocity gyroangle and perpendicular velocity.
In lieu of expanding the transformed Vlasov equation in powers of $\epsilon = \rho_s/L$, where $\rho_s$ is the plasma species' gyroradius and $L$ is the gradient scale length, to include finite Larmor radius effects to some order in $\epsilon$, the PKPM model can simply add Fourier harmonics to some desired order, and we consider it likely that even the first Fourer harmonic is thus a super-set of the physics contained in the approach outlined in \citet{Ramos:2008}.
This argument is not to say that the PKPM model cannot benefit from asymptotic reductions; for example, if desired and appropriate, further numerical savings can be straightforwardly attained by employing reductions of Maxwell's equations, such as the Darwin approximation to eliminate the speed of light as the fastest time scale in the problem \citep{Schmitz:2006, Pezzi:2019a}, because the field-particle couplings are entirely contained within the momentum equation.

The fact that the PKPM model as constructed can handle the interplay between all of the guiding center drifts and develop macroscopic parallel electric fields from the self-consistent evolution of the kinetic equation and corresponding momentum equation can be contrasted with other kinetic MHD-like models in the literature.
For example, kinetic MHD-like models such as the \texttt{kglobal} model \citep{Arnold:2019, Drake:2019} avoid the  numerical difficulties associated with the macroscopic parallel electric field by leveraging physics intuition about how the system attempts to restore parallel pressure balance and zero-net parallel current, introducing a ``cold'' electron fluid to instantaneously maintain quasi-neutrality and provide the expected return current that balances the local parallel flow of the ``hot,'' kinetic electrons. 
A total parallel pressure balance equation can then be defined to give the instantaneous parallel electric field similar to an Ohm's law prescription for the perpendicular electric field. 
In reality of course, because one is solving a kinetic equation, this return current should be self-consistently contained as a sub-population of electrons in the electron distribution function. 
So, while \texttt{kglobal} does successfully apply the KMHD formalism, this formalism is leveraged only under a prescribed model for how the system maintains net zero parallel current. 

The approach in models like \texttt{kglobal} has been highly successful \citep{Arnold:2021, Arnold:2022}, a clear demonstration for why it has arguably been frustrating historically that KMHD resisted such easy discretization.
We thus argue the PKPM formalism presents a step forward in applying the same intuition which guided \citet{Kulsrud:1964} and later \citet{Ramos:2008} to a diverse array of plasma systems. 
The dynamical interplay of the guiding center drifts with the kinetic response of the plasma, the evolution of the macroscopic electromagnetic fields including macroscopic parallel electric fields, and the self-consistent treatment of the distinct particle populations which may arise to, e.g., drive the parallel currents to zero, are all handled by the PKPM formalism at reduced computational cost compared to solving the Vlasov equation in full generality. 

\subsection{Comparison to other (hybrid) spectral method approaches}

Given the widespread popularity of spectral methods for velocity space discretizations of kinetic equations, from fully kinetic \citep{Holloway:1996, Delzanno:2015, Parker:2015, Vencels:2016, Roytershteyn:2018, Koshkarov:2021, Pagliantini:2023, Issan:2024, Schween:2024, Schween:2025} to gyrokinetics \citep{Mandell:2018, Frei:2020, Hoffmann:2023a, Frei:2024} to drift-kinetics \citep{Parker:2016} and other reduced kinetic models  \citep{Zocco:2011, Loureiro:2013, Zocco:2015, Loureiro:2016}, it is worth discussing how the PKPM approach compares to these other techniques for discretizing velocity space. 
The key differences are three-fold: the direct optimization of the spectral basis with our coordinate transformations that avoids the difficulty in handling time and spatially dependent shift and normalization factors, the hybrid nature of the PKPM approach, which does not perform a spectral expansion in all the velocity degrees of freedom, and the lack of transformation of configuration space coordinates which distinguishes the gyroangle Fourier harmonic expansion from the gyroaveraging procedure in spectral gyrokinetic codes. 
All of these differences arise due to specific goals of the PKPM model; for example, as discussed in the beginning, we anticipate a spectral basis in $v_\parallel$ can perform quite poorly on the phase space structures which typically arise in magnetized plasma dynamics, such as field-aligned beams or trapped particle distributions. 

Likewise, we focused in Section~\ref{sec:prelim} on why we pursued a path of transforming the velocity coordinates of the Vlasov equation to optimize the spectral basis.
We reiterate that not only do we avoid any assumptions on the temporal or spatial variation of the flow velocity with our coordinate transformation to move with the local flow velocity, but that the linear combination of Laguerre coefficients to produce the second kinetic equation for $\mathcal{G}$, \eqr{\ref{eq:G-coeff}}, in Section~\ref{sec:pkpm} eliminates the need for an auxiliary equation for the time and spatially dependent Laguerre normalization. 
Thus, we avoid the restriction in other spectral approaches which commonly assume a fixed and/or uniform shift and normalization in the spectral expansion \citep{Vencels:2016,Koshkarov:2021,Frei:2023b,Frei:2024}\footnote{We draw a distinction here between spectral methods utilized in $\delta f$ equations, such as those used to discretize the $\delta f$ gyrokinetic equation \citep{Mandell:2018, Hoffmann:2023a} compared to ``full'' $f$ discretizations of either the Vlasov equation or gyrokinetic equation since in a $\delta f$ approach it is very natural to assume a fixed shift and normalization to the spectral basis either in both space and time or just time.}. 
And, we will show in Section~\ref{sec:demo} test cases that illustrate the utility of not performing the spectral expansion in $v_\parallel$ due to the non-trivial structure which can arise in a variety of magnetized systems, such as magnetic reconnection, similar to other groups use of a hybrid approach mixing spherical harmonics with finite element methods to optimize their simulations of cosmic ray transport \citep{Schween:2024, Schween:2025}. 

We thus use this section to draw a particular contrast between the spectral gyrokinetic approach and the PKPM model for the handling of the Laguerre expansion. 
There is an important consequence from the transformation from particle position to gyrocenter position in spectral velocity representations of the gyrokinetic equation: the Laguerre representation of the Bessel function utilized to perform the gyroaveraging of various quantities, such as in the electrostatic potential, requires a sum over Laguerres \citep{Zocco:2015}:
\begin{align}
    J_0\left(k_\perp \rho_i v_\perp\right) = e^{-\frac{1}{4}k_\perp^2 \rho_i^2} \sum_{n=0}^{\infty} \frac{\left(k_\perp^2 \rho_i^2 / 4\right)^n}{n!} L_n\left(v_\perp^2\right).
\end{align}
Here, $J_0$ is the zeroth order Bessel function of the first kind, $k_\perp$ is the wavenumber perpendicular to the magnetic field, and $\rho_i$ is the ion gyroradius. 
An accurate treatment of gyroaveraging for large $k_\perp \rho_i$ thus requires an increasing number of Laguerre coefficients---see Appendix B of \citet{Mandell:2018}. 
While accuracy improvements for low Laguerres have been utilized in the gyrofluid approach \citep{Dorland:1993}, and a number of nonlinear calculations have shown reasonable results in the fusion context, with works such as \citet{Hoffmann:2023b} and \citet{Mandell:2024} utilizing as few as 3-4 Laguerres, in other cases such as the unstable entropy mode in \citet{Hoffmann:2023a}, more Laguerre resolution is needed to properly account for the FLR effects impact on the transport. 

No such coupling of Laguerres occurs in the PKPM model because we have kept the configuration space coordinates untransformed; we obtain the distribution function at the particle position, not the gyrocenter position. 
In the PKPM formalism, each successive Fourier harmonic in velocity gyroangle will have its own Laguerre expansion. 
We can thus identify the physics of each Fourier harmonic Laguerre coefficient by Laguerre coefficient. 
For example, the first Fourier harmonic gives the evolution of $\mvec{M}_\perp$, and the zeroth Laguerre coefficient of $\mvec{M}_\perp$ can be utilized to obtain the component of the agyrotropic pressure tensor proportional to $\mvec{M}_\perp$ in \eqr{\ref{eq:Pi-agyro}}. 
Likewise, the first Laguerre coefficient of $\mvec{M}_\perp$ can be utilized to obtain the heat flux of the perpendicular temperature perpendicular to the magnetic field in \eqr{\ref{eq:vec_q}}, the $v_\perp^2 \mvec{M}_\perp$ term. 
We defer a systematic comparison of the accuracy of FLR effects in the PKPM approach compared to a spectral gyrokinetic code to a future publication, but we note for now that because the PKPM model does not transform configuration space coordinates to gyrocenter coordinates or perform any gyroaveraging, the Laguerre couplings remain local at every $k_\perp$, and thus there is no obvious need for large Laguerre resolution at large $k_\perp$ in the PKPM approach. 
What specific impact this representation of the distribution function in perpendicular velocity and gyroangle, instead of gyrocenter, perpendicular velocity, and gyrophase, has on the physics of e.g., the entropy cascade \citep{Schekochihin:2009, Tatsuno:2009} is as-yet undetermined, but at an initial glance the couplings which can require high Laguerre resolution are not present in the PKPM approach. 

%% file: pkpm-demo.tex
\section{A Brief Demonstration of the Model}\label{sec:demo}

We now seek to demonstrate the new PKPM model in a handful of non-trivial nonlinear problems which clearly show the utility of the approach. 
We emphasize here that the following benchmarks are merely meant to exhibit the successful numerical implementation of the model, and we defer a systematic comparison to theory and fully kinetic simulations to the second paper in this two-part series. 
Nevertheless, it is the goal of this section to illustrate that the significant reduction in the number of degrees of freedom yields a cost-effective model for understanding plasma systems which are inaccessible with traditional asymptotic approaches. 
The PKPM model is implemented within the \texttt{Gkeyll} simulation framework utilizing a discontinuous Galerkin finite element method \citep{ReedHill:1973,Cockburn:1998b,Cockburn:2001,Hesthaven:2007} for the spatial discretization of all components of the system: the kinetic equations for the $F_0$ and $\mathcal{G}$ distribution functions, \eqr{\ref{eq:reducedF0}} and \eqr{\ref{eq:reducedG}}, the conservation of momentum equation coupled to the pressure tensor computed from these kinetic equations, \eqr{\ref{eq:finalMomentum}}--\eqr{\ref{eq:finalpperp}}, and Maxwell's equation. 
Time integration is handled with an explicit strong stability preserving third-order Runge-Kutta method \citep{Shu:2002}.
These choices of spatial discretization and time integration follow exactly the same procedures as other \texttt{Gkeyll} kinetic equation implementations \citep{Juno:2018,HakimJuno:2020,Mandell:2020}. 
All the simulations performed in these demonstrations of the model also utilize the same discontinuous Galerkin method for a conservative Lenard-Bernstein operator for self-collisions---the extension of the Lenard-Bernstein collision operator for the PKPM model is given here in Appendix~\ref{app:LBO}, and the numerical details of discretizing the collision operator can be found in \citet{Hakim:2020}.  

\subsection{Parallel Electrostatic Shock}

Shock waves, specifically collisionless shock waves, are omnipresent in our universe. 
In astrophysical systems where the collisional mean-free-path is  large, the dynamics of the nonlinear wave steepening and rapid conversion of bulk kinetic energy into other forms of energy such as particle acceleration and heating in these shock waves are not mediated by collisions, but a myriad of collisionless processes such as wave-particle interactions and kinetic instabilities.
To demonstrate the PKPM model, we consider the case of a parallel shock, where the incoming supersonic flow is aligned with the local magnetic field. 
In this shock geometry, on short time scales, there is the potential for electrostatic shocks driven by nonlinear steepening of ion acoustic modes, but only up to a critical Mach number \citep{Forslund:1970, Sorasio:2006}. 
If the incoming flows are sufficiently large, there will be no slow down of the supersonic flow, and the fast plasma can propagate freely, smoothly transitioning around obstacles or interpenetrating the ambient medium through which the plasma is propagating. 

To demonstrate that the PKPM model reproduces this transition from shocked flows to interpenetrating flows and thus contains a complete description of the collisionless physics of electrostatic shocks, we perform two simulations of colliding plasma flows, similar to previous studies of electrostatic shocks.
We initialize an electron-proton plasma in a box of length $L_x = 3584 \lambda_D$, where $\lambda_D = v_{th_e}/\omega_{pe}$ is the electron Debye length, $v_{th_e} = \sqrt{T_e/m_e}$ is the electron thermal velocity, and $\omega_{pe} = \sqrt{e^2 n_0/\epsilon_0 m_e}$ is the electron plasma frequency. 
Here, $T_e$ is the electron temperature, $m_e$ is the electron mass, $e$ is the elementary charge, $n_0$ is the reference electron density, and $\epsilon_0$ is the permittivity of free space. 
For the purposes of these simulations, all of these quantities are normalized to values of 1.0 so that all length scales are normalized with respect to the electron Debye length, all velocities are normalized with respect to the electron thermal velocity, and all time scales are normalized with respect to the inverse electron plasma frequency. 
This electron-proton plasma is initialized with Maxwellian distributions and a supersonic flow for both species in the negative $x$-direction towards a reflecting wall at $x=0$, which leads to a shock wave or interpenetrating plasma propagating in the positive $x$-direction. 
A continuous supply of plasma is provided with a copy boundary condition at $x=L_x$\footnote{Within \texttt{Gkeyll}, this boundary condition corresponds to copying all quantities: both PKPM distribution functions $F_0$ and $\mathcal{G}$ and momentum $\rho \mvec{u}$ for both species, as well as the electromagnetic fields, in the grid cell just abutting the boundary, the ``skin'' cell, into the ghost or halo layer of cells. In this case, because the momentum is initialized in the negative $x$-direction, we then have a continuous injection of plasma from the boundary at $x=L_x$.} in exact analogy to the ``reflecting-wall'' setup commonly employed in particle-in-cell simulations of collisionless shocks and identical to the initial conditions of previous continuum simulations of collisionless shocks.

Specific shock parameters are as follows: we utilize the real proton-electron mass ratio $m_p/m_e = 1836$, a proton-electron temperature ratio $T_p/T_e = 0.25$, a reference magnetic field strength $\mvec{B} = B_0 \hat{\mvec{x}} = \hat{\mvec{x}}$ so the magnetic field points in the x-direction the entirety of the simulation on the time scale of the electrostatic dynamics, and an electron-electron collisionality $\nu_{ee} = 10^{-6} \omega_{pe}$, with the ion-ion collisionality commensurately decreased by the square root of the mass ratio and temperature ratio to the 3/2 power. 
We consider two Mach numbers, one below and one above the critical Mach number, where $M_s = u_{x_0}/c_s$ is the Mach number, $u_{x_0}$ is the magnitude of the upstream flow velocity, and $c_s = \sqrt{T_e/m_p}$ is the sound speed.
The two simulations have upstream Mach numbers of $M_s = 3.0$ and $M_s = 5.0$. 
We choose to utilize a colder proton population to reduce the ion acoustic damping and thus generate stronger shock waves in the cases when the plasma produces a shock. 
The critical Mach number is $M_s \sim 3.0$, but we note that this critical Mach number was determined for plasmas with only one velocity dimension \citep{Forslund:1970}, and the PKPM model is constructed to be three-dimension in velocity-space. 
As such, we expect with this definition of the sound speed, without any additional $\mathcal{O}(1)$ factors for the adiabatic index of the plasma, that the $M_s = 3.0$ simulation will shock, while the $M_s = 5.0$ simulation will not shock\footnote{For a plasma with three velocity dimensions, the sound speed in the cold ion limit is formally $c_s = \sqrt{\gamma_e T_e/m_p}$, where $\gamma_e = 5/3$. With this definition of the sound , our two simulations have upstream Mach numbers of $M_s = 3.0/\sqrt{5/3} \sim 2.32$ and $M_s = 5.0/\sqrt{5/3} \sim 3.87$. Because our protons are colder than the electrons, even if we include the contribution to the sound speed from the ion temperature, $c_s = \sqrt{(\gamma_e T_e + \gamma_i T_i)/m_p} = \sqrt{25/12} \sqrt{T_e/m_p}$, the $M_s  = 3.0/\sqrt{25/12} \sim 2.04$ and $M_s  = 5.0/\sqrt{25/12} \sim 3.46$, we still satisfy the conditions for the two simulations being below and above the critical Mach number.}. 
This subtlety of the conditions under which a one-velocity dimensional plasma shocks compared to a three-velocity dimensional plasma has already been proven to be a use case for the PKPM model, as an early version of the PKPM model was utilized to understand measurements of shock formation, and lack of shock formation compared to fluid model predictions, in the Plasma-Jet Driven Magneto-Inertial Fusion experiment \citep{Cagas:2023}. 

The final input file specifications are our velocity space extents, grid resolutions, and polynomial orders for the discontinuous Galerkin finite element method we utilize. 
The velocity grid extents are $[-8 v_{th_e}, 8 v_{th_e}]$ for the electrons and $[-64 v_{th_p}, 64 v_{th_p}]$ for the protons for the $M_s = 3.0$ simulation and $[-128 v_{th_p}, 128 v_{th_p}]$ for the protons for the $M_s = 5.0$ simulation\footnote{Note that because this simulation is electrostatic and the only coupling between the plasma and electromagnetic fields is Ampere's Law in 1D where $\nabla_{\mvec{x}} \times \mvec{B} = 0$, we do not need to specify $v_{th_e}/c$ for how non-relativstic the simulation is, as there are no light waves in this simulation.}. 
We utilize $N_x = 1792$ grid points in configuration space, $\Delta x = 2 \lambda_D$, with linear polynomials in configuration space, and $\Delta v_\parallel = 0.0625 v_{th_s}$ for both species with quadratic polynomials in velocity space, so that the electrons have $N_{v_\parallel} = 256$ and the protons have $N_{v_\parallel} = 2048$ and $N_{v_\parallel} = 4096$ velocity grid points for the $M_s = 3.0$ and $M_s = 5.0$ simulations respectively. 
These simulations utilized 32 56-core compute nodes on the Frontera cluster at the Texas Advanced Computing Center; the $M_s = 3.0$ ran for half an hour for a total of 16 node hours, while the $M_s = 5.0$ used an hour of computing time for a total of 32 node hours. 

Currently we utilize large velocity space extents for the protons in shock simulations due to the significant transient generated by the large $\partial u_x/\partial x$ at early times, since with this reflecting wall setup  $\partial u_x/\partial x$ is infinite at $t=0, x = 0$. 
The result of this large $\partial u_x/\partial x$ is the generation of a small amount of reflected particles at very high velocity in the hydrodynamic frame which propagate upstream and have no bearing on the evolution of the plasma. 
Once the transient has relaxed, the vast majority of the plasma is contained in a velocity space volume a quarter to an eighth in size, and we consider it worthwhile future work to explore an optimized initial setup of shock simulations in this model to reduce the need to resolve this transient high velocity reflection. 
Nevertheless, we wish to emphasize that the fact that this numerical implementation can successfully evolve this transient stably is evidence of the robustness of this implementation, and thus suggests that the PKPM model is not by any means a numerically brittle model despite its complexity. 

\begin{figure}
    \centering
    \includegraphics[width=.88\textwidth]{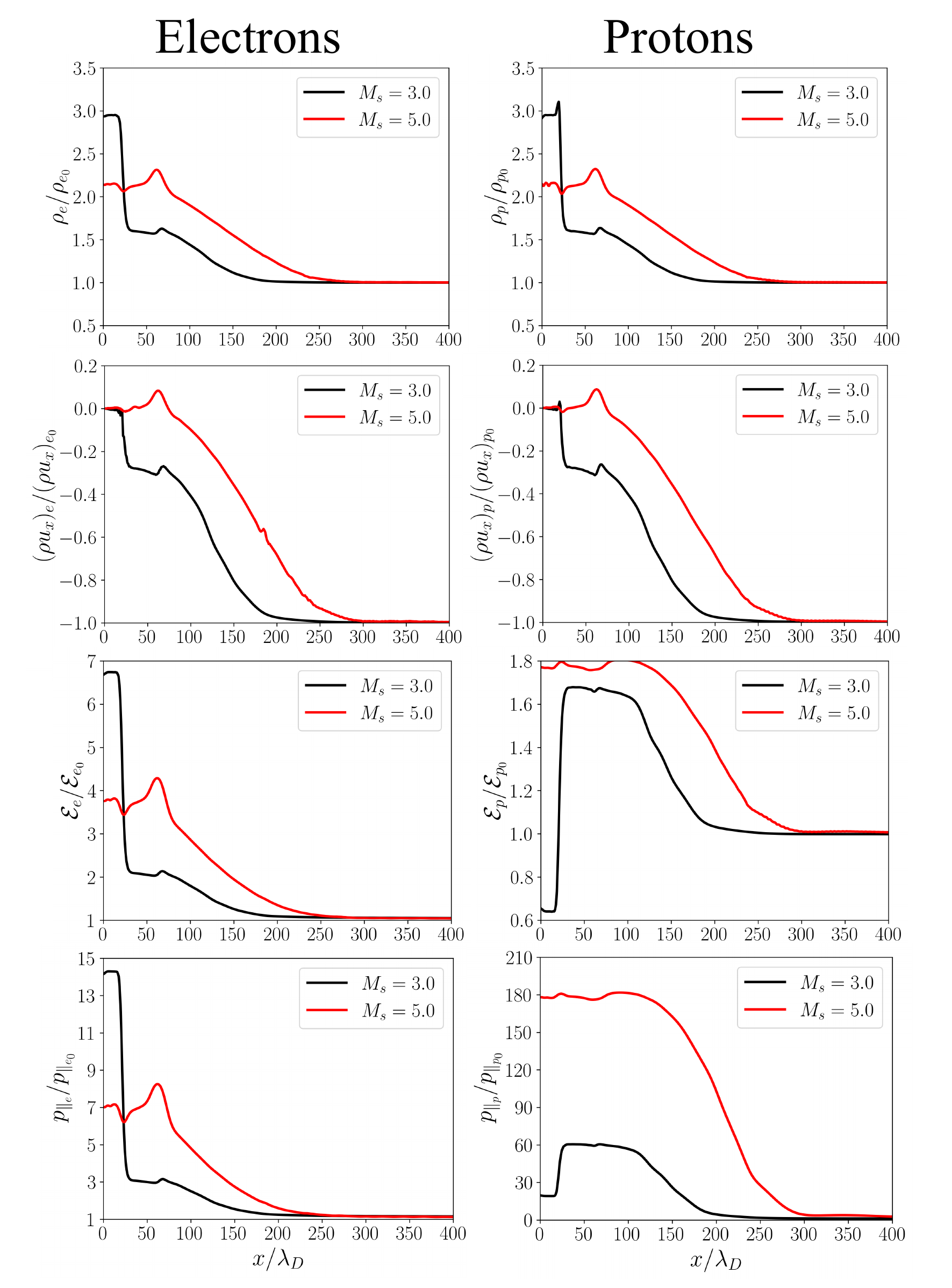}
    \vspace{-0.5cm}    
    \caption{Electron (left column) and proton (right column) mass density (top row), momentum density (middle top row), total energy density (middle bottom row), and parallel pressure (bottom row) at $t=1500\omega_{pe}^{-1}$ for upstream Mach number $M_s = 3.0$ (black) and $M_s = 5.0$ (red). All quantities are normalized to their upstream values for ease of comparison between the $M_s = 3.0$ and $M_s = 5.0$ cases since their upstream flows and energies are different. The characteristics of a shock wave are clearly identifiable in the $M_s = 3.0$ simulation: a sharp pile-up of the density, a rapid stagnation of the flow, significant electron heating over the same length scale, and a decrease in the ion energy from the rapid conversion of ion energy into both electron heating and electromagnetic energy. On the other hand, the $M_s = 5.0$ case exhibits no such sharp transitions, with a smooth gradient up to a total mass density $\rho \sim 2$ and total momentum $\rho u_x \sim 0.0$ for both the electrons and protons, corresponding to two interpenetrating beams of plasma.}
    \label{fig:shock_moms}
\end{figure}

We show in Figure~\ref{fig:shock_moms} the evolution of the fluid moments at $t=1500 \omega_{pe}^{-1}$ for the two upstream Mach numbers. 
All fluid moments are normalized to their upstream values to facilitate cross comparison between the $M_s = 3.0$ and $M_s = 5.0$ simulation since the upstream flows and energies are different for these two simulations. 
We note the key features of the $M_s = 3.0$ indicative of a shock: a sudden density pile-up, a sharp stagnation of the flow in conjunction with this density pile-up, rapid electron heating with the electron parallel pressure increasing by nearly a factor of 15 compared to the density increasing by a factor of 3, and a commensurate decrease in the total ion energy and ion pressure corresponding to the conversion of ion energy into electron heating and electromagnetic energy. 
In contrast, the $M_s = 5.0$ simulation shows a smoother transition to the downstream region that is more characteristic of two interpenetrating beams of plasma: the mass density and ion energy density downstream are $\rho_s \sim 2, \mathcal{E}_p \sim 2$, and so the two colliding plasma beams are simply adding their density and energy. 

\begin{figure}
    \centering
    \includegraphics[width=.495\textwidth]{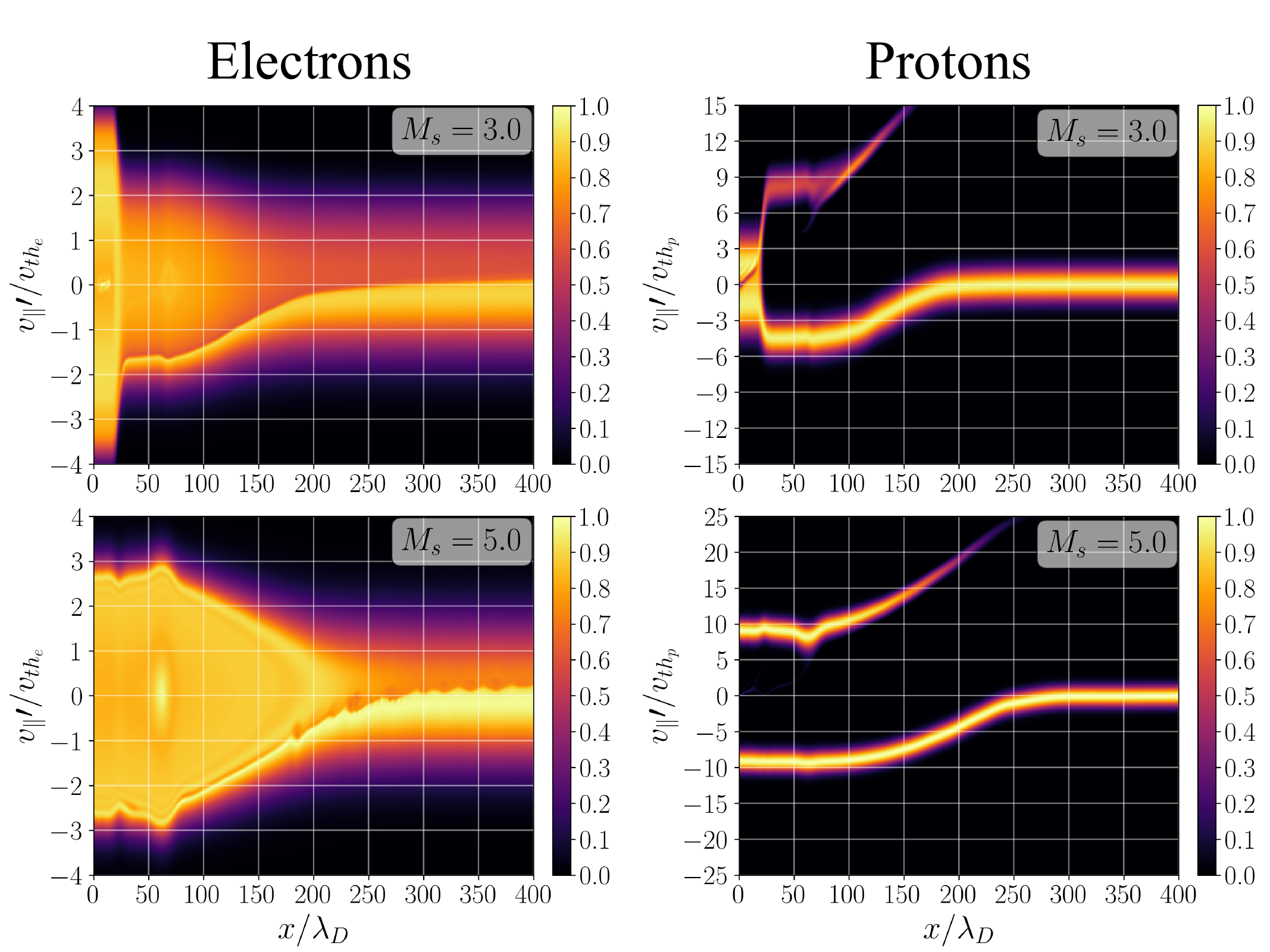}
    \includegraphics[width=.495\textwidth]{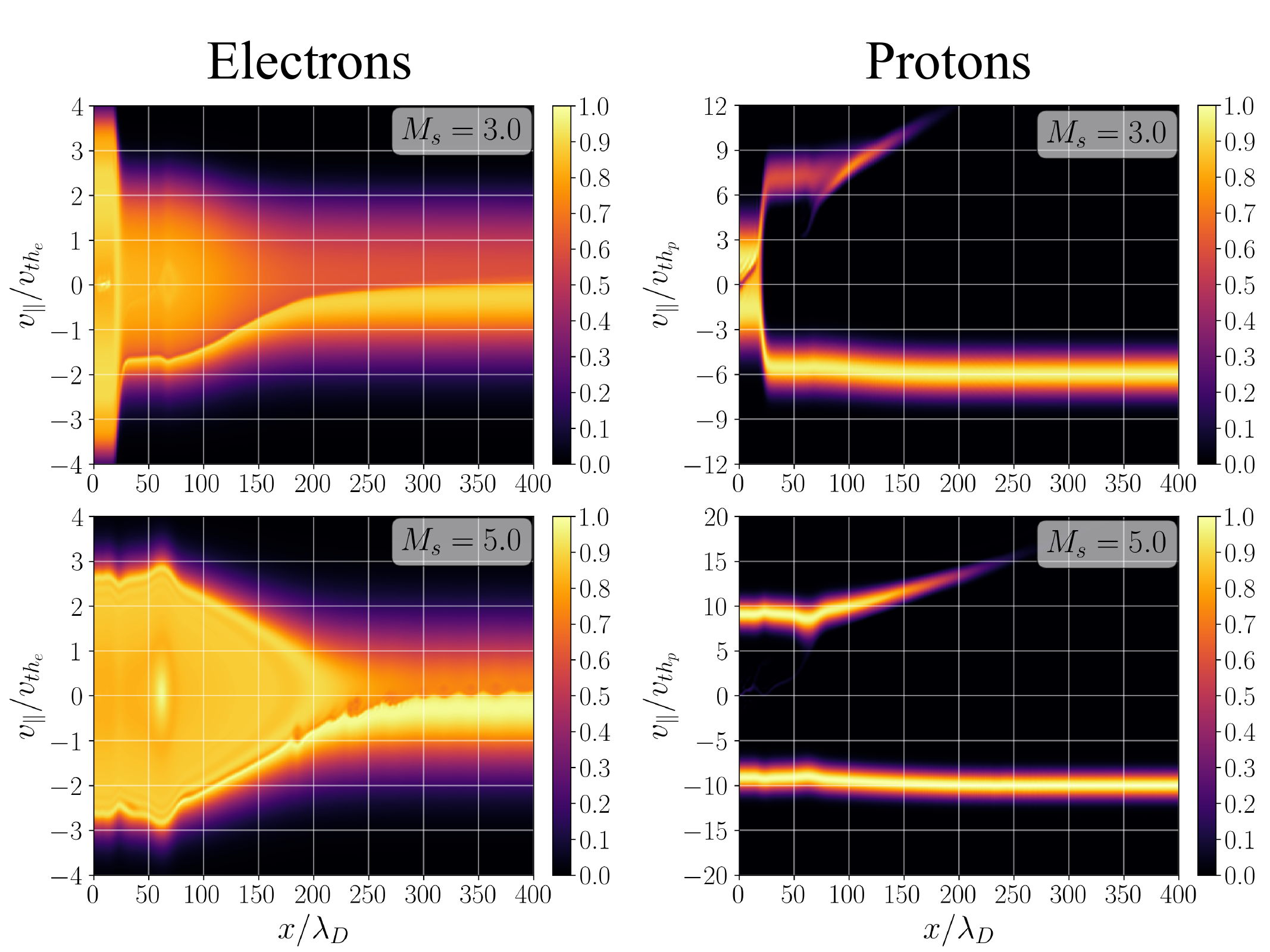}
    \caption{The $F_0$ distribution function in the local fluid flow frame for the electrons (left column) and protons (left middle column), and the $F_0$ distribution function in the lab frame for the electrons (right middle column) and protons (right column) for the $M_s = 3.0$ (top row) and $M_s = 5.0$ (bottom row) simulations. In the lab frame, the incoming proton beam is centered at the upstream velocity, $u_{x_0} = 6.0 v_{th_i}$ and $u_{x_0} = 10.0$, as we expect, and the characteristics of the shock with the trapped electron and ion populations are identifiable in the $M_s = 3.0$ simulation, while the $M_s = 5.0$ simulation shows only two distinct ion beams propagating through each other. We also draw attention to the form of the proton distribution function in the local fluid flow frame and emphasize that these are the distribution functions which are are directly solved for by the numerical method. As expected, the distribution function adjusts in the local fluid flow frame to preserve the identity that the first moment is zero. Note that these distribution function plots are normalized to their respective maximum values on the grid, e.g., $F_{0_s} = F_{0_s}/\max(F_{0_s})$. }
    \label{fig:shock_dist}
\end{figure}

We can see further evidence of the shocked versus unshocked flows examining the distribution functions for these two upstream Mach numbers, specifically the $F_0$ coefficient corresponding to integrating the distribution function over both $v_\perp$ and $\theta$. 
We plot in Figure~\ref{fig:shock_dist} the electron and proton $F_0$ distribution functions in both the local fluid flow frame provided by the numerical solution of the PKPM model, and in the lab frame as would traditionally be obtained from the solution of the Vlasov-Maxwell system of equations---see, for example, \citet{Juno:2018} for simulations of electrostatic shocks with \texttt{Gkeyll}'s Vlasov-Maxwell solver. 
Note that these distribution function plots are normalized to their respective maximum values on the grid, e.g., $F_{0_s} = F_{0_s}/\max(F_{0_s})$. 
For the $M_s = 3.0$ simulation, we can more clearly identify the characteristics of a shock, specifically the trapping of both electrons and protons in the downstream region around $x \sim 25 \lambda_D$. 
On the other hand, the $M_s = 5.0$ simulation shows no significant broadening of the electron distribution function. 
Further, the proton distribution function in the $M_s = 5.0$ simulation is simply two interpenetrating beams of protons at close to the upstream flow velocity of $u_{x_0} = 10 v_{th_p}$. 
We draw particular attention to the local fluid flow frame distribution functions for the protons as evidence of the successful nontrivial numerical implementation of this model. 
We observe that as the reflected population propagates upstream, the fluid flow frame distribution function naturally adjusts due to the pressure forces to produce an approximately even distribution function with only high odd moments, such as heat fluxes.
The first moment of the $F_0$ distribution function thus remains 0 as we expect.

This initial electrostatic evolution only couples the $F_0$ and $\mathcal{G}$ kinetic equations through the collision operator; and in both of these simulations, but especially the $M_s = 3.0$ case where a true shock develops, a significant temperature anisotropy develops, with $T_\parallel > T_\perp$.
Therefore, as the shock propagates, electromagnetic instabilities can be excited which will generate large transverse fluctuations and change the shock from being a purely parallel shock to a quasi-parallel, or even locally quasi-perpendicular, shock. 
The fastest growing modes in this case, such as the electron firehose and Alfv\'en ion cyclotron instabilities, require finite Larmor radius effects to properly capture their growth and saturation, and thus only retaining the zeroth Fourier harmonic is inadequate for accurately simulating the transition from an electrostatic to an electromagnetic parallel shock \citep{Gary:2001}.
Nevertheless, recent work using extended fluid models to model the proton parallel firehose instability suggests that accounting for the agyrotropic components of the pressure tensor provides an accurate model for these temperature anisotropy-driven instabilities via the finite Larmor radius effects approximated by evolving the full pressure tensor \citep{Walters:2024}.
Thus, adding one, or at most two, Fourier harmonics will be sufficient to capture the necessary finite Larmor radius effects to simulate both where in wavenumber space the fastest growing modes exist, and their overall saturation due to the generated magnetic field structure, making the PKPM approach a potentially powerful tool for modeling of collisionless shocks in these particular parameter regimes.

We note that at higher Mach numbers where significant particle acceleration occurs due to processes such as shock-drift acceleration \citep{Paschmann:1982, Sckopke:1983, Anagnostopoulos:1994, Anagnostopoulos:1998, Ball:2001, Anagnostopoulos:2009} and diffusive shock acceleration \citep{Fermi:1949, Fermi:1954, Blandford:1978, Ellison:1983, Blandford:1987, Decker:1988, Malkov:2001}, the PKPM model is likely to be inefficient for representing the kinetic response of the plasma. 
The particle distribution functions which result from these shock acceleration processes are typically highly agyrotropic, and would thus necessitate a large number of Fourier harmonics and Laguerre coefficients to resolve. 
But, we emphasize that in the case considered here, at relatively low Mach number, the relative inexpensiveness of these simulations means that, once the necessary finite Larmor radius effects are included for correctly modeling the saturation of the excited temperature anisotropy instabilities, the multiscale nature of these subcritical, collisionless shocks which transition from electrostatic to electromagnetic shocks will be possible with the PKPM model. 
Further, because the PKPM model is ultimately defined per species, with the transformation of the velocity coordinates to move with the local flow velocity employing that particular species flow velocity, we can imagine unique hybrid modeling approaches where a fully kinetic proton treatment could be coupled to a PKPM electron treatment, thus reducing the computational expense of modeling electrons at modest Mach numbers where the electrons stay magnetized through the shock.

\subsection{Moderate Guide-Field Magnetic Reconnection}
Equally ubiquitous to collisionless shocks as a mechanism by which plasma's rearrange their energy budget is the phenomenon of magnetic reconnection, whereby magnetic fields change their topology to a lower energy state and transfer this excess energy to the plasma. 
In collisionless plasma systems where the resistivity is very small, the onset of magnetic reconnection is considered to be a fundamentally kinetic process. 
The plasma can only break field lines at microscopic length scales where the plasma demagnetizes and the plasma particles are no longer constrained to follow the magnetic field. 

Importantly though, magnetic reconnection is commonly observed component-wise; many plasma systems are undergoing what is referred to as ``guide-field'' reconnection. 
In the geometry of these systems, we can divide the dynamical fields into a set of planar reconnecting components and a component perpendicular to the plane known as the guide field that the plasma particles will still try to ``stick'' to even as the in-plane reconnecting components vanish at the point of magnetic reconnection where the magnetic field is changing its topology. 
We can thus ask: even with retaining only the zeroth Fourier harmonic and thus constraining the plasma to be gyrotropic, how well does the PKPM model capture guide-field reconnection?

A number of studies have shown gyrokinetic models of reconnection compare favorably to fully kinetic simulations in the limit that the guide field is strong \citep{TenBarge:2014, Munoz:2015}, $B_g \gg B_0$ where $B_g$ is the magnitude of the guide field and $B_0$ is the magnitude of the reconnecting field. 
Alternatively, one can think of this limit as the $\delta B/B \ll 1$ limit---a natural limit for standard derivations of gyrokinetics that assume a strong background field and only weak perturbations. 
But, we have continually emphasized that the PKPM approach is not an asymptotic one, and thus we need not restrict ourselves to a strong guide field limit. 
So long as there is a guide field of at least modest strength, do the plasma particles stay magnetized through the layer? 
The answer to this question is yes: a number of studies \citep{Swisdak:2005, Le:2013, Egedal:2013} have shown that even at $B_g \approx 0.1-0.2 B_0$, or $\delta B/B \sim 5-10$, the electrons are still magnetized through the reconnection layer, provided the electrons are sufficiently light and a realistic proton-electron mass ratio is utilized. 

We consider two moderate-guide field cases: $B_g = B_0$ and $B_g = 0.5 B_0$, or $\delta B/B_g = 1$ and $\delta B/B_g = 2$. 
We initialize a force-free current sheet in 2X1V, $(x,y,v_\parallel)$, geometry of the form 
\begin{align}
    J_{x_e} & = -\frac{B_0}{w_0} \frac{\tanh \left (\frac{y}{w_0} \right ) \sech^2 \left (\frac{y}{w_0} \right )}{\sqrt{ \left (\frac{B_g}{B_0} \right )^2 + \sech^2 \left (\frac{y}{w_0} \right )}}, \\
    J_{z_e} & = \frac{B_0}{w_0} \sech^2 \left (\frac{y}{w_0} \right ), \\ 
    B_x & = -B_0 \tanh \left (\frac{y}{w_0} \right ), \\
    B_z & = B_0 \sqrt{\left (\frac{B_g}{B_0} \right )^2 + \sech^2 \left (\frac{y}{w_0} \right )},
\end{align}
where current is given entirely to the electrons to satisfy Ampere's Law. 
A $\sim 1\%$ GEM-like perturbation is utilized \citep{Birn:2001}, along with perturbations to the first 20 wave modes with $\sim 1\%$ noise in $B_x, B_y,$ and $J_z$ to break the symmetry of the GEM-like perturbation and accelerate the development of the reconnection, similar to \citet{TenBarge:2014}.
These simulations utilize the standard GEM reconnection challenge proton-electron temperature ratio $T_p/T_e = 5$, a realistic proton-electron mass ratio, $m_p/m_e = 1836$, and a similar box size as the original GEM reconnection challenge: $L_x = 8 \pi d_p, L_y = 4 \pi d_p$, where $d_p = c/\omega_{pp}$ is the proton inertial length, and $\omega_{pp} = \sqrt{e^2 n_0/\epsilon_0 m_p}$ is the proton plasma frequency respectively. 

Other parameters are as follows: $\beta_e = 2\mu n_0 T_e/B_0^2 = 1/6$ defined in terms of the upstream in-plane magnetic field, velocity space extents $[-8 v_{th_s}, 8 v_{th_s}]$ for both the electrons and protons, $v_{th_e}/c = 1/16$, $\nu_{ee} = 10^{-2} \Omega_{cp}$ where $\Omega_{cp} = eB_0/m_p$ is the proton cyclotron frequency defined in terms of the upstream in-plane magnetic field strength. 
$\nu_{pp}$ is commensurately smaller by the square root of the mass ratio\footnote{Note that the proton-proton collision frequency should also formally be $(T_e/T_p)^{(3/2)}$ smaller, but these collisionalities are chosen principally to provide some velocity space regularization for finite velocity space resolution without modifying the collisionless dynamics and so this additional factor is ignored.}. 
We show the results of three different simulations, $B_g = B_0$ and $B_g = 0.5 B_0$ both with $N_x \times N_y \times N_{v_\parallel} = 896 \times 448 \times 32$ corresponding to $\Delta x \sim 1.2 d_e, \Delta v_\parallel = 0.5 v_{th_s}$, and one higher resolution simulation with $B_g = 0.5 B_0$ and $N_x \times N_y \times N_{v_\parallel} = 1792 \times 896 \times 32$, corresponding to $\Delta x \sim 0.6 d_e$.
We utilize periodic boundary conditions in $x$, a reflecting wall for the plasma and conducting walls for the electromagnetic fields in $y$, and zero-flux boundary conditions in $v_\parallel$.
Like the parallel electrostatic shock simulations, these simulations also utilize linear polynomials in configuration space and quadratic polynomials in velocity space. 
We employ a small amount of hyper-diffusion in the momentum equation on scales comparable to the electron inertial length, $\nu_{Hyp} = 10^{-2} d_e^4$ for the lower resolution simulations and $\nu_{Hyp} = 10^{-3} d_e^4$ for the higher resolution simulation. 
The total cost of these simulations is relatively modest; the $896 \times 448$ resolution simulations required 24 hours on 128 56-core nodes on the Frontera cluster at the Texas Advanced Computing Center, $\sim 3,000$ node hours in total, and the higher resolution simulation cost the expected $\sim 25,000$ Frontera node hours from the doubling of the resolution and halving of the size of the time step. 

\begin{figure}
    \centering
    \includegraphics[width=\textwidth]{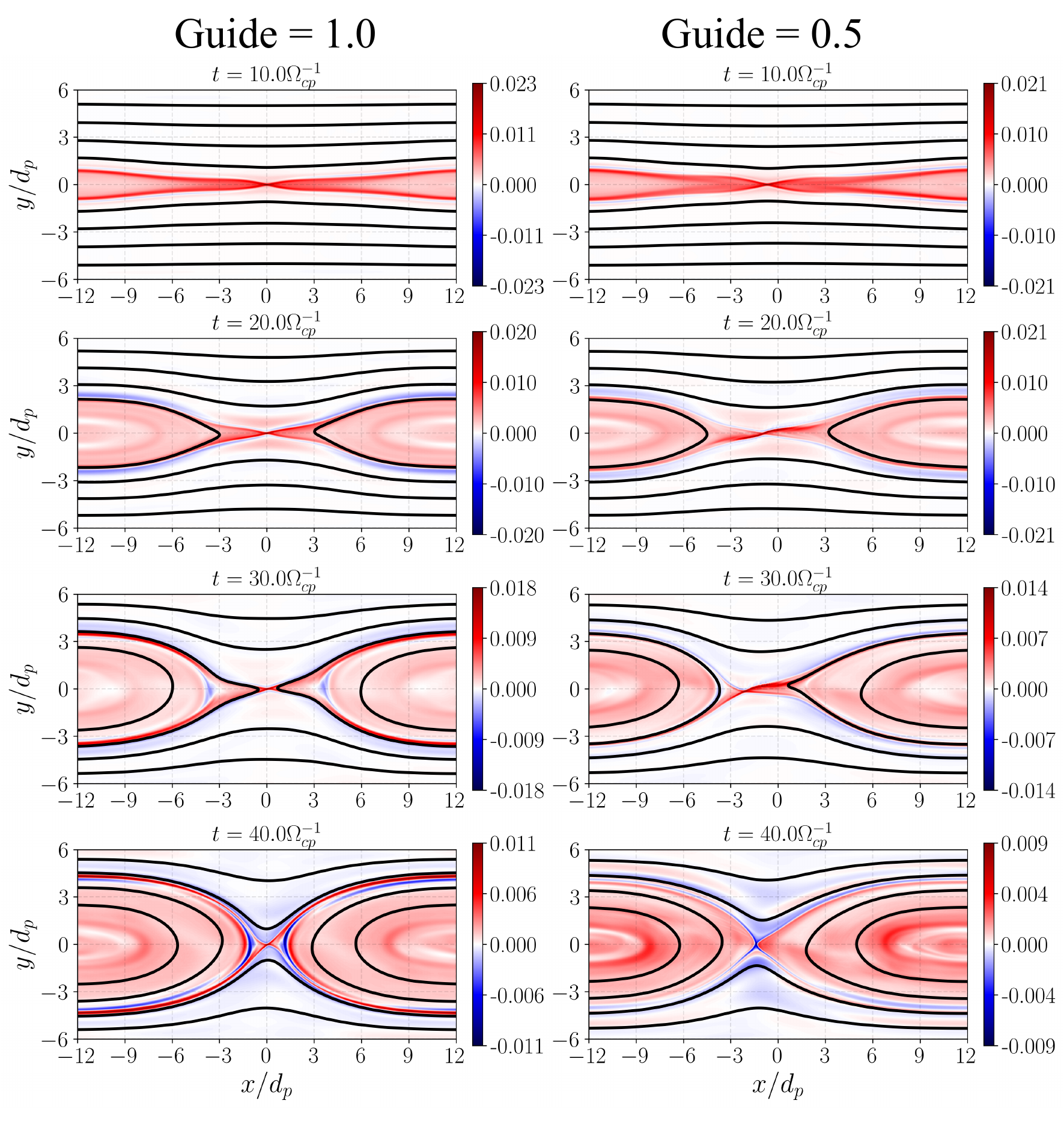}
    \caption{Evolution of the out-of-plane current density, $J_z$ with contours of the in-plane magnetic field super-imposed by computing $A_z$, the out-of-plane vector potential, from the in-plane $B_x$ and $B_y$ for the $B_g = B_0$ (left) and $B_g = 0.5 B_0$ (right) lower resolution simulations. We observe morphologies of the current layer consistent with \citet{Le:2013}, which found at lower electron $\beta_e$ a transition from a regime at lower guide field in which an extended current layer forms from the magnetized electrons developing strong anisotropy and driving a perpendicular current across field lines, to a regime in which the magnetic tension in the guide field causes the current and density to peak near the diagonally opposed separator field lines and negate the impact of the electron anisotropy on the magnetic field's overall tension---see Figure~\ref{fig:pkpm_comp_temp_aniso}. This contrast is especially clear at $t=20 \Omega_{ci}^{-1}$ and $30 \Omega_{ci}^{-1}$ as the reconnection rate reaches its peak values---see Figure~\ref{fig:recon_rate}---and we can see a more concentrated current layer in the $B_g = B_0$ simulation compared to the extended current layer in the $B_g = 0.5 B_0$ simulation.}
    \label{fig:pkpm_Jz}
\end{figure}

\begin{figure}
    \centering
    \includegraphics[width=\textwidth]{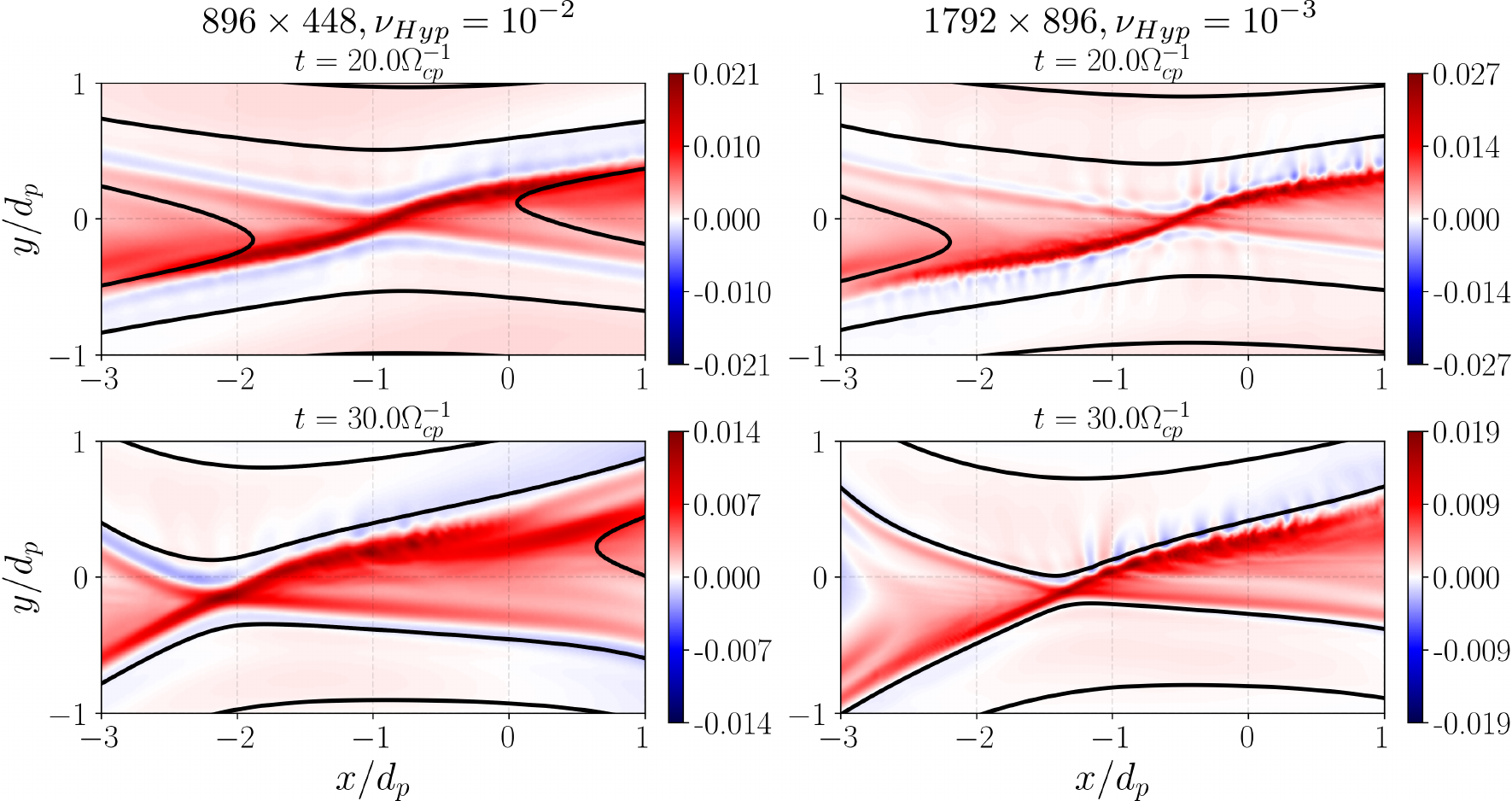}
    \caption{Zoom in of the $B_g = 0.5 B_0$ simulation with lower resolution and larger hyper-diffusion (left) and higher resolution and smaller hyper-diffusion (right). While the mode is identifiable in the lower resolution simulation, the secondary instability is especially prominent at increased resolution.}
    \label{fig:pkpm_Jz_zoom}
\end{figure}

We show in Figure~\ref{fig:pkpm_Jz} the evolution of the out-of-plane current density $J_z$ with contours of the in-plane magnetic field super-imposed by computing $A_z$, the out-of-plane vector potential, from the in-plane $B_x$ and $B_y$ for the lower resolution $896 \times 448$ simulations for the two guide fields. 
Irrespective of the guide field, we observe the typical reconnection morphology from the thinning of the current layer: the magnetic field lines pinch towards a centrally located X-point where the field is reconnecting, and the lower energy state of the field's rearranged topology leads to an acceleration of the plasma on either side of the X-point. 
In fact, the $B_g = 0.5 B_0$ simulation also seems to be developing a secondary instability in the current sheet, and we plot in Figure~\ref{fig:pkpm_Jz_zoom} a zoom-in of the current sheet at both the lower resolution, $896 \times 448$, larger hyper-diffusion simulation and the higher resolution, $1792 \times 896$, smaller hyper-diffusion simulation. 

\begin{figure}
    \centering
    \includegraphics[width=\textwidth]{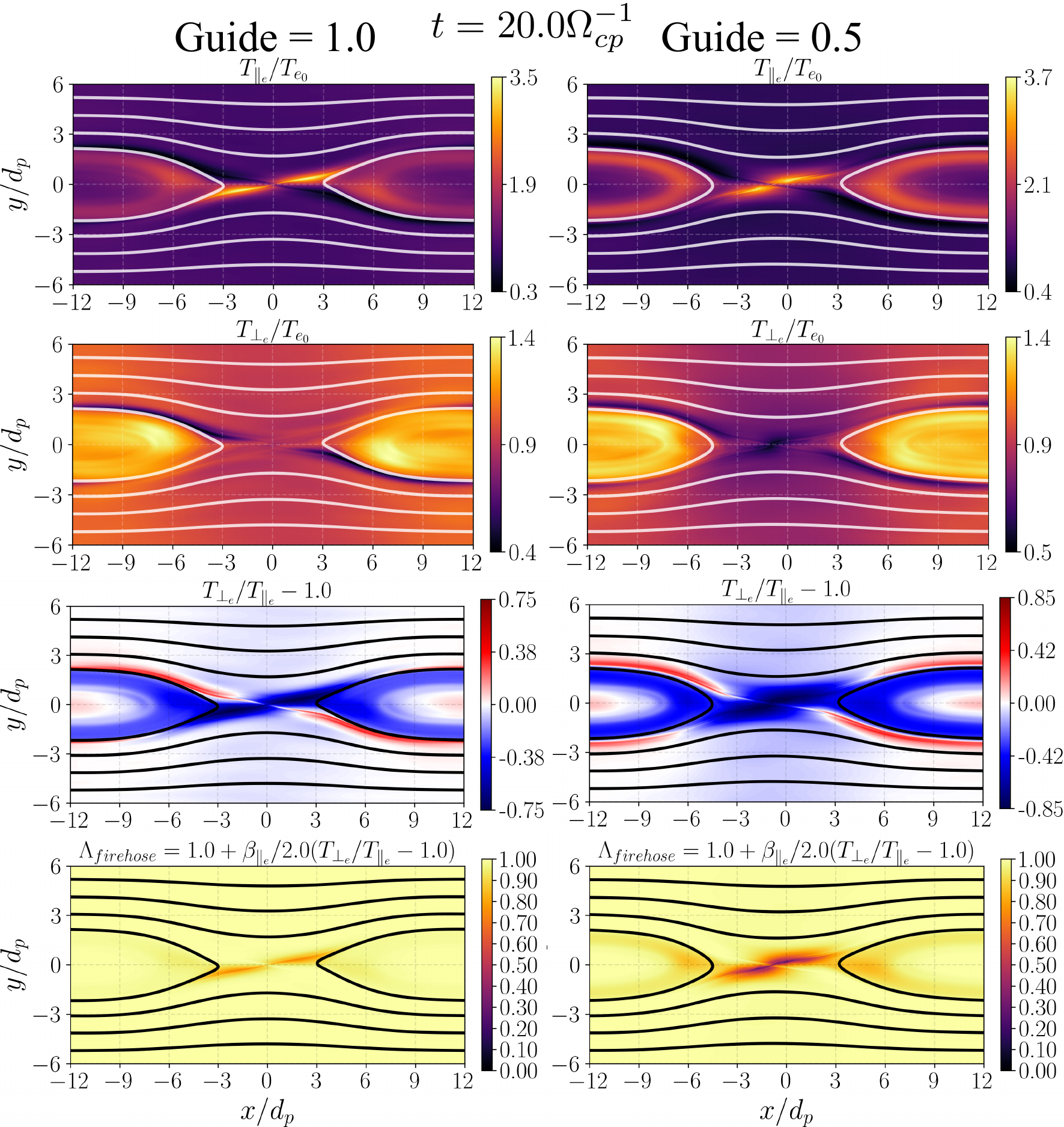}
    \caption{Comparison of the electron parallel temperature normalized to the initial electron temperature (top), electron perpendicular temperature normalized to the initial electron temperature (middle top), electron temperature anisotropy (middle bottom), and electron firehose criteria (bottom) at $t = 20 \Omega_{ci}^{-1}$ for the $B_g = B_0$ simulation (left) and $B_g = 0.5 B_0$ simulation (right). In both cases, a significant electron anisotropy from an excess of parallel pressure develops in the current layer, but a depletion of electron perpendicular pressure in the $B_g = 0.5 B_0$ simulation further increases the electron anisotropy in the layer. Combined with the lower guide field and thus a weaker magnetic field at the X-point, the electron firehose criteria is much closer to marginal stability $\Lambda_{firehose} \sim 0$ for the $B_g = 0.5 B_0$ simulation. Thus, the electrons more significantly modify the tension in the magnetic field at the reconnecting X-point compared to the higher guide field simulation, driving a perpendicular current that spreads the current layer into the exhaust.}
    \label{fig:pkpm_comp_temp_aniso}
\end{figure}

We draw particular attention to details of the reconnection morphology in Figures~\ref{fig:pkpm_Jz} and \ref{fig:pkpm_Jz_zoom}, which are consistent with previous fully kinetic simulations of guide field reconnection at realistic mass ratio: at lower guide field the current layer is extended into the exhaust \citep{Le:2013}.
The explanation for this extended exhaust in the $B_g = 0.5 B_0$ simulation compared to the more peaked current density in the $B_g = B_0$ simulation is identical to the physics of the fully kinetic simulations in \citet{Le:2013}, as we show in Figure~\ref{fig:pkpm_comp_temp_aniso}. 
While both simulations self-consistently develop a reasonably large temperature anisotropy in the layer, the temperature anisotropy in the lower guide field, $B_g = 0.5 B_0$, simulation is large enough that, combined with the lower magnitude guide field, the magnetic tension at the X-point is reduced, driving a large perpendicular current that broadens the layer into the exhaust, $\mvec{J}_\perp \sim (p_\perp - p_\parallel) \nabla_{\parallel} \mvec{b} \times \mvec{B}/|\mvec{B}|^2$, where $\nabla_\parallel = \mvec{b} \cdot \nabla_{\mvec{x}}$. 
We observe that the electron parallel firehose criterion \citep{Li:2000}, 
\begin{align}
    \Lambda_{firehose} = 1 + \frac{\beta_{\parallel_e}}{2} \left (\frac{T_{\perp_e}}{T_{\parallel_e}} - 1 \right ),
\end{align}
where $\beta_{\parallel_e} = 2 \mu_0 n_e T_{\parallel_e}/|\mvec{B}|^2 = 2 \mu_0 p_{\parallel_e}/|\mvec{B}|^2$ is the electron parallel plasma $\beta$, is much closer to marginal stability, $\Lambda_{firehose} \sim 0$ in the $B_g = 0.5 B_0$ simulation. 
On the other hand, the modest values observed in the $B_g = B_0$ simulation correspond to only minor modifications of the magnetic tension due to the electron anisotropy; the larger guide field is able to maintain the overall tension in the field in the exhaust. 

The exact transition to this extended current layer when the firehose criterion becomes sufficiently small, but not necessarily unstable, is consistent with other studies which have examined the impact of proton pressure anisotropy on the overall magnetic field tension and the propagation of Alfv\'en waves in anisotropic plasmas \citep{Bott:2021, Bott:2025}. 
In these studies, an effective Alfv\'en speed must be defined which decreases with increasing pressure anisotropy, corresponding to a reduced capability of the magnetic field to regulate the motion of a magnetized plasma and enhanced perpendicular transport, usually modeled as a Braginskii-like viscous stress \citep{Squire:2017a}. 
In fact, at lower guide field, we would likely observe an analogous phenomena to the nonlinear interruption of Alfv\'en waves \citep{Squire:2016, Squire:2017b}, but from the electrons destabilizing themselves due to the enhanced anisotropy in the extended layer. 
Such simulations could potentially extend the results of previous studies which have found the firehose criterion to be a constraint on the outflows in fully kinetic simulations of low guide field reconnection \citep{Haggerty:2018} into regimes of even moderate guide field where, as we find, the electrons are still magnetized through the current layer. 
We emphasize though that, as with the results of the parallel electrostatic shock, at least the first Fourier harmonic would be required to approximate the finite Larmor radius effects which govern the saturation of firehose modes. 

\begin{figure}
    \centering
    \includegraphics[width=\textwidth]{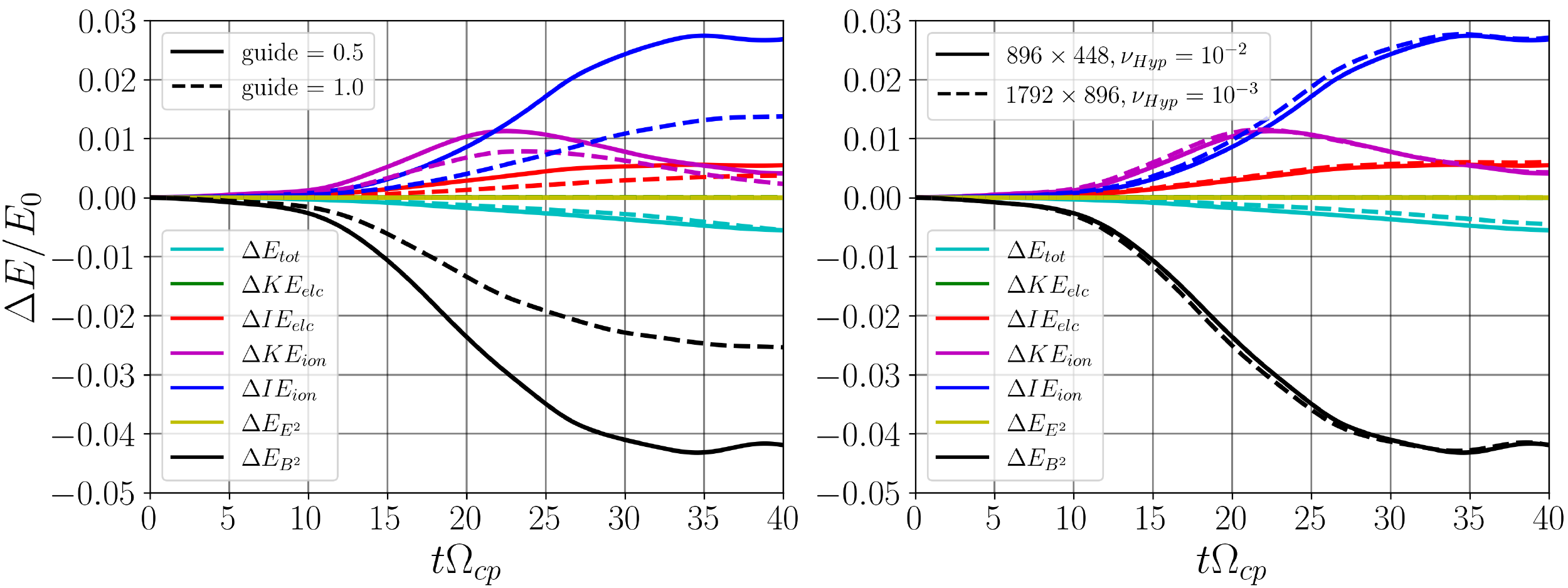}
    \caption{Evolution of the different components of the energy normalized to the total energy at $t=0$ including the total kinetic energy, $\rho_s |\mvec{u}_s|^2/2$, for the electrons and protons, the total internal energy, $3p_s/2 = p_{\parallel_s}/2 + p_{\perp_s}$, for the electrons and protons, the electric field energy $\epsilon_0 |\mvec{E}|^2/2$, and the magnetic field energy, $|\mvec{B}|^2/2\mu_0$ comparing both different guide fields (left) and different resolutions for the $B_g = 0.5 B_0$ simulation (right). We observe a conversion of magnetic energy into initially proton kinetic energy at the onset of magnetic reconnection, followed by heating of the plasma as both the electron and proton internal energies increase. Consistent with \citet{Shay:2014}, we observe that the overall electron internal energy increase is relatively insensitive to the guide field strength, and consistent with \citet{Rowan:2019}, we observe that the relative heating of the protons versus the electrons is reduced at larger guide field, as less magnetic energy is converted to plasma energization in a stronger guide field for the moderate plasma $\beta$ case considered here.}
    \label{fig:recon_energy_comp}
\end{figure}

\begin{figure}
    \centering
    \includegraphics[width=0.5\textwidth]{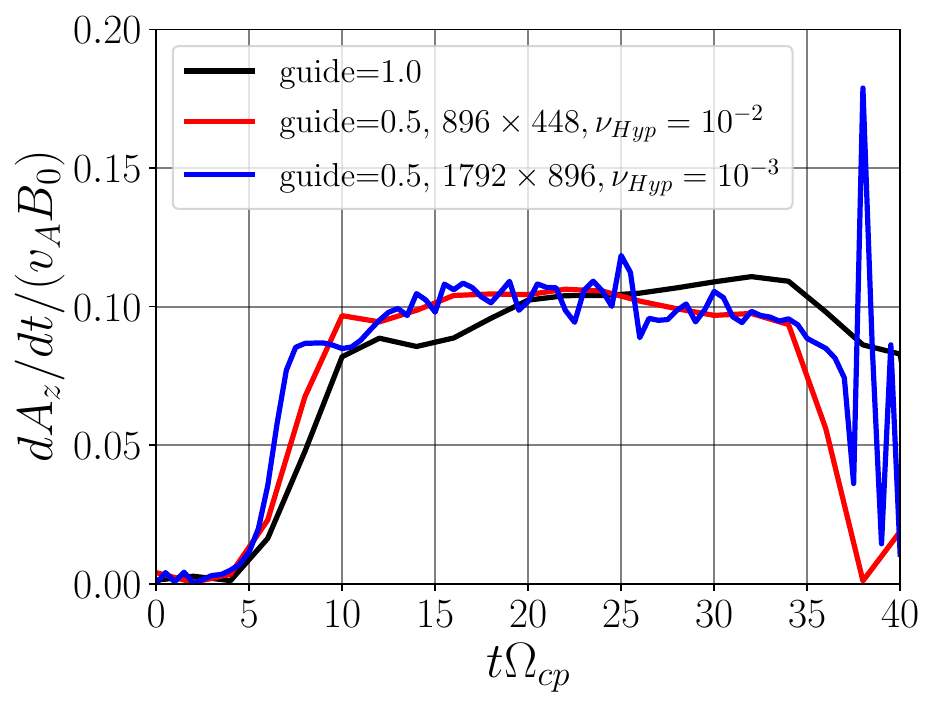}
    \caption{Reconnection rate as a function of time computed from the time rate of change of the out-of-plane vector potential, $d A_z/dt$, at the location of maximum parallel electric field, $E_\parallel = \mvec{E} \cdot \mvec{b}$. Regardless of resolution or guide field, we observe a steady peak value of $\sim 0.1$ in the standard normalized units dividing $d A_z/dt$ by the initial, upstream in-plane magnetic field strength multiplied by the initial, upstream in-plane Alfv\'en speed \citep{Shay:1999, Liu:2017, Cassak:2017, Liu:2022}.}
    \label{fig:recon_rate}
\end{figure}

We next examine the macroscopic evolution of these simulations, plotting the overall energy evolution and reconnection rate in Figures~\ref{fig:recon_energy_comp} and \ref{fig:recon_rate} respectively for different guide fields and resolutions. 
The energy evolution of these moderate guide field simulations is consistent with previous fully kinetic simulations: the overall heating of the electrons has only a weak dependence on guide field strength \citep{Shay:2014}, and as the guide field strength increases, the amount of energy which the protons receive relative to the electrons decreases \citep{Rowan:2019}. 
The reconnection rate, computed as the time rate of change of the out-of-plane vector potential, $dA_z/dt$, at the location of the maximum parallel electric field, $E_\parallel = \mvec{E} \cdot \mvec{b}$, peaks at the expected normalized value of $\sim 0.1$ \citep{Shay:1999, Liu:2017, Cassak:2017, Liu:2022}, where the normalization is defined in the standard way to the initial upstream, in-plane magnetic field strength multiplied by the initial upstream, in-plane Alfv\'en velocity. 
Additionally, outside of a slightly faster reconnection onset in the higher resolution $B_g = 0.5 B_0$ simulation, we find no sensitivity to our results with increasing resolution and lowering the hyper-diffusion, suggesting that the kinetic response of the plasma is not modified by the hyper-diffusion model and the macroscopic dynamics of the reconnection are insensitive to this hyper-diffusion. 

\begin{figure}
    \centering
    \includegraphics[width=\textwidth]{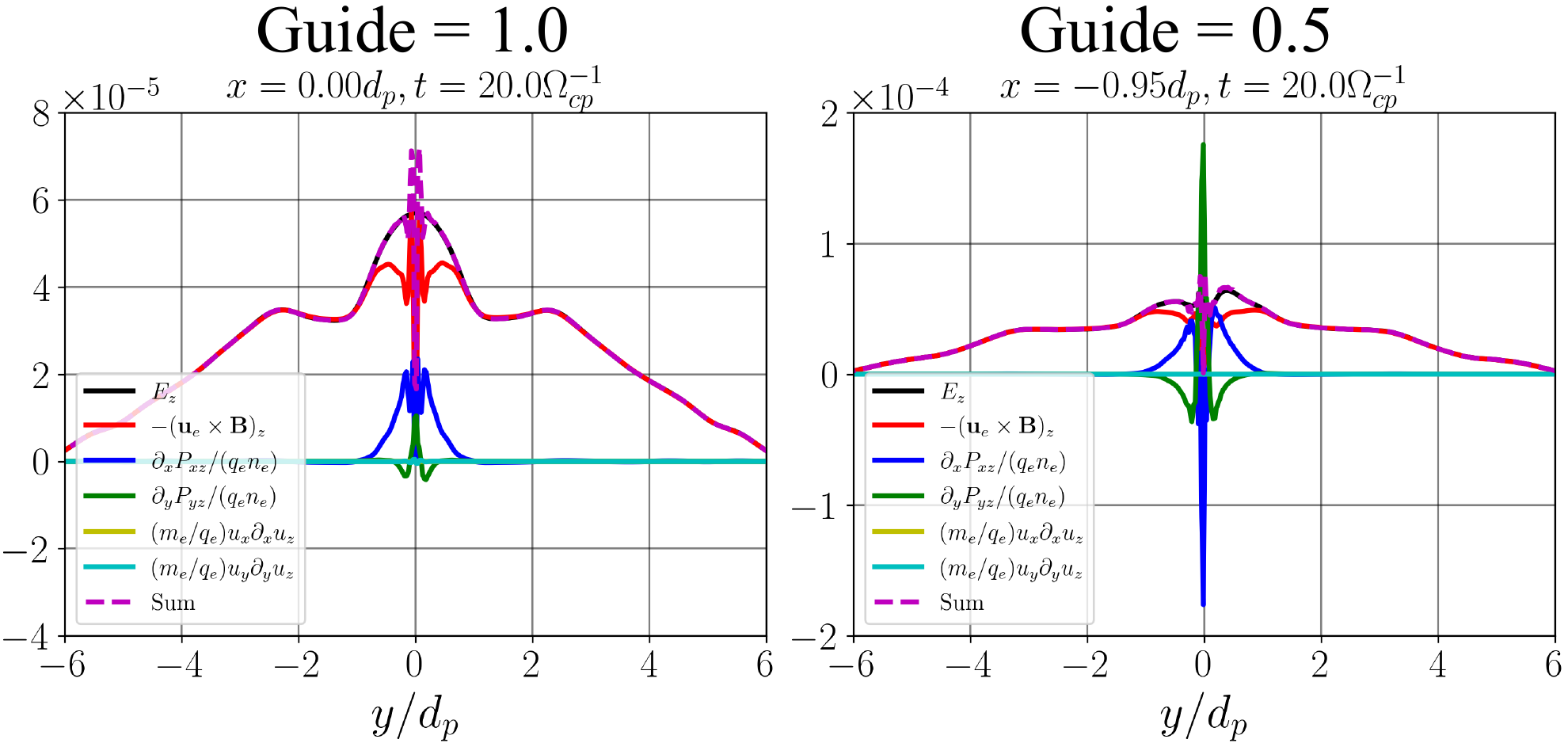}
    \caption{Comparison of the reconnecting electric field, $E_z$, and the individual components of the generalized Ohm's Law \eqr{\ref{eq:ohm}} for the $B_g = B_0$ (left) and $B_g = 0.5 B_0$ (right) simulations at $t = 20 \Omega_{cp}^{-1}$ at a cut in $x$ through the current sheet approximately where the current density peaks in Figure~\ref{fig:pkpm_Jz}. As expected, the Hall term supports the electric field away from the current sheet, but the Hall term goes to zero neat the X-point where both $u_{x_e}$ and $u_{y_e}$ go to zero from the stagnation of the flow. The reconnecting electric field's dynamics are then governed by derivatives of the off-diagonal pressure tensor, which in this case is the gyrotropic electron pressure tensor. The combination of electron pressure anisotropy and the changing magnetic field geometry drive perpendicular currents which break the frozen-in condition as the electrons are no longer advecting with the magnetic field, thus allowing for the conversion of magnetic energy to plasma energy.}
    \label{fig:ohm}
\end{figure}

\begin{figure}
    \centering
    \includegraphics[width=\textwidth]{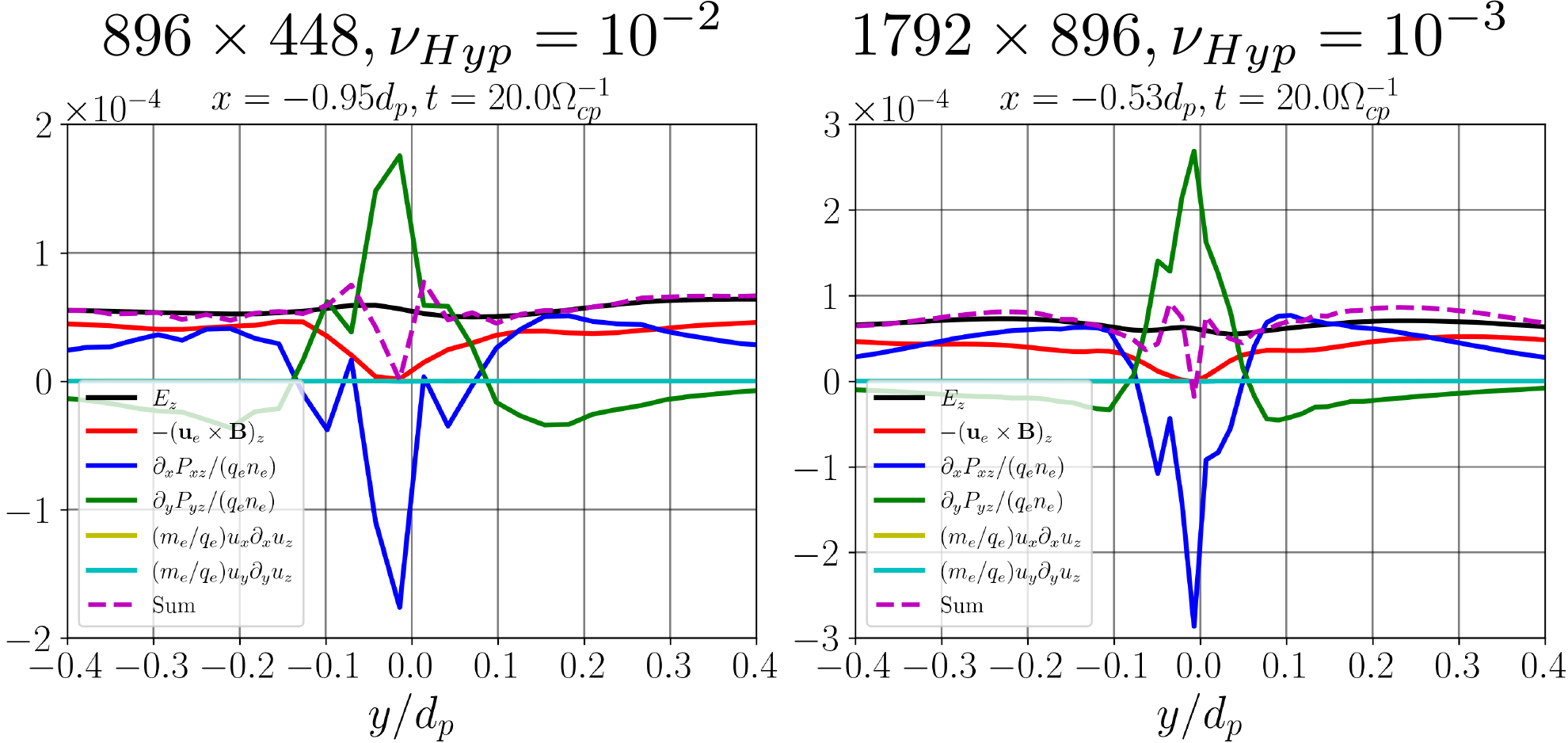}
    \caption{A zoom-in of the $B_g = 0.5 B_0$ simulations with lower resolution and larger hyper-diffusion (left), and higher resolution and lower hyper-diffusion (right). We note that the overall layer width is only marginally affected by resolution and hyper-diffusion, $\Delta \sim 0.2 d_p \sim 8.5 d_e$, where the proton and electron inertial lengths are defined with respect to the initial uniform density. The physics of the reconnecting electric field is completely unchanged qualitatively; the competition between $\partial_x P_{xz}$ and $\partial_y P_{yz}$ ultimately governs the reconnecting electric field evolution.}
    \label{fig:ohm_zoom}
\end{figure}

To further determine how accurately the PKPM model captures the kinetic response of the plasma at the X-point, we can determine the physics of the out-of-plane electric field that governs the reconnection dynamics by rearranging the electron momentum equation:
\begin{align}
    E_z = \frac{m_e}{q_e} \left (\underbrace{\frac{\partial u_{z_e}}{\partial t} + u_{x_e} \frac{\partial u_{z_e}}{\partial x} + u_{y_e} \frac{\partial u_{z_e}}{\partial y}}_{Inertia} + \frac{1}{\rho_e} \left [\frac{\partial P_{xz_e}}{\partial x} + \frac{\partial P_{yz_e}}{\partial y} \right ] \right ) - \underbrace{\left (u_{x_e} B_y - u_{y_e} B_x \right )}_{Hall}. \label{eq:ohm}
\end{align}
Here, we have labeled components of this generalized Ohm's Law which correspond to the electron inertia and Hall terms \citep{Vasyliunas:1975}.
Note that because the reconnection is fairly steady at $t=20\Omega_{cp}^{-1}$---see Figure~\ref{fig:recon_rate}---we will ignore $\partial u_{z_e}/\partial t$ in our analysis, as we expect this term to be small if the reconnection rate is relatively constant. 
We confirm in Figure~\ref{fig:ohm} and Figure~\ref{fig:ohm_zoom} the expected behavior for this generalized Ohm's Law following the results of previous guide field reconnection studies \citep{Swisdak:2005, Le:2013, Egedal:2013}.

The electrons stay magnetized through the layer and thus provide the necessary support for the reconnection electric field through the electron's off-diagonal pressure tensor components, $P_{xz} = (p_\parallel - p_\perp) b_x b_z$ and $P_{yz} = (p_\parallel - p_\perp) b_y b_z$. 
In this case, the off-diagonal pressure tensor components that have historically been found to be the important components do not come from complex particle orbits but simply from the changing magnetic geometry and anisotropy of the electrons which develops as they stay magnetized through the layer\footnote{We refer readers to \citet{Liu:2025} for a recent review of Ohm's Law analysis in simulations, laboratory experiments, and spacecraft observations.}. 
Instead of the fundamental agyrotropy that develops in the zero or low guide field case \citep{Hesse:2018}, it is the gyrotropic electron pressure tensor that can break the frozen-in condition \citep{Egedal:2002}. 
In other words, the electrons no longer simply advect with the magnetic field, but directly change the magnetic field due the perpendicular currents driven by the gyrotropic electron pressure tensor. 
We note that this change in the field driven by perpendicular currents is not necessarily true field line ``breaking.'' 
Figure~\ref{fig:ohm_zoom} shows that while the overall current sheet width is relatively insensitive to resolution and hyper-diffusion, there is a region in which the sum of these various components of Ohm's Law and the reconnecting electric field diverge which does become smaller with increasing resolution. 
Thus, there is a region inside the current layer where the final field line topology changes, $b_x = b_y = 0$, which is not captured in the lowest order PKPM model with no Fourier harmonics.

\begin{figure}
    \centering
    \includegraphics[width=\textwidth]{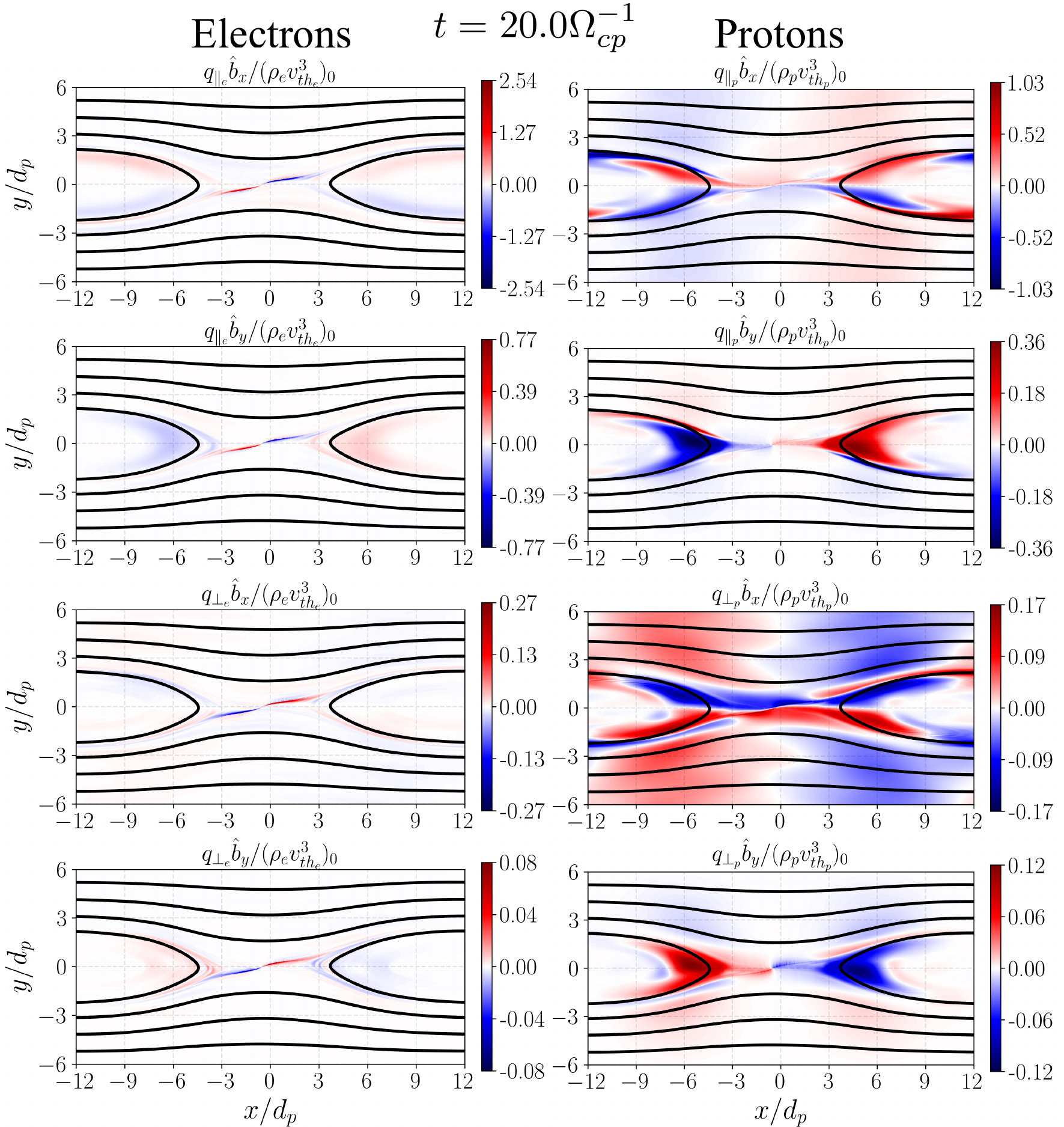}
    \caption{Comparison of the different components of the parallel heat flux normalized to the initial values of a thermally streaming plasma, $\rho_e v_{th_e}^3$, for both the electrons (left) and protons (right). We multiply both heat fluxes by the individual components of the magnetic field unit vector to separate the flow of $T_\parallel$ due to $q_\parallel$ and $T_\perp$ due to $q_\perp$ in the $x$ versus $y$ direction. We draw particular attention to how large the values of $q_{\parallel_e}$ are, suggesting that the exact evolution of $T_{\parallel_e}$ is strongly influence by collisionless heat fluxes, and the fact that $q_{\parallel_e}$ and $q_{\perp_e}$ are the opposite sign in the current layer, which further explains the heightened temperature anisotropy in the $B_g = 0.5 B_0$ simulation. And as large as the electron heat fluxes are, the ion heat fluxes compared to thermal streaming are not small either, with significant energy fluxes in the exhaust as the ions mix and heat in the outflows.}
    \label{fig:pkpm_q}
\end{figure}

Importantly, while a number of fluid models have been developed to exploit this fact that electron anisotropy can break the frozen-in condition and determine the current sheet morphology \citep{Le:2009, Ohia:2012, Cassak:2015}, we reiterate that this PKPM model employed here is fundamentally kinetic. 
The results of, e.g., how the electron anisotropy develops and affects the current sheet morphology are handled self-consistently without the need for any artificial limiters on the electron firehose condition \citep{Ohia:2015}, and recent studies such as \citet{Walters:2024} suggest that with only one or two Fourier harmonics, a self-consistent saturation of firehose modes can be included with this approach. 
As it is, we can clearly identify the impact of the kinetic physics by examining higher velocity moments than typically included in fluid models, such as the total parallel heat flux plotted in Figure~\ref{fig:pkpm_q} for electrons and protons, normalized to the initial thermal streaming values $\rho_s v_{th_s}^3$, for the higher resolution, $1792 \times 896$, $B_g = 0.5 B_0$ simulation at $t=20 \Omega_{cp}^{-1}$.

These heat fluxes are plotted component-wise so that we can separate the flow of $T_\parallel$ due to $q_\parallel$ and $T_\perp$ due to $q_\perp$ in the $x$ versus $y$ direction.
These heat fluxes are not small compared to thermally streaming particles, especially $q_{\parallel}$ for both the electrons and ions. 
In other words, these collisionless heat fluxes are likely strongly influencing the dynamical evolution of $T_\parallel$ for each species. 
More importantly, $q_{\parallel_e}$ and $q_{\perp_e}$ are the opposite signs, towards the X-point and away from the X-point respectively, in the extended current layer of this $B_g = 0.5 B_0$ simulation.
These oppositely directed heat fluxes, in addition to the conservation of magnetic moment $\mu_e \propto T_{\perp_e}/B$ in the layer where the electrons are magnetized, further explain why there is a cooling in $T_{\perp_e}$ observed in Figure~\ref{fig:pkpm_comp_temp_aniso} and that this cooling is more pronounced in the lower guide field case where the magnetic field strength decreases more at the X-point.
Thus, the collisionless heat fluxes are enhancing the very temperature anisotropy in the layer that is modifying the tension in the magnetic field and driving the extended current layer. 

\begin{figure}
    \centering
    \includegraphics[width=\textwidth]{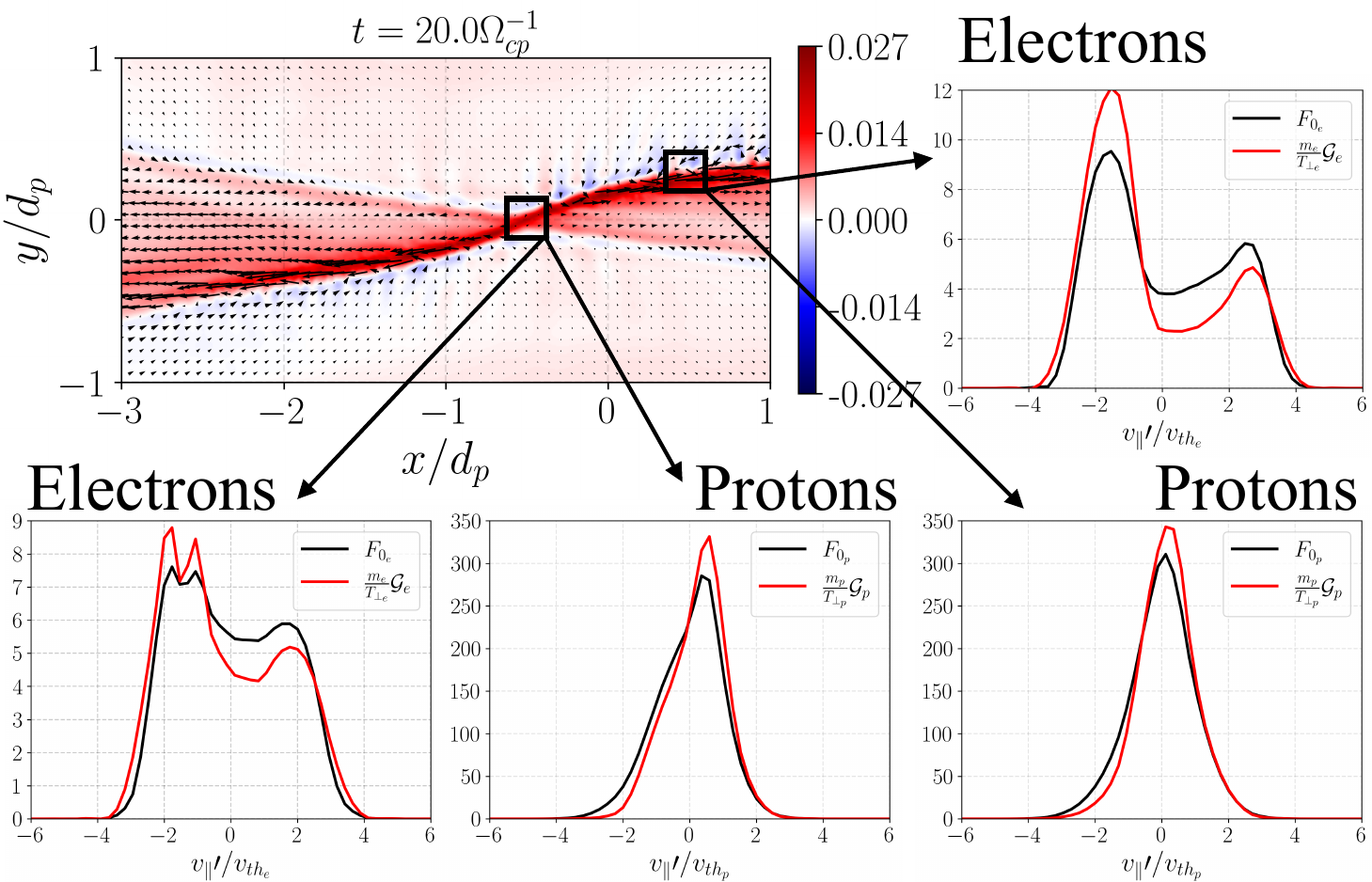}
    \caption{Zoomed in out-of-plane current density for the $B_g = 0.5 B_0$ simulation with superimposed vectors denoting the direction of the in-plane flow direction and corresponding plots of the electron and proton distribution functions at the peak of the current density and inside one of the excited fluctuations from the secondary instability in the extended current sheet. Note that the distribution functions are plotted in their respective flow frames, so the $F_0$ distribution functions have zero first moment, and any skewness or multi-modality to the $F_0$ distribution functions are manifestations of heat fluxes or counter-streaming field-aligned flows. While we need the sign of the magnetic field unit vector to determine which direction the vector heat flux points, we can clearly identify from the overall structure of the electron's $F_0$ and $\mathcal{G}$ distribution functions that $q_{\parallel_e}$ and $q_{\perp_e}$ point in opposite directions, consistent with Figure ~\ref{fig:pkpm_q}. In both cases, for the electrons there is a depletion of $F_0$ relative to $\mathcal{G}$ for negative $v_\parallel'$ particles, and an excess of $F_0$ relative to $\mathcal{G}$ for positive $v_\parallel'$ particles. Likewise, in the location of the excited secondary instability, we have counter-streaming super-thermal beams along the field line, in addition to vortical structures in the in-plane flow velocity, which suggests that the electrons are unstable to a shearing instability. We identify this instability as the electron Kelvin-Helmholtz mode from its wavelength $k d_e \sim 1$ and its similarity to previous reconnection studies which excited secondary Kelvin-Helmholtz instabilities in the layer \citep{Fermo:2012}.}
    \label{fig:pkpm_dist}
\end{figure}

In addition, at this same time of $t=20 \Omega_{cp}^{-1}$, we plot in Figure~\ref{fig:pkpm_dist} the zoom-in of the out-of-plane current density, now with superimposed vectors denoting the direction of the in-plane flow, to examine the secondary instability which we observed to be developing in the extended current sheet in Figure~\ref{fig:pkpm_Jz_zoom}. 
We also show in Figure~\ref{fig:pkpm_dist} one-dimensional plots of the velocity distribution function within both the peak current density and one of the fluctuations of the secondary instability in the extended current layer. 
We identify this secondary instability now from its wavelength, $k d_e \sim 1$, the vortical structure present in the in-plane velocity, and the counter-streaming superthermal electron beams as the electron Kelvin-Helmholtz instability.
This instability has been previously measured in fully kinetic simulations at larger guide field, $B_g = 2 B_0$ and lower proton-electron mass ratio, $m_p/m_e = 25$\citep{Fermo:2012}. 
The exact conditions under which this Kelvin-Helmholtz instability can be excited are likely a consequence of both mass ratio and guide field, as the excitement of this secondary instability seems most prevalent in these extended magnetized current layers and did not show up in the $B_g = B_0$ simulation with our chosen realistic proton-electron mass ratio. 

The distribution function structure observed in this $B_g = 0.5 B_0$ simulation includes a number of other interesting features. 
While we need the sign of the magnetic field unit vector to obtain the sign of the vector heat flux, the depletion of negative velocity particles in $F_0$ compared to $\mathcal{G}$, and likewise the excess of positive velocity particles in  $F_0$ compared to $\mathcal{G}$, easily explains the fact that $q_{\parallel_e}$ and $q_{\perp_e}$ have the opposite sign in Figure~\ref{fig:pkpm_q}. 
Further, the broadened electron distribution function at the X-point is consistent with trapped electron distribution functions in past studies of electron energization in guide-field reconnection \citep{Egedal:2013}. 
Finally, the proton distribution functions also show clearly how the protons develop non-trivial heat fluxes, even if these heat fluxes are appreciably smaller than the electron heat fluxes. 

Thus, even with moderate guide field, because the electrons stay magnetized through the layer, the PKPM model provides a powerful cost-effective tool for kinetic simulations of magnetic reconnection. 
Self-consistent temperature anisotropies and parallel heat fluxes can develop which modify the current layer morphology in agreement with previous fully kinetic simulations. 
Secondary instabilities which may be excited via field-aligned beams are likewise included in this approach. 
Given the sensitivity of the kinetic evolution of the electrons to the proton-electron mass ratio, the PKPM model may provide a breakthrough in understanding the dynamics of guide-field reconnection, even in regimes beyond which magnetized models such as gyrokinetics can be applied. 

%% file: summary.tex
\section{Summary}\label{sec:conclusions}

In this paper, we have derived a novel approach to modeling kinetic plasmas based on classical intuition about collisionless, magnetized plasma dynamics. 
This parallel-kinetic-perpendicular-moment (PKPM) model separates the parallel and perpendicular motion to the local magnetic field through the combination of coordinate transformations and optimized spectral expansions in these transformed coordinates.
Via velocity-space coordinate transformations to the Vlasov equation, first to velocity coordinates moving with the local flow and then to field-aligned velocity coordinates, we can reduce the number of degrees of freedom needed to model magnetized plasma with a truncated Laguerre-Fourier spectral expansion in the perpendicular velocity coordinates, both $v_\perp$ and velocity gyroangle, $\theta$. 
In this regard, the six-dimensional Vlasov-Maxwell dynamics were reduced to some number of four-dimensional equations for spectral coefficients in $x,y,z$ configuration space and the remaining $v_\parallel$ velocity coordinate, where the exact number of equations solved dictates the amount of perpendicular velocity-space resolution one expects to need for the problem of interest. 

The final equations, even in their simplest form where two kinetic equations for the zeroth Fourier harmonic and zeroth and first Laguerre coefficient are retained, contain a multitude of physics. 
While this lowest-order PKPM system only evolves the gyrotropic component of the distribution function, by virtue of how we constructed our Laguerre basis, we have significantly more flexibility than other spectral approaches to kinetic equations which discretize the untransformed Vlasov equation. 
Indeed, there are no assumptions made on the strength of the flow velocity or the variation of the perpendicular temperature; the PKPM model as implemented handles supersonic flows and large, $\mathcal{O}(1)$ variations in the perpendicular temperature as well as the resulting temperature anisotropies which develop self-consistently from this perpendicular temperature variation. 
By transforming the velocity coordinates to the local flow frame, the restriction of a general shift vector in a spectral expansion is avoided, and we do not need to assume spatially or temporally independent background flows. 
By defining the second kinetic equation in terms of a linear combination of Laguerre coefficients, we can absorb the Laguerre normalization into the second kinetic equation and avoid the need for an auxiliary equation for the perpendicular temperature. 

We have demonstrated this flexibility and the underlying physics fidelity of the simplest PKPM model on two non-trivial, nonlinear problems: a parallel electrostatic collisionless shock and moderate guide field reconnection. 
The PKPM model handles the transition between shocked and unshocked flows when the inflowing plasma sonic Mach number is sufficiently large that no nonlinear steepening of ion acoustic waves occurs to trap electrons. 
Likewise, the PKPM model correctly accounts for the impact of magnetized electrons in moderate guide field reconnection, where electron pressure anisotropy can break the frozen-in condition for the electrons and convert the stored magnetic energy into plasma energy, both kinetic and internal. 
The uniqueness of the PKPM model is thus immediately apparent: to the authors' knowledge there are no implementations in any codes of asymptotic, magnetized, kinetic models which can handle supersonic flows or $\delta B/B \sim 1$ fluctuations. 
Yet the PKPM model is leveraging the same physics intuition that has informed these historical asymptotic models. 
For example, it is because the electrons continue to follow field lines, even if there is $\mathcal{O}(1)$ variation in the magnetic field, that the PKPM model compares favorably with past fully kinetic simulations of moderate guide field reconnection, including the self-consistent generation of extended current layers due to the electrons reducing the field line tension via temperature anisotropy and driving macroscopic perpendicular currents. 

We reiterate that this paper is but the first part of a multi-part series, as we have not yet discussed how we discretize even the simplest PKPM model with a discontinuous Galerkin (DG) finite element method. 
Preserving the properties of the continuous PKPM system, most especially minimizing errors in the conservation of energy from the coupling of the bulk kinetic energy from the momentum equation with the evolution of the internal energy given by the kinetic equations, are nontrivial endeavors, and for maximum clarity of the narrative, we have elected to separate the theoretical and numerical discussions. 
Still, we emphasize the demonstration of the PKPM model in Section~\ref{sec:demo} is not only meant to communicate the physics fidelity of the PKPM model, but the successful numerical implementation of this unique hybrid discretization strategy mixing a Laguerre-Fourier spectral expansion with DG. 
Indeed, while philosophically similar hybrid discretizations using a combination of spherical harmonics and DG in ``test-fluid'' approaches which do not feedback on the flow velocity and magnetic field evolution have been derived and implemented\citep{Schween:2024, Schween:2025}, the demonstration of the PKPM model in this paper is the first use of this kind of hybrid discretization utilizing an optimized Laguerre-Fourier spectral basis for magnetized plasma dynamics. 

The successful numerical implementation and demonstration of the PKPM model here is also in one respect the culmination of the theoretical work which inspired the PKPM approach, but has resisted easy numerical discretization. 
Kinetic MHD \citep{Kulsrud:1964, Kulsrud:1983} and Finite Larmor Radius Kinetic theory \citep{Ramos:2008, Ramos:2016} both transform the velocity coordinates of the Vlasov equation to velocity coordinates which are both field-aligned and moving at a local flow velocity, the $\mvec{E} \times \mvec{B}$ velocity in the case of Kinetic MHD and the total flow velocity in the case of Finite Larmor Radius Kinetic theory. 
However, the stiff constraint equations which these models derive, such as the $E_\parallel$ equation in Kinetic MHD, may not be satisfied at any given instant in time if one evolves kinetic equations for each individual plasma species that thus generate non-zero parallel currents. 
Therefore, while there have been recent successes in Kinetic MHD-like approaches that leverage physics intuition and add fluid equations which instantaneously satisfy equilibrium relations, such as cold electron return currents which balance the currents driven by the kinetic species \citep{Drake:2019, Arnold:2019, Arnold:2021, Arnold:2022}, the PKPM model self-consistently handles the relaxation to these constraint equations by dynamically evolving, e.g., parallel currents. 
As a result, in simulations of phenomena such as magnetic reconnection, the distribution functions naturally develop distinct particle populations that can then evolve to states predicted by these asymptotic models, or on the other hand, drive secondary instabilities from these counter-streaming kinetic populations. 

We emphasize at this stage that this manuscript constitutes only the beginning of a larger program of research with this PKPM model. 
A systematic convergence study in Laguerre resolution of subcritical collisionless shocks, guide field reconnection, or other phenomenon such as turbulence and transport in laboratory and astrophysical plasmas, is of immediate interest. 
For example, how accurately this optimized Laguerre basis in $v_\perp$ resolves the loss cone of a magnetic mirror across a range of mirror force strengths will inform rules-of-thumb of how many Laguerre spectral coefficients are needed for a variety of applications. 
Further, there will likely be cases in which the amount of Laguerre resolution needed for the physics of interest is different in different regions of configuration space, and may even be different per-species. 
We can thus imagine further optimization of this approach to refine our perpendicular resolution only where we need it. 
And in cases where one plasma species may be poorly modeled by the PKPM approach, such as in a high Mach number collisionless shock where we often observe the protons to be highly agyrotropic due to reflection off the compressed magnetic field while we observe that the electrons stay mostly gyrotropic through the shock, significant computational gains could still be realized by a flexible coupling of the full Vlasov model in \gke\ \citep{Juno:2018,HakimJuno:2020} to the PKPM model presented here.

Perhaps most important  for examining the physics fidelity of the PKPM model as more perpendicular moments are added is the impact of the Fourier harmonics on the model.
As we showed in Section~\ref{sec:pkpm} with \eqr{\ref{eq:Pi-agyro}} and discussed in Section~\ref{sec:demo}, the first two Fourier harmonics give the self-consistent evolution of the remaining off-diagonal pressure tensor components, in addition to other higher moments such as the heat fluxes perpendicular to the field. 
Indeed, the fact that adding only one or two Fourier harmonics confers this increased physics is a direct consequence of the fact that this approach does not transform to gyrocenter coordinates and instead keeps the configuration space coordinates at the particle position. 
Adding a few Fourier harmonics thus allows the PKPM model to directly connect to the success of multi-fluid models which include the full pressure tensor \citep{Wang:2015, Ng:2015, Wang:2018, Dong:2019, Ng:2019, TenBarge:2019, AllmannRahn:2021, Walters:2024, Kuldinow:2024b}.
With just a few more equations, we would obtain a hybrid fluid-kinetic approach with all of the higher order fluid effects of the cutting edge of extended fluid modeling for the dynamics perpendicular to the local magnetic field, with a complete kinetic description parallel to the field capable of handling a myriad of kinetic phenomena: parallel heat fluxes, field-aligned beams, trapped particles, and more.

In fact, we consider it likely that the inclusion of the first one or two Fourier harmonics to be not just a sweet spot in terms of the physics fidelity of this reduced approach, but optimal computationally as well. 
The evolution of the first two Fourier harmonics can be encoded in the evolution of the vector $\mvec{M}_\perp$ and tensor $\mvec{F}_{\perp\perp}$, given by \eqr{\ref{eq:Mperp}} and \eqr{\ref{eq:F-perp-perp}} respectively, and the evolution of these quantities can be determined by multiplying the Vlasov equation in our transformed coordinates, \eqr{\ref{eq:vlasov-cgl}}, by $\mvec{v}_\perp$ and $\mvec{v}_\perp \otimes \mvec{v}_\perp$ and integrating over velocity gyroangle. 
These evolution equations for $\mvec{M}_\perp$ and tensor $\mvec{F}_{\perp\perp}$ could then be expanded in our Laguerre basis in $v_\perp$. 
So for example, a ``lowest-order'' approached that was extended to the first Fourier harmonic through the $\mvec{M}_\perp$ equation would go from solving kinetic equations for $F_0$ and $\mathcal{G}$, to include $\mvec{M}_{\perp_0}$ and $\gvec{\mathcal{N}}_\perp = T_\perp/m (\mvec{M}_{\perp_0} - \mvec{M}_{\perp_1})$, where $\mvec{M}_{\perp_{0,1}}$ are the zeroth and first Laguerre coefficients of $\mvec{M}_\perp$ respectively and $\gvec{\mathcal{N}}_\perp$ is the linear combination of first and zeroth Laguerre coefficients which would appear similar to the $\mathcal{G}$ term in our PKPM expansion---a modest increase from two total kinetic equations to eight total kinetic equations.
Higher Fourier harmonics could become burdensome to represent as tensors of increasing rank, especially in combination with the Laguerre expansion in $v_\perp$, and we defer a careful examination of computational efficiency and physics fidelity with increasing Fourier harmonics and Laguerre coefficients to future work.

Even still, the utility of the PKPM approach has already been demonstrated in studies as diverse as the feasibility of novel fusion configurations which require collisionless shock generation to compress the target fuel \citep{Cagas:2023}, to novel energization studies of electromagnetic waves which utilize the PKPM perspective to understand how different kinetic populations of particles ultimately lead to heating via ``pressure-work'' \citep{Conley:2024}. 
Further, the success of the PKPM model in modeling moderate guide-field reconnection with a realistic proton-electron mass ratio suggests it is an ideal model for kinetic studies of new reconnection regimes, such as shear-flow suppression of the tearing mode \citep{Mallet:2025a, Mallet:2025b}. 
With the necessary modifications to the PKPM model to solve the equations in curvilinear coordinates, further direct modeling of fusion reactor geometries could be performed as well. 
Given recent discoveries in the importance of kinetic effects as reactor-relevant temperatures are simulated \citep{Shukla:2025}, we may yet find benefit in transport and macrostability modeling with this approach for self-consistent simulations all the way to the plasma-material interface, where the fact that this model includes the physics of the plasma sheath will allow for more detailed analysis of the heat deposition on the plasma-facing components. 
We conclude that the PKPM model derived in this paper has widespread applicability and its cost-effectiveness in even its simplest form presents unique opportunities for modeling diverse weakly collisional, magnetized plasma systems. 

%% file: ap-notation.tex
\section{Coordinate-free Tensor Notation}\label{app:notation}

Throughout this paper we have used a coordinate free notation based on
an extended form of the notation adopted in \citet{Thorne:2017}. 
In this appendix, we present an overview of our notation for ease of following the derivations in the main text of the paper.

We work in flat 3D space and denote the vector space as $\vecspace$. 
Given two or more vectors in $\vecspace$, we will denote their \emph{tensor product} with the $\otimes$ symbol. 
For example, given $\mvec{u}, \mvec{v} \in \vecspace$, we can write their
tensor-product as $\mvec{u}\otimes\mvec{v}$. 
The tensor-product creates a \emph{multilinear mapping} from $n$ vectors, where $n$ is the number of vectors in the product, to a real number. 
Given $\mvec{u}\otimes\mvec{v}$, the mapping is $\mvec{u}\otimes\mvec{v} : \vecspace\times \vecspace \rightarrow \mathbb{R}$, and we can evaluate it for the vectors $\mvec{a}$ and
$\mvec{b}$ as follows
\begin{align}
  (\mvec{u}\otimes\mvec{v})(\mvec{a}, \mvec{b}) =
  (\mvec{u}\cdot\mvec{a}) (\mvec{v}\cdot\mvec{b}).
\end{align}

A tensor of rank $n$ is a \emph{multilinear function} that takes $n$ vectors and maps them to a real number. 
For example, a second order tensor $\mvec{T}(\mvec{a},\mvec{b})$ will take two input vectors
($\mvec{a}$ and $\mvec{b}$ in this case) and produce a single scalar. 
As the mapping is multilinear we have
\begin{align}
  \mvec{T}(\alpha\mvec{a} + \delta\mvec{d},\mvec{b})
  =
  \alpha\mvec{T}(\mvec{a},\mvec{b})
  +
  \delta\mvec{T}(\mvec{d},\mvec{b}).
\end{align}
In this sense, a scalar is a rank-0 tensor, that simply evaluates to itself. 
A vector $\mvec{u}$ is a rank-1 tensor mapping an input vector $\mvec{a}$ to
\begin{align}
  \mvec{u}(\mvec{a}) = \mvec{u}\cdot\mvec{a}.
\end{align}
Defined in this manner, rank-$n$ tensors (including vectors) are \emph{geometric} quantities, hence independent of the basis vectors used to represent them.

We can introduce a set of basis vectors $\basis{i}$, $i=1,2,3$, for $\vecspace$. 
These basis vectors need not be orthogonal (or even unit) vectors. 
Given these basis vectors, we can construct a set of \emph{dual basis} $\dbasis{i}$ defined such that $\basis{i}\cdot\dbasis{j} = \delta_i^j$. 
In flat space, we can always introduce a set of orthonormal basis $\cbas{i}$, with $i=1,2,3$. 
These orthonormal basis are their own duals.

Now, if we feed a vector with one of the basis $\basis{i}$ or $\dbasis{i}$, we will get
\begin{align}
  \mvec{u}(\basis{i}) & = \mvec{u}\cdot\basis{i} \equiv u_i, \\
  \mvec{u}(\dbasis{i}) & = \mvec{u}\cdot\dbasis{i} \equiv u^i.
\end{align}
Hence, with the basis as an input, the vector mapping produces the \emph{component} of the vector along that basis. 
Analogously, we will define the components of a higher-rank tensor as the real numbers produced when the input vectors are the basis. 
So
\begin{align}
  T_{ij} & \equiv \mvec{T}(\basis{i},\basis{j}), \\
  T^{ij} & \equiv \mvec{T}(\dbasis{i},\dbasis{j}). 
\end{align}

Since tensors are multilinear mappings, in a specific basis, we can write them as linear combinations of the tensor products of the selected basis. 
For example,
\begin{align}
  \mvec{T} =
  T^{ij} \basis{i}\otimes\basis{j}
  =
  \mvec{T}(\dbasis{i},\dbasis{j}) \basis{i}\otimes\basis{j}
  = T_{ij} \dbasis{i}\otimes\dbasis{j}
  =
  \mvec{T}(\basis{i},\basis{j}) \dbasis{i}\otimes\dbasis{j}.
\end{align}
Hence, using the components, we can write an explicit formula for the evaluation of the multilinear form in terms of its components as
\begin{align}
  \mvec{T}(\mvec{a},\mvec{b}) = T^{ij}
  (\mvec{a}\cdot \basis{i}) (\mvec{b}\cdot \basis{j})
  =
  T^{ij} a_i b_j
  =
  T_{ij}
  (\mvec{a}\cdot \dbasis{i}) (\mvec{b}\cdot \dbasis{j})
  =
  T_{ij} a^i b^j.
\end{align}
Note that throughout we are using the Einstein summation convention: repeated upstairs/downstairs indices are to be summed.

We can also compute the partial evaluation of the tensor by filling up one or more of its slots with vectors. 
The resulting function (taking fewer input parameters) is also a tensor, but a  lower rank tensor. 
For example $\mvec{T}(\mvec{a},\_)$ results in a vector (rank-1 tensor). 
In a specific representation,
\begin{align}
  \mvec{T}(\mvec{a},\_) = T^{ij} \mvec{a}\cdot
  (\bbasis{i}\otimes\basis{j})
  = T^{ij} (\mvec{a}\cdot\basis{i}) \basis{j},
\end{align}
where we have used the ``breve'' marker on the $\basis{i}$ to indicate with which of the vectors making up the tensor product the dot-product must be taken. 
Similarly,
\begin{align}
  \mvec{T}(\_,\mvec{a}) = T^{ij} \mvec{a}\cdot
  (\basis{i}\otimes\bbasis{j})
  = T^{ij} \basis{i}(\mvec{a}\cdot\basis{j}).
\end{align}
Of course, we do not need to use the tangent or their reciprocals to represent tensors. 
As they are geometric objects, any basis will do, for example, the normalized tangent vectors. 
Once the components are known in one set of basis, we can simply compute them in another basis by evaluating, for example,
\begin{align}
  \mvec{T}(\nbasis{i},\nbasis{j}) \equiv \hat{T}_{ij}
  = {T}_{mn} (\nbasis{i}\cdot\dbasis{m}) (\nbasis{j}\cdot\dbasis{n}).
\end{align}

The \emph{trace} of a tensor product is a linear operator defined as, for example,
\begin{align}
  \Tr(\mvec{u}\otimes\mvec{v}) \equiv \mvec{u}\cdot\mvec{v}.
\end{align}
For products with more than two vectors, we need to indicate the pair of vectors on which the trace operator acts. 
For example,
\begin{align}
  \Tr(\bmvec{u}\otimes\mvec{v}\otimes\bmvec{w}) =
  (\mvec{u}\cdot\mvec{w}) \mvec{v}.
\end{align}
Here, we used the ``breve'' marker on the vectors we wish to participate in the trace. 
Just like partial evaluation, the trace operator is also a rank-reducing operation: the resulting object has \emph{two} ranks lower than the original tensor.

The dot-product operator can be extended to act on a pair of tensors. 
To define this operation, consider we want to compute the dot-product between a vector $\mvec{u}$ and a second-order tensor $\mvec{T}$. 
We first need to select the slot of the tensor with which we wish to take the product. 
For example, we will denote the dot-product with the first slot as $\mvec{u}\cdot\mvec{T}(\bdum,\_)$ and define this to be just the partial evaluation $\mvec{T}(\mvec{u},\_)$. 
Similarly, $\mvec{u}\cdot\mvec{T}(\_,\bdum) = \mvec{T}(\_,\mvec{u})$.

The \emph{metric tensor} is a special bilinear mapping
\begin{align}
  \mvec{g}(\mvec{a},\mvec{b}) = \mvec{a}\cdot\mvec{b}.
\end{align}
From this definition, we see that the partial evaluation of the metric tensor is particularly simple:
\begin{align}
  \mvec{a}\cdot\mvec{g}(\bdum,\_) = \mvec{g}(\mvec{a},\_) = \mvec{a}.
\end{align}
We can compute the \emph{components} of the metric tensor as
\begin{align}
  \mvec{g}(\basis{i},\basis{j}) &= \basis{i}\cdot\basis{j} \equiv g_{ij}, \\
  \mvec{g}(\dbasis{i},\dbasis{j}) &= \dbasis{i}\cdot\dbasis{j} \equiv g^{ij}.  
\end{align}
There are two useful alternative expressions for the metric tensor, first in terms of the basis and their duals as
\begin{align}
  \mvec{g} = \basis{i}\otimes\dbasis{i} = \dbasis{i}\otimes\basis{i}.
\end{align}
The second useful expression is
\begin{align}
  \mvec{g} = \nabla\otimes\mvec{x},
\end{align}
where $\nabla$ is the vector derivative operator, and $\mvec{x}$ is the position-vector in space. 
Note that all these expressions are independent of dimensions and applicable to not just 3D space. 
These expressions also show that for Euclidean space
\begin{align}
  \Tr(\mvec{g}) = \sum_{i=1}^D \basis{i}\cdot\dbasis{i} = \sum_{i=1}^D \delta_i^i = D,
\end{align}
where $D$ is the dimension of the space.

Now consider any vector $\mvec{u}$ and write
\begin{align}
  u^i = \mvec{u}\cdot\dbasis{i} = (\mvec{u}\cdot\basis{j})
  (\dbasis{j}\cdot\dbasis{i})
  = g^{ij} u_j.
\end{align}
Similarly, we have
\begin{align}
  u_i = \mvec{u}\cdot\basis{i} = (\mvec{u}\cdot\dbasis{j})
  (\basis{j}\cdot\basis{i})
  = g_{ij} u^j.
\end{align}
This process is sometimes called \emph{raising} and \emph{lowering} of indices, and extends to tensors of any rank. 
In fact, we can also easily show that
\begin{align}
  \dbasis{i} &= g^{ij}\basis{j}, \\
  \basis{i} &= g_{ij}\dbasis{j}.
\end{align}
These expressions are useful to replace the basis vectors for their reciprocals (and vice-versa).

Notice that we must distinguish between a tensor $\mvec{T}$, its \emph{definition}, for example $\mvec{T} = \mvec{u}\otimes\mvec{v}$, and its \emph{evaluation} $\mvec{T}(\mvec{a},\mvec{b})$. 
It is helpful to think of tensors as functions in a programming language: there, also one must distinguish between the \emph{name}, the \emph{definition}, and its \emph{evaluation}.

Finally, we remark that tensors are a very special, but important, class amongst general scalar-valued functions. 
In general, an arbitrary function $f : \vecspace\rightarrow \mathbb{R}$ need not be linear. 
For example, $f(\mvec{u}) = \mvec{u}\cdot\mvec{u}$ is a quadratic function of its input vector and hence is not a tensor.

Second-order tensors, also called \emph{dyads}, appear frequently in mathematical physics. 
We shall define a \emph{dyadic product} as follows. 
Let $\mvec{T}$ be a second-order tensor and $\mvec{u}$ and $\mvec{v}$ be vectors. 
Then the dyadic product is denoted by the $:$ symbol and is defined as
\begin{align}
  \mvec{T} : \mvec{u}\otimes \mvec{v} \equiv
  \mvec{T}(\mvec{u}, \mvec{v}).
\end{align}
In particular, if $\mvec{T} = \mvec{a}\otimes \mvec{b}$, then
\begin{align}
  \mvec{a}\otimes \mvec{b} : \mvec{u}\otimes \mvec{v} =
  (\mvec{a}\cdot\mvec{u}) (\mvec{b}\cdot\mvec{v}).
  \label{eq:dyadic-prod}
\end{align}
Now, let $\mvec{T}$ and $\mvec{G}$ be two dyads. 
Then the above definitions can be used to write the dyadic product in term of the dyad representation in a particular basis as
\begin{align}
  \mvec{T} : \mvec{G}
  =
  \mvec{T} : G_{ij} \dbasis{i}\otimes\dbasis{j}
  =
  \mvec{T}(\dbasis{i},\dbasis{j}) G_{ij}
  =
  T^{ij} G_{ij}.
\end{align}
This operation also shows that $\mvec{T} : \mvec{G} = \mvec{G} : \mvec{T}$.  
The dyadic product is a \emph{rank reducing} operator: it takes two second-order tensors and produces a scalar.

From the definitions, we can also see that the dyadic product with the metric-tensor is particularly simple
\begin{align}
  \mvec{g} : \mvec{T}
  =
  \Tr(\mvec{T}).
\end{align}
From this result, it follows that $\mvec{g}:\mvec{g} = D$, where, as defined before, $D$ is the dimension of the space. 

Besides the dot-product (valid in a space of any dimension), we can also define the cross-product in 3D as follow. 
The cross-product of two vectors is denoted by $\mvec{b}\times\mvec{c}$ and results in a vector. 
Here we will consider a different approach to the cross product by defining a \emph{third}-order tensor $\veps(\mvec{a},\mvec{b},\mvec{c})$ as
\begin{align}
  \veps(\mvec{a},\mvec{b},\mvec{c}) =
  \mvec{a}\cdot(\mvec{b}\times\mvec{c}).
\end{align}
The partial evaluation of this tensor with two of its slots filled gives
\begin{align}
  \veps(\_,\mvec{b},\mvec{c}) = \mvec{b}\times\mvec{c},
\end{align}
that is, the resulting vector has the same components as the cross-product of $\mvec{b}$ and $\mvec{c}$. 
Hence, the second-order tensor that results from filling in the last slot, $\veps(\_,\_,\mvec{c})$, has the property that
\begin{align}
  \mvec{b}\cdot\veps(\_,\bdum,\mvec{c})
  = \veps(\_,\mvec{b},\mvec{c}) = \mvec{b}\times\mvec{c}.
\end{align}
We have thus written the cross-product as a dot-product between a vector and a special second-order tensor. 
As with the dot-product, we can take the cross-product between tensors. 
For example, consider the cross-product between a vector $\mvec{u}$ and a second-order tensor $\mvec{a}\otimes\mvec{b}$. 
We need to specify which of the vectors making up the second-order tensor with which we wish to cross. 
For example,
\begin{align}
  \mvec{u}\times(\bmvec{a}\otimes\mvec{b}) = (\mvec{u}\times\mvec{a})\otimes\mvec{b},
\end{align}
is the cross-product with the first slot of the tensor, and 
\begin{align}
  \mvec{u}\times(\mvec{a}\otimes\bmvec{b}) =\mvec{a}\otimes(\mvec{u}\times\mvec{b}),
\end{align}
is the cross-product with the second slot of the tensor. Applying this operation to a general second order tensor $\mvec{P}$, we can write
\begin{align}
  \mvec{u}\times\mvec{P}(\bdum,\_) =
  P^{mn} (\mvec{u}\times\basis{m})\otimes\basis{n}.
\end{align}
Notice that the cross-product of a vector with a second-order tensor is a second-order tensor. 
If $\mvec{P}$ is \emph{symmetric}, then the second-order tensor
\begin{align}
  \mvec{u}\times\mvec{P}(\bdum,\_) + \mvec{u}\times\mvec{P}(\_,\bdum),
\end{align}
is also symmetric.

%% file: ap-lag-poly.tex
\section{Useful Relations for Laguerre polynomials}

We use normalized Laguerre polynomials, for which the
orthogonality relation is
\begin{align}
  \int_0^\infty e^{-x} L_n(x) L_m(x) \dvol{x} = \delta_{n m}.
\end{align}
The first few Laguerre polynomials are
\begin{align}
  L_0(x) &= 1 \\
  L_1(x) &= 1-x \\
  L_2(x) &= \frac{x^2}{2} - 2x +1.
\end{align}
A useful relation is
\begin{align}
  x \frac{dL_n}{dx} = n L_n(x) - n L_{n-1}(x).
  \label{eq:d-lag}
\end{align}
The recurrence relation for the polynomial is
\begin{align}
  (n+1) L_{n+1}(x) = (2n+1-x) L_n(x) - n L_{n-1}(x).
\end{align}

%% file: ap-agyro.tex
\section{The Agyrotropic Terms in the Evolution of the Gyrotropic
  Distribution Function}\label{app:f0-agyro}

For completeness, we list below the agyrotropic terms in the evolution of the gyrotropic distribution function. 
These terms are denoted by $\mathrm{AG}$ in \eqr{\ref{eq:gyro-f-f0}}. 
One would need to include these terms to add the effects of agyrotropy in the lowest-order PKPM system. 
However, we note that in the examples given in this paper, we do not include these terms, leaving the implementation of these terms for future work. 
We have
\begin{align}
  \mathrm{AG} = \delx\cdot\mvec{M}_\perp
  +
\frac{1}{v_\perp}
  \frac{\partial}{\partial v_\parallel}
  \left( v_\perp
    \frac{1}{2\pi}
    \int_0^{2\pi} 
    a_a^\parallel \bdf d{\theta}
    \right)
  +
  \frac{1}{v_\perp}
  \frac{\partial}{\partial v_\perp}
  \left( v_\perp
    \frac{1}{2\pi}
  \int_0^{2\pi} 
  a_a^\perp \bdf d{\theta}
  \right).
\end{align}
Here, we have defined
\begin{align}
  \frac{1}{2\pi}\int_0^{2\pi} 
  a_a^\parallel \bdf d{\theta}  
  &=
  \mvec{M}_\perp\cdot
  \left[
  \pfrac{\buni}{t}
  +
  (v_\parallel\buni + \mvec{u})\cdot\delx\buni
  \right] 
  - \buni\cdot\left[\mvec{M}_\perp\cdot\delx\mvec{u}\right] \notag \\
  &-
  \frac{v_\perp^2\bdf_2^c}{4}
  \buni\cdot\left( 
  \gvec{\tau}_1\cdot\nabla\gvec{\tau}_1 - \gvec{\tau}_2\cdot\nabla\gvec{\tau}_2
  \right)
  -
  \frac{v_\perp^2\bdf_2^s}{4}
  \buni\cdot\left( 
  \gvec{\tau}_1\cdot\nabla\gvec{\tau}_2 + \gvec{\tau}_2\cdot\nabla\gvec{\tau}_1
  \right),
\end{align}
and
\begin{align}
  v_\perp
  \frac{1}{2\pi}
  \int_0^{2\pi} 
  a_a^\perp \bdf d{\theta}  
  =
  \mvec{M}_\perp
  \cdot
  \left[
  -v_\parallel \pfrac{\buni}{t}
  -v_\parallel(v_\parallel\buni + \mvec{u})\cdot\delx\buni
  +
  \frac{1}{\rho} \delx\cdot\mvec{P}
  - v_\parallel\buni\cdot\delx\mvec{u}
  \right] \notag \\
  -
  \left[
  \frac{v_\perp^2\bdf_2^c}{4}
  \left( 
  \gvec{\tau}_1\otimes\gvec{\tau}_1 - \gvec{\tau}_2\otimes\gvec{\tau}_2
  \right)
  +
  \frac{v_\perp^2\bdf_2^s}{4}
  \left( 
  \gvec{\tau}_1\otimes\gvec{\tau}_2 + \gvec{\tau}_2\otimes\gvec{\tau}_1
  \right)  
  \right]
  : (v_\parallel\delx\otimes\buni + \delx\otimes\mvec{u}).  
\end{align}
Note that the first two Fourier harmonics appear in the agyrotropic terms. 
Hence, to get the lowest order agyrotropic effects, we would need to evolve additional equations, one for each of the $f_{1,2}^{s,c}$ along with evolution equations for the plane perpendicular to the magnetic field through $\gvec{\tau}_1$ and $\gvec{\tau}_2$. 
Alternatively, as we argue in the main text, we could evolve a vector equation for $\mvec{M}_\perp$ and tensor equation for $\mvec{F}_{\perp\perp}$ to fold in the evolution of $\gvec{\tau}_1$ and $\gvec{\tau}_2$ directly into the evolved quantities. 
The procedure for obtaining equations for $\mvec{M}_\perp$ and $\mvec{F}_{\perp\perp}$ would involve multiplying the Vlasov equation in our transformed coordinates, \eqr{\ref{eq:vlasov-cgl}}, by $\mvec{v}_\perp$ and $\mvec{v}_\perp \otimes \mvec{v}_\perp$ and integrating over velocity gyroangle.
These evolution equations for $\mvec{M}_\perp$ and $\mvec{F}_{\perp\perp}$ could then be further expanded in the Laguerre basis in $v_\perp$ to obtain the final coupled system of equations.

%% file: ap-lbo.tex
\section{The PKPM Dougherty Collision Operator}\label{app:LBO}

For the simulations shown here we use the Dougherty Fokker-Planck (D-FPO) collision operator as presented in
\cite{Lenard:1958, Dougherty:1964} that approximates the collisions as a combination of drag and diffusion in velocity space. 
However, unlike the Rosenbluth-Landau Fokker-Planck collision operator, the collision frequency in the D-FPO is independent of velocity. 
Though this assumption of velocity-independent collision frequency is an approximation, the essential features of a collision operator: particle, momentum and energy conservation, an H-theorem, and balance between drag and diffusion, are contained in the D-FPO---see \citet{Hakim:2020} and \citet{JunoThesis:2020} for the theory of the D-FPO in the context of the numerical discretization of this operator. 

The D-FPO operator for self-collisions is: 
\begin{align}
    C[f] = \nu \nabla_{\mvec{v}} \cdot \left ( \left [\mvec{v} - \mvec{u}\right ] f + \frac{T}{m} \nabla_{\mvec{v}} f \right ),
\end{align}
where $\nu$ is the velocity-independent collision frequency, usually taken to be a constant or the Spitzer collisionality \citep{Braginskii:1965}, and as before, we are suppressing species subscripts. 
To manipulate this collision operator into a form which can be included in the PKPM model, we first transform the velocity coordinates to the local flow frame, 
\begin{align}
    C[\bdf] = \nu \nabla_{\mvec{v}'} \cdot \left ( \left [\mvec{v}' + \mvec{u} - \mvec{u}\right ] \bdf + \frac{T}{m} \nabla_{\mvec{v}'} \bdf \right ) = \nu \nabla_{\mvec{v}'} \cdot \left ( \mvec{v}' \bdf + \frac{T}{m} \nabla_{\mvec{v}'} \bdf \right ). 
\end{align}
And if we focus exclusively on the gyrotropic piece of the distribution function, $\ell = 0$, the transformation to $(v_\parallel, v_\perp, \theta)$ coordinates is also reasonably straightforward, 
\begin{align}
    C[\bdf] = \nu \pfrac{}{v_\parallel} \left ( v_\parallel \bdf + \frac{T}{m} \pfrac{\bdf}{v_\parallel} \right ) + \nu \pfrac{}{\xi} \left (2 \xi \left [\bdf + \frac{T}{T_\perp} \pfrac{\bdf}{\xi} \right ]\right ),
\end{align}
where we have transformed the perpendicular velocity derivative to $\xi = m v_\perp^2/2 T_\perp$, the coordinates of the Laguerre expansion. 
This coordinate transformation in the perpendicular velocity is very similar to the transformation to the coordinate $\mu$ utilized in gyrokinetic variations of the D-FPO \citep{Francisquez:2020}. 

With the orthogonality of Laguerre polynomials, the first term evaluates to simply,
\begin{align}
    C_{v_\parallel} [F_{n,0}] & = \int d\xi \exp\left (-\xi\right ) \nu \pfrac{}{v_\parallel} \left ( v_\parallel L_n\left (\xi \right ) \bdf + \frac{T}{m} \pfrac{}{v_\parallel} L_n\left (\xi \right ) \bdf \right ), \notag \\
    & = \nu \pfrac{}{v_\parallel} \left ( v_\parallel F_{n,0} + \frac{T}{m} \pfrac{F_{n,0}}{v_\parallel} \right ).
\end{align}
The second term can be integrated by parts to obtain,
\begin{align}
    \nu\int & d\xi \exp(-\xi) L_n(\xi) \pfrac{}{\xi} \left (2 \xi \left [\bdf + \frac{T}{T_\perp} \pfrac{\bdf}{\xi} \right ]\right ),  \notag \\
    & = -2 \nu\int d\xi \exp(-\xi) \xi \pfrac{L_m(\xi)}{\xi} \left (\left [\bdf + \frac{T}{T_\perp} \pfrac{\bdf}{\xi} \right ]\right ), \notag \\
    & = -2 \nu\int d\xi \exp(-\xi) \left (n L_n (\xi) - n L_{n-1}(\xi) \right ) \left (\left [\bdf + \frac{T}{T_\perp} \pfrac{\bdf}{\xi} \right ]\right ), \notag \\
    & = -2 n \nu \left [ (F_{n,0} - F_{n-1,0}) - \int d\xi \exp(-\xi) \pfrac{}{\xi} \left ( L_n (\xi) - L_{n-1}(\xi) \right ) \frac{T}{T_\perp} \bdf \right ], \notag \\ 
    & = -2 n \nu \left [ (F_{n,0} - F_{n-1,0}) + \frac{T}{T_\perp}  F_{n-1,0} \right ], 
\end{align}
so that the collision operators for the lowest order PKPM system are: 
\begin{align}
    C[F_0] & = \nu \pfrac{}{v_\parallel} \left ( v_\parallel F_{0} + \frac{T}{m} \pfrac{F_{0}}{v_\parallel} \right ), \label{eq:F0-LBO} \\
    C[\mathcal{G}] & = 2\nu \left (\frac{T}{m} F_{0} - \mathcal{G} \right ) + \nu \pfrac{}{v_\parallel} \left ( v_\parallel \mathcal{G} + \frac{T}{m} \pfrac{\mathcal{G}}{v_\parallel} \right ) \label{eq:G-LBO}. 
\end{align}
We can take the $1/2 \thinspace m v_\parallel^2$ moment of \eqr{\ref{eq:F0-LBO}} and the mass-weighted zeroth moment of \eqr{\ref{eq:G-LBO}} to determine how the collision operators affect the evolution of $p_\parallel$ and $p_\perp$, 
\begin{align}
    \frac{1}{2} \pfrac{p_\parallel}{t} = \nu \left (\rho \frac{T}{m} - p_\parallel \right ), \\
    \pfrac{p_\perp}{t} = 2\nu \left (\rho \frac{T}{m} - p_\perp \right ),  
\end{align}
which when summed to obtain the total internal energy equation is
\begin{align}
    \pfrac{}{t} \left ( \frac{3}{2} p \right ) = \nu \left (3 \rho \frac{T}{m} - p_\parallel - 2 p_\perp \right ) = \nu \left (3 p - 3 p \right ) = 0, 
\end{align}
as expected. 

Note that this expression does not imply PKPM lacks viscosity and viscous heating. 
In the high collisionality limit, the D-FPO in these velocity coordinates will modify the pressure tensor, which then couples to the momentum equation. 
This equation merely implies that there is no additional heating due to the collision operator in these velocity coordinates, only scattering between $p_\parallel$ and $p_\perp$. 

We can also include cross-species collision of the form,
\begin{align}
    C[f]_{rs} = \nu_{rs} \nabla_{\mvec{v}} \cdot \left ( \left [\mvec{v} - \mvec{u}_{rs}\right ] f + v_{t,rs}^2 \nabla_{\mvec{v}} f \right ),     
\end{align}
where the subscript $rs$ denotes the collisions of species $s$, the species we are evolving, with species $r$. 
Here, the intermediate flow $\mvec{u}_{rs}$ and intermediate thermal velocity squared $v_{t,rs}^2$ are determined from the constraints that energy and momentum are conserved, along with the relaxation rates such as the Morse relaxation rates \citep{Morse:1963}.
For the purposes of inclusion in the PKPM model, these two parameters can be any of the forms derived in previous works \citep{Greene:1973, Francisquez:2022} and need not be specified at this stage.

Two important changes arise due to the inclusion of cross species collisions.
The first is that cross-species collisions include the effects of momentum-exchange between the two species, i.e., the first velocity moment of the cross-species collision operator is:
\begin{align}
    \int m_s \mvec{v} C[f]_{rs} \thinspace d^3\mvec{v} & = \int m_s \mvec{v}  \nu_{rs} \nabla_{\mvec{v}} \cdot \left ( \left [\mvec{v} - \mvec{u}_{rs}\right ] f + v_{t,rs}^2 \nabla_{\mvec{v}} f \right ) \thinspace d^3\mvec{v}, \notag \\
    & = \nu_{rs} \left (\rho_s \mvec{u}_{rs} - \rho_s \mvec{u} \right ).
\end{align}
Thus, in the boost to the local flow frame, the previous substitution of the pressure forces must be modified to
\begin{align}
    \frac{q}{m} \left (\mvec{E} + \mvec{u} \times \mvec{B} \right ) - \pfrac{\mvec{u}}{t} - \mvec{u} \cdot \nabla \mvec{u} = \frac{1}{\rho} \nabla \cdot \mvec{P} - \nu_{rs} \left (\mvec{u}_{rs} - \mvec{u} \right ).
\end{align}
In addition, this collision operator's velocity coordinates also need to be transformed to the local flow frame,
\begin{align}
    C[f]_{rs} = \nu_{rs} \nabla_{\mvec{v}'} \cdot \left ( \left [\mvec{v}' + \mvec{u} - \mvec{u}_{rs}\right ] \bdf + v_{t,rs}^2 \nabla_{\mvec{v}'} \bdf \right ). 
\end{align}
Thus, there is a cancellation of the forces arising due to the momentum exchange such that,
\begin{align}
    \nabla_{\mvec{v}'} \cdot \left ( \left [ \frac{1}{\rho_s} \nabla \cdot \mvec{P} - \nu_{rs} \left (\mvec{u}_{rs} - \mvec{u} \right ) \right ] \bdf \right ) & =  \nu_{rs} \nabla_{\mvec{v}'} \cdot \left ( \left [\mvec{v}' + \mvec{u} - \mvec{u}_{rs}\right ] \bdf + v_{t,rs}^2 \nabla_{\mvec{v}'} \bdf \right ),  \notag \\
    \nabla_{\mvec{v}'} \cdot \left ( \left [ \frac{1}{\rho_s} \nabla \cdot \mvec{P} \right ] f \right ) & =  \nu_{rs} \nabla_{\mvec{v}'} \cdot \left (  \mvec{v}' f + v_{t,rs}^2 \nabla_{\mvec{v}'} \bdf \right ).
\end{align}
The inclusion of cross-species collisions using the D-FPO only requires the sum over collision frequencies and intermediate thermal velocities. 
The same substitution of the divergence of the pressure tensor cancels the momentum exchange, and therefore, as we expect, the momentum exchange from cross-species collisions is contained solely in the momentum evolution.